\documentclass[structabstract]{aa}
\usepackage{graphicx}
\usepackage{txfonts}
\usepackage{natbib}
\usepackage{url}
\usepackage{color}
\usepackage{rotating}

\begin{document}

\newcommand{\Alii}{\ion{Al}{ii}}
\newcommand{\Aliii}{\ion{Al}{iii}}
\newcommand{\Ari}{\ion{Ar}{i}}
\newcommand{\ci}{\ion{C}{i}}
\newcommand{\cie}{\ion{C}{i}$^\star$}
\newcommand{\ciee}{\ion{C}{i}$^{\star\star}$}
\newcommand{\cii}{\ion{C}{ii}}
\newcommand{\ciie}{\ion{C}{ii}$^\star$}
\newcommand{\oiv}{\ion{O}{iv}}
\newcommand{\ciii}{\ion{C}{iii}}
\newcommand{\pii}{\ion{P}{ii}}
\newcommand{\pv}{\ion{P}{v}}
\newcommand{\nii}{\ion{N}{ii}}
\newcommand{\niii}{\ion{N}{iii}}
\newcommand{\Niii}{\ion{Ni}{ii}}
\newcommand{\civ}{\ion{C}{iv}}
\newcommand{\cuii}{\ion{Cu}{ii}}
\newcommand{\oi}{\ion{O}{i}}
\newcommand{\oiii}{\ion{O}{iii}}
\newcommand{\siiv}{\ion{Si}{iv}}
\newcommand{\siii}{\ion{Si}{ii}}
\newcommand{\feii}{\ion{Fe}{ii}}
\newcommand{\znii}{\ion{Zn}{ii}}
\newcommand{\Sii}{\ion{S}{ii}}
\newcommand{\crii}{\ion{Cr}{ii}}
\newcommand{\mgi}{\ion{Mg}{i}}
\newcommand{\MgII}{\ion{Mg}{ii}}
\newcommand{\hi}{\ion{H}{i}}
\newcommand{\htwo}{\mbox{H$_2$}}
\newcommand{\hd}{\mbox{HD}}
\newcommand{\deut}{\element[][]{D}}
\newcommand{\Lya}{Lyman-$\alpha$}
\newcommand{\Lyb}{Lyman-$\beta$}
\newcommand{\Lyd}{Lyman-$\delta$}
\newcommand{\Lye}{Lyman-$\epsilon$}
\newcommand{\qso}{\textsc{Q~J\,0643$-$5041}}
\newcommand{\zabs}{\ensuremath{z_{\rm{abs}}}}
\newcommand{\zem}{\ensuremath{z_{\rm{em}}}}

\def\dmm{$\Delta \mu/\mu$}
\def\kms{km\,s$^{-1}$}
\def\ms{m\,s$^{-1}$}

\title{Molecular hydrogen in the \zabs=2.66 damped \Lya\ absorber toward \qso\thanks{Based on data obtained with the Ultraviolet and Visual Echelle Spectrograph (UVES) at the European Southern Observatory Very Large Telescope (ESO-VLT), under program~ID~080.A-0288(A) and archival data.}}
\subtitle{Physical conditions and limits on the cosmological variation of the proton-to-electron mass ratio}
\titlerunning{\htwo\ in \qso}
\author{D. Albornoz V\'asquez\inst{1} \and H. Rahmani\inst{2,3} \and P. Noterdaeme\inst{1} \and P. Petitjean\inst{1} \and R. Srianand \inst{2} \and C. Ledoux\inst{4}}

\institute{Institut d'Astrophysique de Paris, UMR 7095 CNRS, Universit\'e Pierre et Marie Curie, 98 bis Boulevard Arago, Paris 75014, France \and Inter-University Centre for Astronomy and Astrophysics, Post Bag 4, Ganeshkhind, Pune 411 007, India \and School of Astronomy, Institute for Research in Fundamental Sciences (IPM), PO Box 19395-5531, Tehran, Iran \and European Southern Observatory, Alonso de C\'ordova 3107, Casilla 19001, Vitacura, Santiago, Chile}

\date{\today}

\abstract
{
Molecular hydrogen in the interstellar medium (ISM) of high redshift galaxies can be detected directly from its UV absorptions imprinted in the spectrum of background quasars. Associated absorptions from \hi\ and metals allow one to study the chemical enrichment of the gas, while the analysis of excited species and molecules make it possible to infer the physical state of the ISM gas. In addition, given the numerous \htwo\ lines usually detected, these absorption systems are unique tools to constrain the cosmological variation of the proton-to-electron mass ratio, $\mu$.
}
{
We intend to study the chemical and physical state of the gas in the \htwo-bearing cloud at $z_{\rm abs}~=~2.658601$ toward the quasar \qso\ ($z_{\rm em}~=~3.09$) and derive a useful constraint on the variation of $\mu$.
}
{
We use high signal-to-noise ratio, high-resolution VLT-UVES data of \qso\ amounting to a total of more than 23 hours exposure time and fit the \hi, metals and \htwo\ absorption features with multiple-component Voigt profiles. We study the relative populations of \htwo\ rotational levels and the fine-structure excitation of neutral carbon to determine the physical conditions in the \htwo-bearing cloud.
}
{
We find some evidence for part of the quasar broad line emission region not being fully covered by the \htwo-bearing cloud. We measure a total neutral hydrogen column density of $\log~N($\hi$)($cm$^{-2})~=~21.03~\pm~0.08$. Molecular hydrogen is detected in several rotational levels, possibly up to J~=~7, in a single component. The corresponding molecular fraction is $\log~f~=~-2.19^{+0.07}_{-0.08}$, where $f~=~2N($\htwo$)/(2N($\htwo$)+N($\hi$))$. The \htwo\ Doppler parameter is of the order of 1.5~\kms\ for J~=~0, 1 and 2 and larger for J~$>$~2. The molecular component has a kinetic temperature of $T_{kin}~\simeq$~80~K, which yields a mean thermal velocity of $\sim1$~\kms, consistent with the Doppler broadening of the lines. The UV ambient flux is of the order of the mean ISM Galactic flux. We discuss the possible detection of \hd\ and derive an upper limit of $\log~N($\hd$)~\lesssim$~13.65$~\pm~$0.07 leading to $\log$~\hd/(2$\times$\htwo)~$\lesssim$~$-$5.19$~\pm~$0.07 which is consistently lower than the primordial D/H ratio. Metals span $\sim$~210~\kms\ with [Zn/H]~=~$-$0.91$\pm$0.09 relative to solar, with iron depleted relative to zinc [Zn/Fe]~=~0.45$\pm$0.06, and with the rare detection of copper. We follow the procedures used in our previous works to derive a constraint on the cosmological variation of $\mu$, \dmm~$=(7.4\pm4.3_{\rm stat}\pm5.1_{\rm syst})\times10^{-6}$.
}
{}

\maketitle


\section{Introduction}
Damped \Lya\ systems (DLA) are the spectral signature of large column densities of neutral hydrogen ($N(\hi)\geq 2\times10^{20}$~cm$^{-2}$) located along the line of sight to bright background sources like QSOs and GRBs.

Because the involved \hi\ column densities are similar to what is seen through galactic discs \citep[e.g.][]{Zwaan2005} and because of the presence of heavy elements \citep[e.g.][]{Pettini1997}, DLAs are thought to be located close to galaxies and their circumgalactic media in which multi-phase gas is expected to be found. However, most of the gas producing DLAs is diffuse ($n<0.1$~cm$^{-3}$) and warm ($T>3000$~K) as evidenced by the scarcity of molecular \citep[e.g.][]{Petitjean2000} and 21-cm \citep{Srianand2012} absorption lines. Indeed, direct measurements (via disentangling the turbulent and kinetic contributions to Doppler parameters of different species in single component absorption systems) report temperatures of the order of $10^4$~K \citep[][]{Noterdaeme2012,Carswell2012}. Such temperatures prevent molecular hydrogen formation on the surface of dust grains. In addition, ultra-violet (UV) background radiation in DLA galaxies is large enough to photodissociate \htwo\ in low-density gas clouds~\citep{Srianand2005a}. Therefore the {\sl average} environment at DLA hosts is not favourable to the formation of \htwo.

Consequently, \htwo\ is detected at $z>1.8$ via Lyman- and Werner-band absorption lines in only about 10\% of the DLAs~\citep{Noterdaeme2008a}, down to molecular fractions of $f = 2 N($\htwo$)/(2N($\htwo$)+N($\hi$)) \simeq 1.0\times10^{-7}$~\citep{Srianand2010}. To our knowledge, 23 high-$z$ \htwo\ detections in the line of sight of distant QSOs have been reported so far (\citealt{Cui2005}; \citealt{Noterdaeme2008a} and references therein; \citealt{Srianand2008}; \citealt{Srianand2010}; \citealt{Noterdaeme2010}; \citealt{Tumlinson2010}; \citealt{Jorgenson2010}; \citealt{Fynbo2011}; \citealt{Guimaraes2012}; \citealt{Srianand2012}). \citet{Petitjean2006} have shown that selecting high metallicity DLAs increases the probability of detecting \htwo. This is the natural consequence of \htwo\ being more easily formed and shielded in dusty environments, as shown by the relation between the \htwo\ detection rate and the depletion factor \citep{Ledoux2003} and that between the column densities of \htwo\ and that of metals missing from the gas phase \citep{Noterdaeme2008a}. The precise estimation of column densities of different \htwo\ rotational levels via Voigt profile fitting as well as the detection of excited states of neutral carbon allows one to study the physical conditions of the gas \citep{Srianand2005b,Noterdaeme2007b}.

In the context of Grand Unified Theories, the fundamental parameter $\mu=m_{\rm p}/m_{\rm e}$, the proton-to-electron mass ratio could take different values at distant space-time position (for an up-to-date review on the variation of fundamental constants refer to~\citet{Uzan2011}). Originally proposed by~\citet{Thompson1975}, a probe of $\mu$ can be achieved in astrophysical absorbing systems by comparing the measured wavelengths of identified ro-vibrational transitions of molecules to their vacuum laboratory-established values. Relative shifts between different lines could be an evidence of $\mu_{\rm{abs}}\neq\mu_{\rm{lab}}$, where $\mu_{\rm{lab}}$ is the laboratory measured value\footnote{In the laboratory: $\mu_{\rm{lab}}=m_p/m_e\simeq1836.1526675(39)$~\citep{Mohr:2000ie}.} and $\mu_{\rm{abs}}$ is its value at the absorbing system. 

This method has been used several times in the past~\citep{Varshalovich1993,Cowie1995,Levshakov2002,Ivanchik2005,Reinhold2006,Ubachs2007,Thompson2009,Wendt2011,Wendt2012,King2011, Rahmani2013}. At present, measurements of \dmm\ using \htwo\ have been performed in 7 \htwo-bearing DLAs at $z>2$ and suggest that \dmm~$< 10^{-5}$ at $2 < z < 3$. At $z<1.0$ a stringent constraint on \dmm\ is obtained using inversion transitions of NH$_3$ and rotational molecular transitions~\citep{Murphy2008,Henkel2009,Kanekar2011}. The best reported limit using this technique is \dmm~=~(3.5$\pm$1.2)$\times$10$^{-7}$~\citep{Kanekar2011}. \citet{Bagdonaite2013} obtained the strongest constraint till date of \dmm~=~(0.0$\pm$1.0)$\times$10$^{-7}$ at $z = 0.89$ using methanol transitions. Tight constraints can be also obtained using 21-cm absorption in conjunction with UV metal lines and assuming all other constants have not changed. \citet{Rahmani2012} could derive \dmm~=~(0.0$\pm$1.50)$\times$10$^{-6}$ using a sample of four 21-cm absorbers at $z<1.3$ and \citet{Srianand2010} have measured \dmm~=~(1.7$\pm$1.7)$\times$10$^{-6}$ at $z = 3.17$ using the 21-cm absorber towards J1337+3152.

In this work we fit the \htwo\ absorption features to determine the physical conditions in the $\zabs=2.6586$ DLA towards \qso. We discuss the detection of \hd\ and give an upper limit on the deuterium abundance in the molecular component. We present a critical analysis of the data, in particular we search for any partial coverage of the QSO broad line region by the intervening cloud and search with great care systematic errors in wavelength calibration having an impact on the precise estimation of the absorbing wavelength $\lambda_{\rm obs}$ of all \htwo\ transitions. We make use of the numerous \htwo\ transitions detected to derive a limit on \dmm.


\section{Observations and data processing}

\begin{figure*}[t]
\hspace*{-2.27cm}	
\includegraphics[scale=0.2,natwidth=6cm,natheight=6cm]{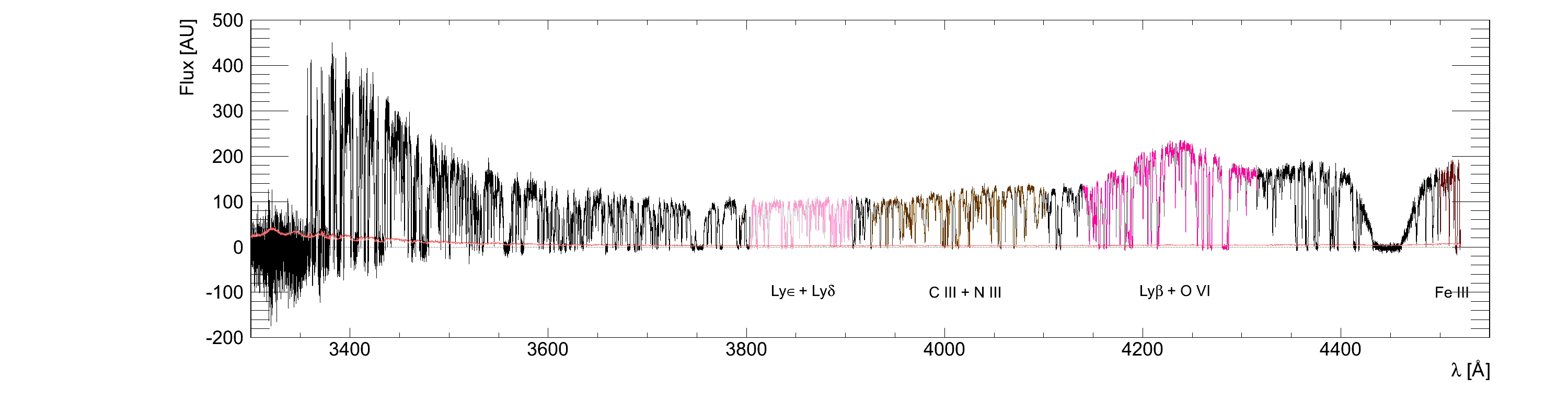}
\hspace*{-2.27cm}	
\includegraphics[scale=0.2,natwidth=6cm,natheight=6cm]{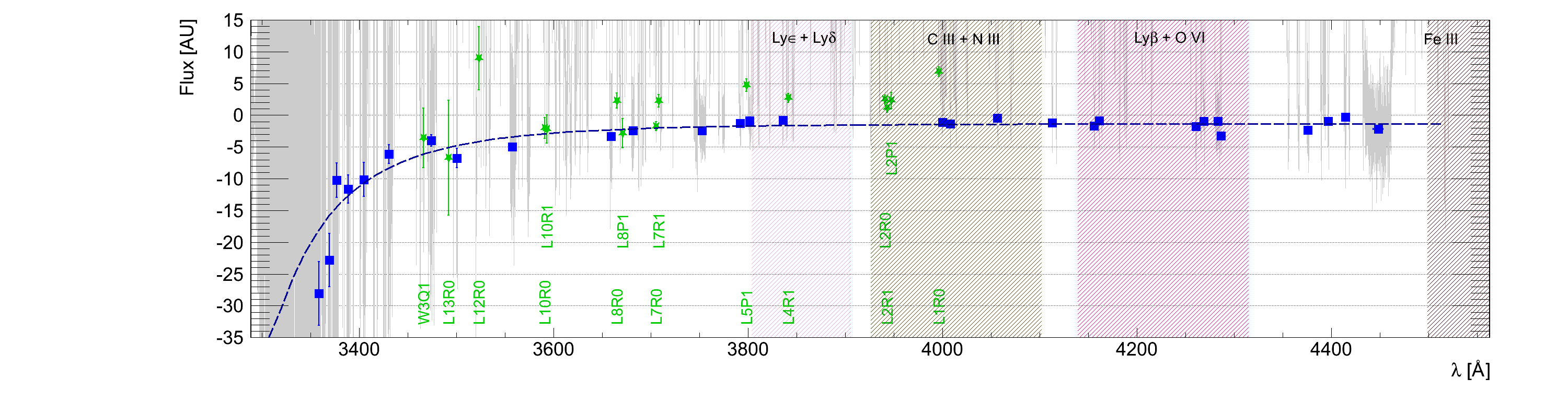}
\caption{Top panel: \qso\ spectrum obtained after combining the 15 exposures with central grating setting at 390~nm and attached ThAr lamp exposure for wavelength calibration. In colors the broad line emission line intervals as expected from the composite emission spectrum of quasars from~\citet{VandenBerk:2001hc} at \zem=3.09. Bottom panel: the median of the selected intervals from saturated \Lya\ forest absorption lines for the analysis of the zero level are shown in blue square points. The error bar on the wavelength is simply the selected interval. The error bar on the flux is the error on the mean measured flux of the selected pixels. The dashed line is the fit to the medians taking into account the errors. Green star points represent the median level at the bottom of saturated and unblended \htwo\ lines. Errors bars are equivalently computed to those of blue points. We show the BLR expected extension as given in ~\citet{VandenBerk:2001hc} in shaded colors, and the flux measured in gray in the background.}
\label{fig:Spec}
\end{figure*}

\begin{table*}[t]
\caption{Summary of \qso\ exposures containing \htwo\ and \hd\ absorption features at \zabs$\simeq2.6586$ using the blue arm of the Ultraviolet and Visual Echelle Spectrograph at ESO-VLT used for analysis in the present work. Other 42 (red arm) exposures with grating central wavelengths of 564, 580, 760 and 850~nm were used for the fits of metallic absorption lines.}
\begin{center}
\begin{tabular}{c c c c c c c c}
\hline
\hline
\# & Program ID & Date & Grating central & DIMM Seeing & Airmass & Exposure time [s] & Wavelength \\
& & & wavelength [nm] & [arcsec]\,$^a$ & & & calibration \\
\hline
1 & 073.A-0071(A) & 17-09-2004 & 390 & 2.3 $\rightarrow$ 1.89 & 1.522 $\rightarrow$ 1.229 & 5500 & ... \\ 
2 & 073.A-0071(A) & 18-09-2004 & 390 & 1.91 $\rightarrow$ 1.9 & 1.561 $\rightarrow$ 1.229 & 6000 & ... \\ 
3 & 073.A-0071(A) & 19-09-2004 & 390 & 0.87 $\rightarrow$ 1.0 & 1.545 $\rightarrow$ 1.222 & 6000 & ... \\ 
4 & 074.A-0201(A) & 09-10-2004 & 390 & ... $\rightarrow$ 0.6 & 1.316 $\rightarrow$ 1.143 & 5800 & ... \\ 
5 & 074.A-0201(A) & 10-10-2004 & 390 & ... $\rightarrow$ 0.37 & 1.291 $\rightarrow$ 1.139 & 5500 & ... \\ 
6 & 080.A-0288(A) & 11-12-2007 & 390 & 0.87 $\rightarrow$ ... & 1.165 $\rightarrow$ 1.279 & 3725 & ThAr attached \\ 
7 & 080.A-0288(A) & 03-01-2008 & 390 & 1.04 $\rightarrow$ 0.91 & 1.182 $\rightarrow$ 1.122 & 3725 & ThAr attached \\ 
8 & 080.A-0288(A) & 04-01-2008 & 390 & 0.82 $\rightarrow$ 1.77 & 1.233 $\rightarrow$ 1.143 & 3725 & ThAr attached \\ 
9 & 080.A-0288(A) & 04-01-2008 & 390 & 1.3 $\rightarrow$ 1.66 & 1.140 $\rightarrow$ 1.113 & 3725 & ThAr attached \\ 
10 & 080.A-0288(A) & 04-01-2008 & 390 & 1.66 $\rightarrow$ 1.69 & 1.113 $\rightarrow$ 1.118 & 1389 & ThAr attached \\ 
11 & 080.A-0288(A) & 06-01-2008 & 390 & 1.24 $\rightarrow$ ... & 1.128 $\rightarrow$ 1.113 & 2104 & ThAr attached \\ 
12 & 080.A-0288(A) & 06-01-2008 & 390 & 0.89 $\rightarrow$ 1.24 & 1.173 $\rightarrow$ 1.295 & 3725 & ThAr attached \\ 
13 & 080.A-0288(A) & 06-01-2008 & 390 & 1.1 $\rightarrow$ 1.08 & 1.311 $\rightarrow$ 1.545 & 3725 & ThAr attached \\ 
14 & 080.A-0288(A) & 07-01-2008 & 390 & 1.07 $\rightarrow$ 0.84 & 1.215 $\rightarrow$ 1.135 & 3725 & ThAr attached \\ 
15 & 080.A-0288(A) & 13-01-2008 & 390 & 0.69 $\rightarrow$ 1.0 & 1.347 $\rightarrow$ 1.201 & 3725 & ThAr attached \\ 
16 & 080.A-0288(A) & 16-01-2008 & 390 & 0.6 $\rightarrow$ 0.85 & 1.121 $\rightarrow$ 1.178 & 3725 & ThAr attached \\ 
17 & 080.A-0288(A) & 06-02-2008 & 390 & ... $\rightarrow$ ... & 1.113 $\rightarrow$ 1.136 & 3725 & ThAr attached \\ 
18 & 080.A-0288(A) & 07-02-2008 & 390 & 0.8 $\rightarrow$ 0.9 & 1.125 $\rightarrow$ 1.190 & 3725 & ThAr attached \\ 
19 & 080.A-0288(A) & 09-02-2008 & 390 & ... $\rightarrow$ 0.8 & 1.113 $\rightarrow$ 1.145 & 3725 & ThAr attached \\ 
20 & 080.A-0288(A) & 10-02-2008 & 390 & 1.24 $\rightarrow$ 0.81 & 1.215 $\rightarrow$ 1.374 & 3725 & ThAr attached \\ 
21 & 072.A-0442(A) & 02-11-2003 & 437 & 2.16 $\rightarrow$ 1.65 & 1.251 $\rightarrow$ 1.153 & 3600 & ... \\ 
\hline
\end{tabular}
\end{center}
\footnotesize
$^a$ some values were not measured (...). \\
\normalsize
\label{tab:Exposures}
\end{table*}

The detection of \htwo\ at \zabs$=2.66$ towards \qso\ ($z_{\rm{em}}=3.09$) was first reported in~\citet{Noterdaeme2008a}. This quasar was observed using both spectroscopic VLT-UVES arms~\citep{Dekker2000} in 2003 and 2004 in the course of programs~IDs~072.A-0442(A) (aimed to study DLAs, PI S.~L\'opez), 073.A-0071(A) (aimed to search for \htwo-bearing systems at high redshift, PI C.~Ledoux) and 074.A-0201(A) (aimed to constrain the variation of the fine-structure constant at high redshift using metal absorption lines, PI R.~Srianand). For these observations, a slit of 1.2 arcsec was used and the CCD was binned 2$\times$2 resulting in a spectral resolving power of $R\sim 40000$ or FWHM~=~$7.5$~\kms\ in the blue arm and $R\sim 36000$ or FWHM~=~$8.3$~\kms\ in the red arm. There was no ThAr lamp attached to the observations. We also use more recent (2007-2008) data, program~ID~080.A-0228(A) aimed to constrain the cosmological variation of $\mu$ (PI P.~Petitjean). A UVB spectrum derived from these observations is shown in Fig.~\ref{fig:Spec}. The settings for the exposures were a 1~arcsec slit and 2$\times$2 binning of the CCD pixels, for a resolving power of $R\sim 45000$ or FWHM~=~$6.7$~\kms\ in the blue arm and $R\sim 43000$ or FWHM~=~$7.0$~\kms\ in the red arm. Wavelength calibration was performed with attached ThAr exposures. Here, the data acquisition was solely focused on \htwo\ transitions. Hence, the \htwo\ lines are apparent in all the observing data sets, while metal lines falling in the red arm ($\lambda > 4500 \AA$) are observed only in the early observations. Blue arm exposures --relevant for detecting \htwo\ absorption lines-- are listed in Table~\ref{tab:Exposures}. 

All the data were reduced using UVES Common Pipeline Library (CPL) data reduction pipeline release 5.3.11\footnote{http://www.eso.org/sci/facilities/paranal/instruments/uves/doc} using the optimal extraction method. We used 4th order polynomials to find the dispersion solutions. The number of suitable ThAr lines used for wavelength calibration was usually more than 700 and the rms error was found to be in the range 70--80 \ms\ with zero average. However, this error reflects only the calibration error at the observed wavelengths of the ThAr lines that are used for wavelength calibration. Systematic errors affecting wavelength calibration should be measured by other techniques that will be discussed later in the paper. All the spectra are corrected for the motion of the observatory around the barycenter of the solar system. The velocity component of the observatory's barycentric motion towards the line of sight to the QSO was calculated at the exposure mid point. Conversion of air to vacuum wavelengths was performed using the formula given in~\citet{Edlen1966}. We interpolate these spectra into an common wavelength array and generate the weighted mean combined spectrum using the inverse square of error as the weight. This is the combined spectrum we use to study the physical condition of this absorption system.

To study of the variation of $\mu$ we only consider those exposures that have attached mode ThAr lamp calibration. We follow a more careful procedure to make a new combined spectrum from these exposures. We start from the final un-rebinned extracted spectrum of each order produced by the CPL. We apply the CPL wavelength solution to each order and merge the orders by implementing a weighted mean in the overlapping regions. In the last step we made use of an uniform wavelength array of step size of 2.0 \kms\ for all exposures. As a result no further rebinning is required for spectrum combination. We fitted a continuum to each individual spectrum and generated a combined spectrum using a weighted mean. We further fit a lower order polynomial to adjust the continuum. This spectrum will be used for \dmm\ measurement. In \citet{Rahmani2013} we found this procedure to produce final combined spectrum consistent with that obtained using UVES$\_$POPLER.


\section{Residual zero-level flux: partial coverage?}
Quasi-stellar objects are compact objects emitting an intrinsic and extremely luminous continuum flux. Broad emission lines seen in the spectra of QSOs are believed to originate from an extended region of $\sim$~pc size, the broad line region (BLR). \citet{Balashev2011} reported that an intervening molecular cloud toward Q1232+082 covers only a fraction of the BLR, most probably because of its compact size. This results in a residual flux detected at the bottom of saturated spectral features associated with the neutral cloud at wavelengths that are coincident with the BLR emission.

Here, we search for a similar effect. For this, we estimate the residual flux at the bottom of saturated \htwo\ lines of the molecular cloud at \zabs$=2.6586$ toward \qso. We first verify the instrumental zero flux level of the spectra, bearing in mind that UVES is not flux calibrated. We estimate the residual flux at the bottom of saturated \Lya-forest lines, associated with large scale gas clouds in the intergalactic medium which are extended enough to cover the background source completely. We first consider 29 sets of pixels defined by intervals in wavelength close to the center of saturated \Lya-forest lines, with no interference with \htwo\ transitions. The median flux value for each interval together with its error is represented by blue points in the bottom panel of Fig.~\ref{fig:Spec}. We fit the observed distribution with a simple function of the wavelength, represented by the dotted line in Fig.~\ref{fig:Spec}, bottom panel, and correct the systematic zero error by subtracting it from the flux.

We estimate the residual flux at the bottom of saturated \htwo\ transitions unblended with other absorption features. We normalise the flux and fit the \htwo\ system. For this analysis, all fitting parameters (redshift, column density and Doppler parameter) of a given rotational level are tied for all lines. The fit shows that for some of the most saturated lines the flux measured at the bottom of the lines is clearly above the Voigt profile absorption model. We select the pixels where the model fit falls below $1\%$ of the emitted flux and plot in Fig.~\ref{fig:Spec}, bottom panel, in green the resulting median flux of such selected pixels. It becomes clear that the measured flux at the bottom of saturated \htwo\ lines is systematically larger than that measured at the bottom of the \Lya\ forest saturated lines (in blue in Fig.~\ref{fig:Spec}, bottom panel).

\begin{figure}[t]
\centering
\hspace*{-0.6cm}
\includegraphics[scale=0.21,natwidth=6cm,natheight=6cm]{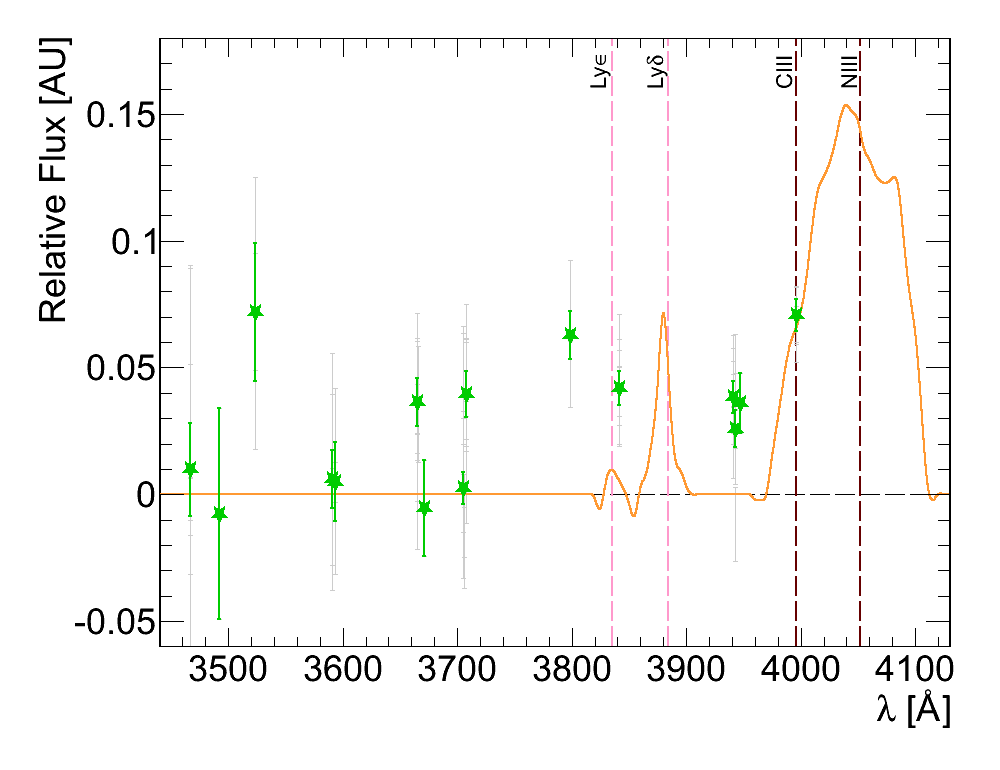}
\caption{Fraction of emitted flux attributed to the broad line region associated with \qso\ is drawn in solid line. We indicate the position of the emission lines at \zem=3.09. Pixels corresponding to saturated \htwo\ transitions are shown in gray, and in green star points their median at each selected line position.}
\label{fig:BLRfit}
\end{figure}

In order to establish whether this residual flux is due to partial coverage of the BLR, we attempt to measure the BLR emission. Unfortunately, because of the difference between emitting and absorbing redshifts, only a few saturated \htwo\ lines fall on top of broad emission lines of \ciii, \niii, \Lyd\ and \Lye. As it can be seen by eye in the top panel of Fig.~\ref{fig:Spec}, the relative amplitude of these features with respect to the total emitted flux is very small. Nonetheless, we estimate the \qso\ continuum emission alone by ignoring the wavelength intervals where the BLR emission is expected as defined in~\citet{VandenBerk:2001hc} (coloured and shaded regions in Fig.~\ref{fig:Spec}). After subtracting this continuum contribution, we are left with the relative BLR emission which is plotted in Fig.~\ref{fig:BLRfit}, solid line. 

In this figure we see that green points, representing the residual flux at the bottom of \htwo\ saturated lines, are systematically above the $1\%$ level. Around the BLR emitting region the effect is possibly larger. Note that the variation does not follow exactly the BLR emission fraction, as is also the case in~\citet{Balashev2011}. The distribution of these points shows roughly two behaviours. Below $\sim3750~\AA$ the points are distributed around $\simeq 1\%$, and above $\sim3750~\AA$ they are distributed around $\simeq 5\%$, with altogether larger dispersion towards the blue where the signal-to-noise ratio (SNR) is lower. A closer look at the distribution of these points shows that there are two distributions, one peaked at zero flux as it should be for a full absorption, and another peaked at $\simeq 4\%$ of the total flux. While all points, on the BLR emission lines or not, contribute to the $\simeq 4\%$ peak, only those not associated with BLR emission peak at zero flux. It is therefore possible that part of the BLR is not covered. The only absorption line located on top of an important broad emission line is L1R0 falling on top of \ciii\ (see Fig.~\ref{fig:BLRfit}). It has the highest residual of all absorption lines. Actually the residual flux matches exactly the relative flux attributed to \ciii, however, this is probably a chance coincidence.

In conclusion, we can say that, although not stricking, there is weak evidence for the BLR to be partly covered by the \htwo\ bearing cloud (see also Section~\ref{sec:CIfine}). A higher SNR spectrum would be required to confirm or refute the partial coverage. The effect being poorly determined and globally small, we do not attempt to take it into account when fitting \htwo\ lines with Voigt profiles. In particular, the saturated lines we use can be fitted in the wings of the absorption profile where partial coverage does not have any impact. Moreover, metal profiles are dominated by strong and saturated components stemming from diffuse extended gas.

\section{Fit of the absorption profiles}
We proceed by studying the various absorption lines associated with the DLA, fitting Voigt profiles with VPFIT 10.0~\footnote{http://www.ast.cam.ac.uk/~rfc/vpfit.html}. The reference absorption redshift (defining the origin of 
our adopted velocity scale) is set to the position of the detected \ci\ component, i.e. $z_{abs}=2.65859(7)$.

\subsection{\hi\ content}
In order to determine the \hi\ absorption profile, we checked the structure of the absorption profiles of other elements thought to be associated to \hi\ such as \cii\ and \oi. Both of these elements feature a multicomponent saturated pattern spanning about 200~\kms\ (see Fig.~\ref{fig:Metals}). It is therefore difficult to have a precise \hi\ absorption decomposition, hence to have a good estimate of the \hi~ column density at the position of the molecular cloud. A main component is fixed at $z_{\rm abs}=2.65859(7)$, and the fit to \Lya\ and \Lyb\ gives $N_{\rm main}(\hi)=10^{21.03\pm0.08}\,$cm$^{-2}$, which is consistent with all other \hi\ transitions (see Fig.~\ref{fig:DLA}). Other nine components (with $N_{\rm other}(\hi) < 10^{17.3}\,$cm$^{-2}$) were added to match the \hi\ profile down to Lyman-7.

\subsection{Metal content \label{sec:metals}}
Low ionisation species --\feii, \crii, \siii, \znii, \Sii, \mgi, \pii, \cuii\ and \Niii-- are detected in a multicomponent absorption pattern spanning about $\sim$210~\kms, with roughly a 125~\kms\ extension towards the blue and 85~\kms\ toward the red, with respect to the carbon and molecular component at $z=2.65859(7)$. The absorption profiles are fitted with 21 velocity components. After a first guess is obtained, we perform a fit with tied redshifts and Doppler parameters between different species. The latter is taken as the combination of a kinematical term for a fixed temperature of 10$^4$~K and a turbulent term which is the same for all species. The resulting profile can be seen in Fig.~\ref{fig:Metals} and the parameters of the fit are given in Table~\ref{tab:metals}. The turbulent term is the major component of the Doppler parameters. We caution that the \znii\ column densities over the range $v=-90$, $-25$~\kms\ are probably overestimated as they appear relatively strong compared to other species. This is likely due to contaminations from telluric features and sky line residuals. Taking into account the noise in the portions of spectra relevant to \znii, the 3$\sigma$ detection limit is at $\log N \simeq 11.47$, showing that the blue components of \znii\ are below the noise level. We note that using the 3\,$\sigma$ upper-limits instead of the fitted column densities has negligible influence on the derived total column density of zinc. Interestingly, we detect weak but significant \cuii\,$\lambda$1358 absorption, consistently with non-detection of the weaker \cuii\,$\lambda$1367. The profile follows well that of other species and we measure a total column density of $\log N($\cuii$) = 12.41~\pm~0.06$. To our knowledge, this is only the second detection of copper in a DLA \citep[see][]{Kulkarni2012}.

\begin{sidewaystable*}
\vspace*{18.61cm}
\begin{center}
\caption{Metal components at the absorbing system at $z_{\rm abs}=2.659$ toward \qso obtained with multicomponent fit using VPFit. Doppler parameters are those of \siii.}
\begin{tabular}{c c c c c c c c c c}
\hline
\hline
& & & \znii & \Sii & \siii & \crii & \feii & \Niii & \cuii \\
\# & $\Delta v$ & $b$ & $\log$(N) & $\log$(N) & $\log$(N) & $\log$(N) & $\log$(N) & $\log$(N) & $\log$(N) \\
& [km\,s$^{-1}$] & [km\,s$^{-1}$] & [cm$^{-2}$] & [cm$^{-2}$] & [cm$^{-2}$] & [cm$^{-2}$] & [cm$^{-2}$] & [cm$^{-2}$] & [cm$^{-2}$] \\
\hline
1 & -122.9~$\pm$~0.1 & 8.5~$\pm$~0.2 & 11.70~$\pm$~0.11 & 13.80~$\pm$~0.18 & 13.10~$\pm$~0.01 & 12.37~$\pm$~0.14 & 12.54~$\pm$~0.03 & 12.05~$\pm$~0.17 & ... \\
2 & -100.0~$\pm$~1.1 & 3.7~$\pm$~0.5 & 11.28~$\pm$~0.23$^a$ & 13.49~$\pm$~0.29 & 12.99~$\pm$~0.21 & 11.61~$\pm$~0.82 & 12.65~$\pm$~0.21 & 11.78~$\pm$~0.40 & ... \\
3 & -96.0~$\pm$~0.4 & 2.8~$\pm$~0.5 & ... & ... & 13.33~$\pm$~0.10 & 11.89~$\pm$~0.42 & 12.99~$\pm$~0.10 & 11.83~$\pm$~0.36 & ... \\
4 & -87.4~$\pm$~1.1 & 5.2~$\pm$~1.1 & 11.77~$\pm$~0.10 & 13.16~$\pm$~0.65 & 13.08~$\pm$~0.07 & ... & 12.68~$\pm$~0.08 & 11.94~$\pm$~0.21 & ... \\
5 & -77.1~$\pm$~0.5 & 2.4~$\pm$~0.6 & 11.29~$\pm$~0.26$^a$ & 13.82~$\pm$~0.19 & 12.73~$\pm$~0.11 & 12.07~$\pm$~0.22 & 12.47~$\pm$~0.08 & ... & ... \\
6 & -68.8~$\pm$~1.2 & 5.3~$\pm$~2.0 & 11.14~$\pm$~0.39$^a$ & 13.73~$\pm$~0.27 & 12.78~$\pm$~0.23 & 11.74~$\pm$~0.55 & 12.30~$\pm$~0.30 & 11.94~$\pm$~0.26 & ... \\
7 & -56.5~$\pm$~3.9 & 8.7~$\pm$~5.5 & ... & 13.55~$\pm$~0.53 & 12.95~$\pm$~0.42 & 12.17~$\pm$~0.44 & 12.61~$\pm$~0.41 & 12.03~$\pm$~0.39 & ... \\
8 & -45.9~$\pm$~1.6 & 8.3~$\pm$~1.3 & 11.37~$\pm$~0.27$^a$ & 13.25~$\pm$~0.90 & 13.26~$\pm$~0.17 & 12.44~$\pm$~0.20 & 12.91~$\pm$~0.17 & 11.76~$\pm$~0.62 & ... \\
9 & -27.7~$\pm$~0.6 & 5.0~$\pm$~0.7 & 11.17~$\pm$~0.37$^a$ & 13.77~$\pm$~0.21 & 13.33~$\pm$~0.08 & 12.10~$\pm$~0.18 & 12.92~$\pm$~0.08 & 12.17~$\pm$~0.10 & ... \\
10 & -20.6~$\pm$~0.3 & 2.4~$\pm$~0.3 & 10.71~$\pm$~0.97$^a$ & 13.50~$\pm$~0.39 & 13.82~$\pm$~0.06 & 11.95~$\pm$~0.24 & 13.24~$\pm$~0.05 & 12.06~$\pm$~0.13 & ... \\
11 & -11.9~$\pm$~0.7 & 8.3~$\pm$~2.2 & 10.93~$\pm$~0.79$^a$ & 13.85~$\pm$~0.21 & 13.64~$\pm$~0.10 & 12.18~$\pm$~0.19 & 13.15~$\pm$~0.10 & 12.21~$\pm$~0.13 & 11.62~$\pm$~0.13 \\
12 & 0.7~$\pm$~0.1 & 2.9~$\pm$~0.1 & 12.10~$\pm$~0.06 & 14.57~$\pm$~0.07 & 14.84~$\pm$~0.02 & 12.53~$\pm$~0.06 & 14.39~$\pm$~0.03 & 12.96~$\pm$~0.02 & 11.44~$\pm$~0.12 \\
13 & 17.0~$\pm$~0.5 & 7.1~$\pm$~0.5 & 11.48~$\pm$~0.31 & 14.37~$\pm$~0.08 & 14.76~$\pm$~0.04 & 12.71~$\pm$~0.07 & 14.43~$\pm$~0.04 & 13.00~$\pm$~0.04 & 11.70~$\pm$~0.09 \\
14 & 25.2~$\pm$~0.7 & 2.7~$\pm$~0.5 & ... & 14.27~$\pm$~0.11 & 14.11~$\pm$~0.15 & 12.34~$\pm$~0.12 & 13.90~$\pm$~0.12 & 12.56~$\pm$~0.10 & 11.19~$\pm$~0.23 \\
15 & 34.5~$\pm$~0.7 & 3.9~$\pm$~1.4 & 10.87~$\pm$~0.85$^a$ & 13.83~$\pm$~0.16 & 13.68~$\pm$~0.17 & 12.21~$\pm$~0.14 & 13.35~$\pm$~0.15 & 12.10~$\pm$~0.17 & 11.38~$\pm$~0.15 \\
16 & 48.4~$\pm$~1.1 & 7.2~$\pm$~1.0 & 11.80~$\pm$~0.17 & 14.35~$\pm$~0.14 & 14.83~$\pm$~0.09 & 12.90~$\pm$~0.09 & 14.48~$\pm$~0.10 & 13.15~$\pm$~0.09 & 11.64~$\pm$~0.13 \\
17 & 56.7~$\pm$~0.6 & 5.3~$\pm$~1.0 & 11.63~$\pm$~0.21 & 14.54~$\pm$~0.14 & 14.75~$\pm$~0.12 & 12.57~$\pm$~0.25 & 14.29~$\pm$~0.15 & 13.00~$\pm$~0.15 & 10.66~$\pm$~1.35 \\
18 & 66.7~$\pm$~2.5 & 10.5~$\pm$~1.9 & ... & 14.34~$\pm$~0.22 & 13.88~$\pm$~0.23 & 12.69~$\pm$~0.30 & 13.51~$\pm$~0.17 & 12.86~$\pm$~0.17 & 11.56~$\pm$~0.17 \\
19 & 67.1~$\pm$~0.5 & 4.2~$\pm$~0.5 & 11.67~$\pm$~0.19 & ... & 14.46~$\pm$~0.07 & 12.11~$\pm$~0.57 & 14.18~$\pm$~0.05 & 12.59~$\pm$~0.11 & ... \\
20 & 72.5~$\pm$~4.8 & 2.4~$\pm$~1.6 & 11.18~$\pm$~0.47$^a$ & 13.90~$\pm$~0.28 & 13.59~$\pm$~0.24 & ... & ... & 8.91~$\pm$~10.64 & ... \\
21 & 85.6~$\pm$~0.9 & 5.6~$\pm$~0.7 & 10.76~$\pm$~1.44$^a$ & 13.75~$\pm$~0.22 & 13.10~$\pm$~0.08 & 12.49~$\pm$~0.10 & 12.63~$\pm$~0.08 & 11.81~$\pm$~0.19 & 11.18~$\pm$~0.22 \\
\hline
$\sum$ & & & $12.75~\pm~0.05$ & $15.35~\pm~0.05$ & $15.52~\pm~0.03$ & $13.66~\pm~0.06$ & $15.14~\pm~0.04$ & $13.84~\pm~0.04$ & $12.41~\pm~0.06$ \\
\hline
$\left[\frac{\rm{X}}{\rm{H}}\right]$ & & & $-0.91~\pm~0.09$ & $-0.87~\pm~0.09$ & $-1.05~\pm~0.09$ & $-1.02~\pm~0.10$ & $-1.36~\pm~0.09$ & $-1.41~\pm~0.09$ & $-0.88~\pm~0.10$ \\
$\left[\frac{\rm{X}}{\rm{Zn}}\right]$ & & & & $0.03~\pm~0.07$ & $-0.14~\pm~0.06$ & $-0.11~\pm~0.07$ & $-0.45~\pm~0.06$ & $-0.50~\pm~0.06$ & $0.03~\pm~0.08$ \\
$\left[\frac{\rm{X}}{\rm{S}}\right]$ & & &$-0.03~\pm~0.07$ & & $-0.18~\pm~0.06$ & $-0.14~\pm~0.07$ & $-0.49~\pm~0.06$ & $-0.53~\pm~0.06$ & $-0.01~\pm~0.07$ \\
\hline
\end{tabular}
\end{center}
\footnotesize
$^a$ column density below 3$\sigma$ detection.\\
\normalsize
\label{tab:metals} 
\end{sidewaystable*}

Highly ionised species --\civ, \siiv\ and \Aliii-- are detected with a broad profile with maximum optical depth at $\sim$ 45~\kms\ towards the red from the molecular component (corresponding to components number 16 and 12 respectively, see Table~\ref{tab:metals}). The latter component is quite weak compared to the strongest component. Highly ionised species are, hence, much more present redwards of the molecular component. Voigt profile fits to these elements are displayed in Fig.~\ref{fig:Metals_h}.

\subsection{Molecular hydrogen}
\begin{table}[t]
\caption{\htwo, \ci\ and \ciie column densities in the z$_{\rm abs}=2.6586$ \htwo-bearing cloud toward \qso.}
\begin{center}
\begin{tabular}{l c c c c}
\hline
\hline
Transition & $\Delta v\,^a$ & $\log\left(N\right)$ & $\log\left(N_{\lim}\right)$ & $b$ \\
 & [\kms] & [cm$^{-2}$] & [cm$^{-2}$] & [\kms] \\
\hline
\htwo~(J~=~0) & 0.25~$\pm$~0.08 & 18.22~$\pm$~0.01 & ... & 1.59~$\pm$~1.10$\,^d$ \\
\htwo~(J~=~1) & 0.25~$\pm$~0.08 & 18.25~$\pm$~0.01 & ... & 1.39~$\pm$~0.70$\,^d$ \\
\htwo~(J~=~2) & 0.33~$\pm$~0.08 & 16.62~$\pm$~0.12 & ... & 1.41~$\pm$~0.29$\,^d$ \\
\htwo~(J~=~3) & 0.33~$\pm$~0.08 & 14.84~$\pm$~0.05 & ... & 2.14~$\pm$~0.61$\,^d$ \\
\htwo~(J~=~4) & 0.57~$\pm$~0.25 & 13.94~$\pm$~0.02 & ... & 3.45~$\pm$~1.01$\,^d$ \\
\htwo~(J~=~5) & 0.25$\,^b$ & 13.86~$\pm$~0.07 & ... & 9.49~$\pm$~2.37 \\
\htwo~(J~=~6) & 0.25$\,^b$ & 13.40~$\pm$~0.16 & 13.7 & 9.49$\,^b$ \\
\htwo~(J~=~7) & 0.25$\,^b$ & 13.35~$\pm$~0.13 & 13.5 & 9.49$\,^b$ \\
\hline
C I & 0~$\pm$~0.25 & 12.57~$\pm$~0.09 & ... & 0.71~$\pm$~0.29 \\
C I* & 0~$\pm$~0.25$\,^c$ & 12.47~$\pm$~0.06 & ... & 0.71~$\pm$~0.29$\,^c$ \\
C I** & 0~$\pm$~0.25$\,^c$ & 11.48~$\pm$~0.30 & 11.95 & 0.71~$\pm$~0.29$\,^c$ \\
\hline
C II* & 0.82~$\pm$~0.08 & 14.12~$\pm$~0.04 & ... & 1.30~$\pm$~0.04 \\
\hline
\end{tabular}
\end{center}
\footnotesize
$^a$ velocity relative to \ci.\\
$^b$ fixed value.\\
$^c$ tied parameter.\\
$^d$ statistical spread rather than fit error.\\
\normalsize
\label{tab:H2CI}
\end{table}

Absorption features of molecular hydrogen fall exclusively in the blue arm. We use the ThAr attached wavelength calibrated spectrum to fit the molecular transition lines with Voigt profiles.

\htwo\ is detected in a single component for rotational levels J~=~0 to 5. The Voigt profile fit is not always satisfactory, in particular for J~$=0$ and J~$=1$ saturated lines. As was already stated, there is often residual flux at the bottom of the saturated lines. The detection is secured for J~=~0 to 5 with a reliable fit. For J~$=6$, the L4P6 and L5P6 are detected at the 1.7$\sigma$ confidence level only. The 3$\sigma$ level upper limit of the column density is $N($\htwo$,\,$J=$6)\leq10^{13.65}$cm$^{-2}$. Regarding J~$=7$, the L5R7 is detected at the 2.1$\sigma$ confidence level only. The 3$\sigma$ column density upper limit is $N($\htwo$,\,$J=$7)\leq10^{13.51}$cm$^{-2}$. Nonetheless a fit to these rotational levels is performed with fixed Doppler parameter (to the value of the parameter of J~$=5$) which is in agreement with the upper limits. A summary of the fit to the \htwo\ lines is given in Table~\ref{tab:H2CI}, while a selection of Voigt profile fits to the most isolated lines can be found in Figures~\ref{fig:H2J0},~\ref{fig:H2J1},~\ref{fig:H2J2},~\ref{fig:H2J3} and~\ref{fig:H2J45}.

The total column density is $N$(\htwo)~$=10^{18.540\pm0.005}~\rm{cm}^{-2}$ with 99\% of this amount in the first two rotational levels. The molecular fraction $f=2N$(\htwo)/($N$(\hi)+2$N$(\htwo)) is therefore $\log f=-2.19^{+0.07}_{-0.08}$ when taking into account all atomic hydrogen. Since the atomic hydrogen column density at the position of the molecular cloud is probably lower than the total measured column density, we get $f\gtrsim10^{-2.19}$. A molecular fraction of $\log f=-2.19$ is comparable to the average molecular fractions recorded in high redshift DLA systems with $\log f > -4.5$~\citep[see][]{Noterdaeme2008a} (with 10 systems satisfying this condition, the mean is $\langle\log f\rangle \simeq -2.3$).

\subsection{Deuterated molecular hydrogen}
\begin{figure}[t]
\centering
\hspace*{-0.61cm}
\includegraphics[scale=0.23,natwidth=6cm,natheight=6cm]{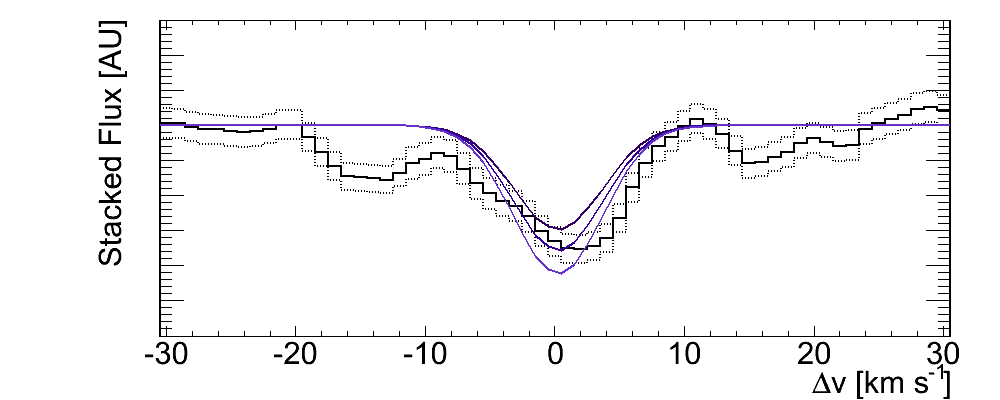}
\hspace*{-0.61cm}
\includegraphics[scale=0.23,natwidth=6cm,natheight=6cm]{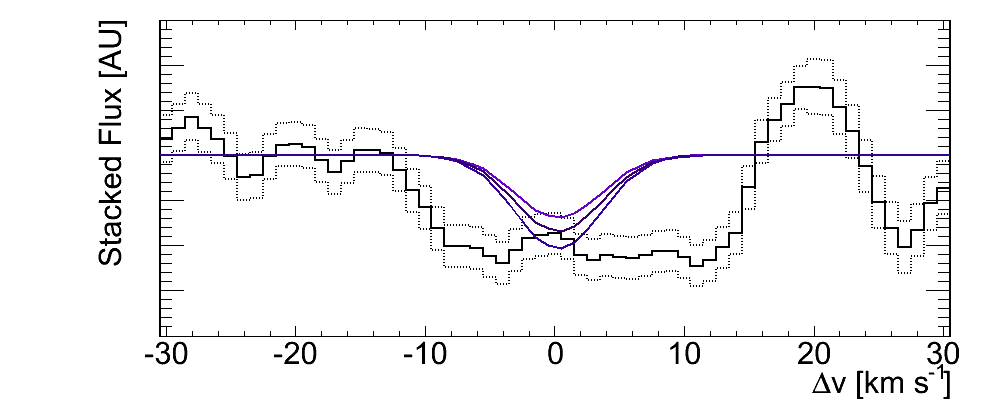}
\caption{Stacked spectra for \hd\ J~$=0$ (7 lines, upper panel) and J~$=1$ (5 lines, lower panel) transitions (solid), together with the error (dotted). We also provide synthetic Voigt profile models with tied Doppler parameters and redshifts to those of \htwo\ J~$=0,\,1$, for $\log N[\rm{cm}^{-2}]=13.3,\,13.4,\,13.5$ and $\log N[\rm{cm}^{-2}]=13.2,\,13.3,\,13.4$ respectively.}
\label{fig:HDstacked}
\end{figure}

We study the best \hd\ transitions available, normalising the flux locally and accounting for obvious contaminations. We select 7 J~$=0$ transitions (L5R0, L7R0, L12R0, L13R0, W0R0, W3R0 and W4R0), and 5 J~$=1$ transitions (L2P1, L4P1, L7R1, L9P1 and W0R1). The spectral portions around these transitions are shown in Figs.~\ref{fig:HDJ0} and~\ref{fig:HDJ1} together with an indicative Voigt profile model. Then we stacked these isolated spectra, weighting the transitions by the local SNR and by the oscillator strength of the transition. Finally, we compare this stack with a synthetic spectrum of the transitions with the same weights, and with redshift and Doppler parameters set to the \htwo\ J~$=0,\,1$. The result is shown in Fig.~\ref{fig:HDstacked}. For J~$=0$ we clearly see a line, while for J~$=1$ the obtained feature seems shapeless and is dominated by unsolved contaminations. The best column densities to match the obtained spectra are $N($\hd$,\,$J=$0)\simeq10^{13.4~\pm~0.1}$cm$^{-2}$ and $N($\hd$,\,$J=$1)\simeq10^{13.3~\pm~0.1}$cm$^{-2}$. These derived column densities should be considered as upper limits, i.e, $N($\hd$)\lesssim10^{13.65~\pm~0.07}$cm$^{-2}$. A higher SNR is needed to confirm the detection.

\subsection{Neutral and singly-ionised carbon}
Neutral carbon is detected in a single component at a redshift close to molecules but 250~\ms\ blueshifted. Slight differences, of the order of less than 1~\kms\ in the position of \htwo\ and \ci\ features have already been observed~\citep[see][]{Srianand2012}. In our case this shift is observed over a large range of wavelengths and for several neutral carbon transitions, hence the effect is real. \ci\ and \htwo\ absorption is expected to arise in the same gas because the two species are sensitive to the same radiation, but this is only true on the first order and self-shielding effects can be subtle so that the absorption profiles can still bear small differences in shape and distribution~\citep[see e.g.][]{Srianand1998}. Therefore, it is not so surprising to observe this small 250~\ms\ shift in the centroid of the two absorbing systems.

The column densities of the three sub-levels of the ground state are respectively, $N($\ci$)=10^{12.56\pm0.07}\,$cm$^{-2}$ for the 2s$^2$2p$^2\,^3$P$^e_0$ level, $N($\cie$)=10^{12.41\pm0.14}\,$cm$^{-2}$ for the 2s$^2$2p$^2\,^3$P$^e_1$ level and $N_{\lim}($\ciee$) < 10^{11.95}\,$cm$^{-2}$ for the 2s$^2$2p$^2\,^3$P$^e_2$ level. From the measured $N($\ci$)$ and $N($\cie$)$ column densities, and for standard conditions, we would expect the \ciee\ level to have $N($\ciee$) \lesssim 10^{11.7}\,$cm$^{-2}$~\citep[see][Figure~8]{Noterdaeme2007a}, which is consistent with the observations. 

Ionised carbon in the excited state \ciie\ (2s$^2$2p$\,^2$P$^o_{3/2}$) is also detected in a single well defined component, yielding $N($\ciie$)=10^{14.11\pm0.04}\,$cm$^{-2}$, redshifted with respect to the neutral component by about 400$\,$m$\,$s$^{-1}$. \cii\ is also detected, however all components with $\Delta v \geq-30$~\kms\ are strongly saturated, hence no fit could be achieved.

The results of the above fits are listed in Table~\ref{tab:H2CI}, and the spectral features can be seen in Figure~\ref{fig:Carbon}.


\section{Physical characteristics of the absorption system at $\zabs\simeq2.6586$}

\subsection{Metallicity, depletion and dust content}\label{sec:depdust}
\begin{figure}[t]
\centering
\hspace*{-0.6cm}
\includegraphics[scale=0.21,natwidth=6cm,natheight=6cm]{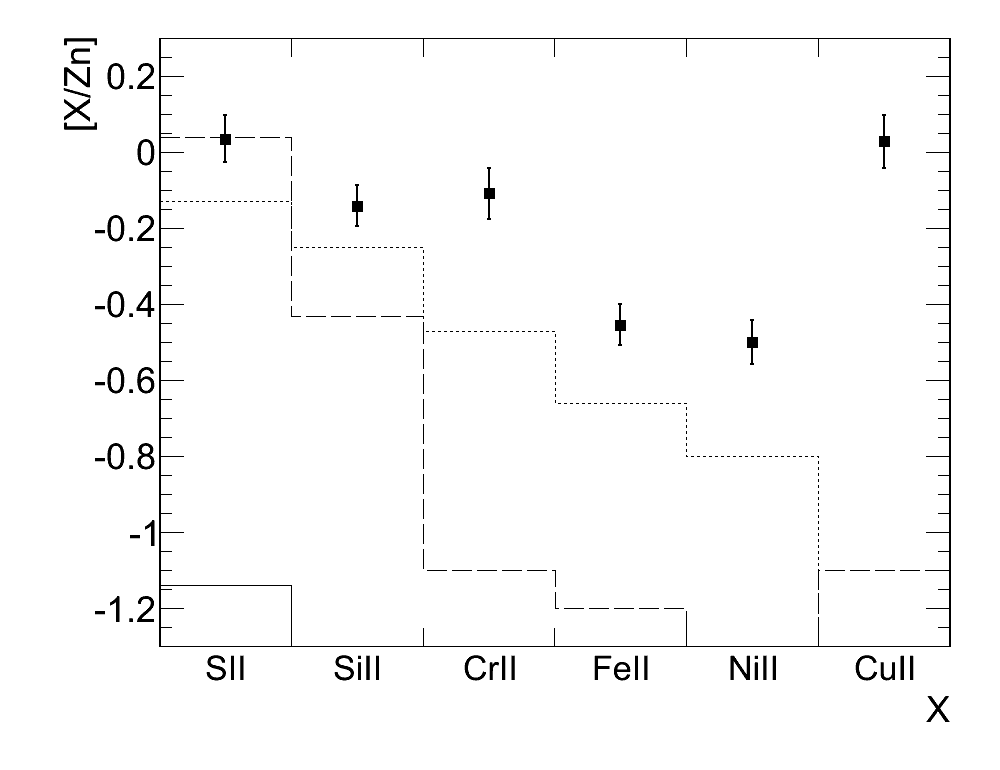}
\caption{Depletion factor, [X/Zn], of different species relative to zinc calculated by integrating the column densities over the whole profile. Dotted, dashed and solid lines correspond to depletion factors observed in, respectively, the halo, the warm and cold ISM of the Galaxy~\citep{Welty1999}. Note that the depletion factor observed in the Galaxy for copper is unavailable.}
\label{fig:Depletion_total}
\end{figure}

\begin{figure}[t]
\centering
\hspace*{-0.61cm}
\includegraphics[scale=0.23,natwidth=6cm,natheight=6cm]{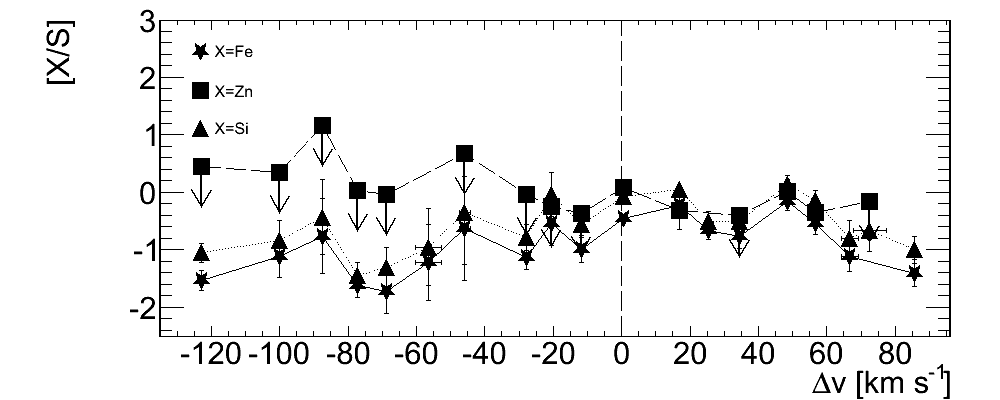}
\hspace*{-0.61cm}
\includegraphics[scale=0.23,natwidth=6cm,natheight=6cm]{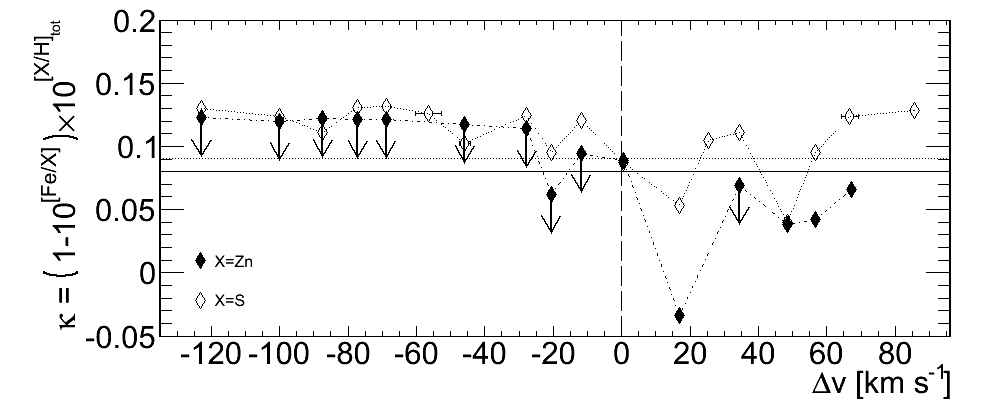}
\hspace*{-0.61cm}
\includegraphics[scale=0.23,natwidth=6cm,natheight=6cm]{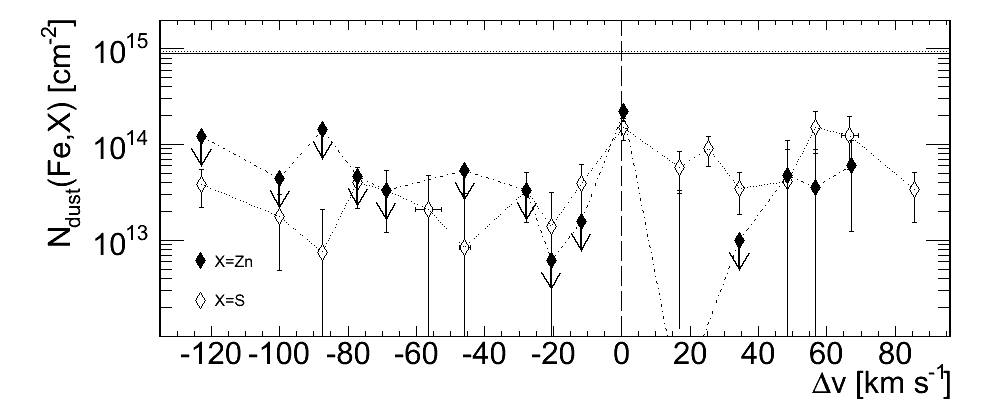}
\caption{Depletion and dust patterns through the absorption profile. Top panel: depletion of Fe, Zn and Si with respect to S. Middle panel: dust to gas ratio computed with respect to Zn and S; the horizontal lines show the value of $\kappa$ obtained when summing all components, using zinc (solid) and sulphur (dotted). Bottom panel: iron column density in the dust phase computed with respect to Zn and S; horizontal lines mark the integrated value over all the profile, using zinc (solid) and sulphur (dotted). Vertical dashed lines show the \ci\ and \htwo\ component at $z=2.65859(7)$.}
\label{fig:Depletion}
\end{figure}

Integrating the column densities over all components, we derive, [Zn/H]=$-$0.91$\pm$0.09\footnote{By definition [X/Y]$\equiv \log [N($X$)/N($Y$)]- \log[N($X$)/N($Y$)]_{\bigodot}$ where all solar values are taken from~\citet{Lodders2003}.} and [S/H]=$-$0.87$\pm$0.09. Again, the very good agreement between these two metallicity estimators shows that the estimated \znii\ total column density is exact. A metallicity of [X/H]$\sim$ $-$0.9 is slightly high for a $z\sim2.5$ DLA~\citep[][]{Prochaska2003}, to be compared to a median metallicity of $\sim$ $-$1.1. This supports the idea that \htwo\ is expected to be found in relatively high metallicity gas~\citep{Petitjean2006}.

We now compare the relative abundances of elements to determine the depletion in this system. The relative amount of dust can be studied by the ratio of volatile versus refractory elements~\citep[e.g.][]{Pettini1994}. [S/Fe] and [Zn/Fe] are usually used for this purpose~\citep[e.g.][]{Vladilo1998,Centurion2000}, bearing in mind that sulphur is an $\alpha$-element hence its relative abundance may suffer some enhancement, and that whether zinc behaves as an $\alpha$-element or an iron-peak element is not fully understood~\citep[see][]{Rafelski2012}. While \znii\ transitions are weaker and blended with \mgi\ and \crii, \Sii\ transitions fall in the \Lya\ forest and in a portion of the spectrum where the SNR is poor. We choose to present the analysis of depletion and dust content using \znii, bearing in mind that the components in the blue side of the absorption profile should be taken as upper limits (see Sect.~\ref{sec:metals}).

Depletion of iron and silicon relative to zinc is mild, [Fe/Zn]$=$-$0.45\pm0.06$ and [Si/Zn]$=$-$0.14\pm0.06$ when summing up all components, in good agreement with [Fe/S] and [Si/S] (see Table~\ref{tab:metals}). The overall depletion pattern is shown in Fig.~\ref{fig:Depletion_total} and is consistent with what is seen in the galactic halo. \cuii\ shows no depletion: [Cu/H]=$-$0.88$\pm$0.10, [Cu/Zn]=$-$0.03$\pm$0.08 and [Cu/S]=$-$0.01$\pm$0.07, similarly to what was observed in the only other copper detection in a DLA system~\citep{Kulkarni2012}, and in contrast to what is observed in the local ISM, with [Cu/X]~$\sim-1.1$ in the warm ISM and [Cu/X]~$\sim-1.4$ in the cold ISM~\citep{Cartledge2006}. Note that the strictly solar relative abundances of copper, zinc and sulphur indicates that we do not observe any $\alpha$-element enrichment of zinc at all, contrary to what was suggested by~\citet{Rafelski2012}. This supports the usual interpretation that sub-solar [Fe/Zn] ratios are rather the result of iron depletion into dust grains.

It has been argued that the depletion could be larger in the components where \htwo\ is found~\citep[see e.g.][]{Petitjean2002}. We plot in Fig.~\ref{fig:Depletion}, top panel, the depletion of S, Si and Fe relative to Zn in each components versus the velocity position of the components with respect to the position of the \ci\ and \htwo\ components located at $z=2.65859(7)$. Notice that zinc and sulphur follow each other with mild differences on each edge of the absorption pattern. In particular, towards the blue sulphur appears to be depleted with respect to zinc, which is unlikely to be a real effect, but rather the result of the overestimation of \znii\ column densities in those components. Hence, in the following the component-by-component decomposition, we must critically compare the results using zinc or sulphur as references.

We have estimated the relative dust content on a single component --or a cloud-- by means of the dust to gas ratio, $\kappa \simeq \left(1-10^{\left[Fe/X \right]} \right)\times 10^{\left[X/H\right]}$, where X=Zn or X=S (see~\citealt{Prochaska2001} and~\citealt{Jenkins2009}). Averaging over the whole profile, we find $\kappa \simeq 0.091$ when using sulphur and $\kappa \simeq 0.080$ when using zinc, hence we estimate $\kappa \simeq 0.08$. This is a typical value for high redshift DLAs~\citep{Srianand2005a}. In order to compute $\kappa$ component by component, we will assume that the metallicity is the same for all components and equal to the DLA mean metallicity. The result is shown in the middle panel of Fig.~\ref{fig:Depletion}.

Finally, we can also estimate the \feii\ equivalent column density trapped into dust grains in each component $k$: $N^k_{\rm{dust}}($Fe,X$)=(1-10^{[\rm{Fe/X}]^k})\,N^k($X$)\,(N($Fe$)/N($X$))^{DLA}$, where $(N($Fe$)/N($X$))^{DLA}$ is the ratio of the total column densities in the DLA and X=Zn or X=S~\citet[see][]{Vladilo2006}. The values obtained at each component are shown in the bottom panel of Fig.~\ref{fig:Depletion}. For reference, the total values are $N^{DLA}_{\rm{dust}}($Fe,Zn$)\simeq10^{14.95}\,$cm$^{-2}$ and $N^{DLA}_{\rm{dust}}($Fe,S$)\simeq10^{14.97}\,$cm$^{-2}$. This system is comparable to most \htwo-bearing absorption systems at high redshift~\citep{Noterdaeme2008a}. It is apparent from Fig.~\ref{fig:Depletion} that the depletion pattern is quite homogeneous through the profile, with only a mild indication for a higher dust column density at zero velocity. This indicates that the molecular component can be easily hidden in the metal profile. We note however that the Doppler parameter of the metal component corresponding to \htwo\ is significantly lower than in the rest of the profile, indicating colder gas. In addition, the profile of highly ionised species such as \civ\ or \siiv\ (Fig.~\ref{fig:Metals_h}) is comparatively weak at the position of \htwo, revealing less ionisation in this component.

\subsection{Thermal properties in the molecular cloud from \htwo\ rotational level populations}
\begin{table}[t]
\caption{\htwo\ excitation temperatures in the $z_{\rm abs}=2.6586$ \htwo-bearing cloud toward \qso (all values in K).}
\begin{center}
\begin{tabular}{c c c}
\hline
\hline
$T_{ij}$ & $i=0$ & $i=3$ \\
$j=1$ & 80.6$^{+1.1}_{-1.0}$ & 97.9$\pm 1.5$ \\
$j=2$ & 96.8$^{+5.6}_{-5.0}$ & 92.4$^{+6.1}_{-7.1}$ \\
$j=3$ & 94.5$\pm 1.2$ & ... \\
$j=4$ & 142$\pm 1$ & 560$^{+93}_{-70}$ \\
$j=5$ & 189$\pm 3$ & 569$^{+67}_{-54}$ \\
$j=6$ & 263$\pm 7$ & 907$^{+186}_{-132}$ \\
$j=7$ & 80.6$^{+1.0}_{-1.1}$ & 897$^{+97}_{-80}$ \\
\hline
\end{tabular}
\end{center}
\normalsize
\label{tab:Texc}
\end{table}

\subsubsection{Excitation temperatures}
The excitation temperature $T_{ij}$ between two rotational levels $J=i,\,j$ is defined by
\begin{equation}
\frac{{N}_j}{{N}_i}=\frac{g_j}{g_i}\exp\left(-\frac{E_{ij}}{k_B\,T_{ij}}\right),
\end{equation}
where $g_J$ is the statistical weight of the $J$ level and $E_{ij}\simeq(j(j+1)-i(i+1))\times B$ is the energy difference between the rotational levels and $B$ is the rotational constant of the \htwo\ molecule ($B/k_B=85.3\,$K). Using this relation we compute the excitation temperatures $T_{ij}$ with $i=0,\,3$ and $j=1\,...\,7$. See the results in Table~\ref{tab:Texc}.

It is known that $T_{01}$ is related to the temperature of formation on the surface of dust grains when the dominant thermalisation process is the formation of \htwo, and to the kinetic temperature $T_{kin}$ when the dominant thermalisation process is collisions. In thick, self-shielded molecular clouds where $N($\htwo$)\geq10^{16.5}\,$cm$^{-2}$, $T_{01}$ is a good tracer of the kinetic temperature~\citep[see][and~references~therein]{Srianand2005b,Roy2006}. Here, $T_{kin} \simeq T_{01} = (80.6^{+1.1}_{-1.0})$~K, as in most high redshift \htwo-bearing clouds detected so far. Such a temperature is also very close to the that of the ISM of the MW or the Magellanic Clouds, where the average temperature is found to be $(77\pm17)$~K~\citep{Savage1977} and ($82\pm21$)~K~\citep{Tumlinson2002}, respectively, suggesting a cold neutral medium.

\subsubsection{Ortho-para ratio and local thermal equilibrium}
The ortho-para ratio (OPR), i.e the ratio of the total population (column densities) of ortho (even J) levels to that of para (odd J) levels, is another tracer of the state of the gas in the molecular gas, and is given by
\begin{equation}
OPR = \frac{\sum_{J=odd}N(J)}{\sum_{J=even}N(J)}.
\end{equation}
Here we measure OPR$=1.06$. If we assume the \htwo-bearing gas is at local thermal equilibrium (LTE), then we can relate the OPR to the equilibrium distribution with a single temperature for the whole gas and estimate $T^{LTE}$:
\begin{equation}
OPR_{LTE} \simeq 3\frac{\sum_{J=odd} (2J+1)\exp\left(\frac{-B\,J\,(J+1)}{k_BT^{LTE}}\right)}{\sum_{J=even} (2J+1)\exp\left(\frac{-B\,J\,(J+1)}{k_BT^{LTE}}\right)},
\end{equation}
where $k_B$ is the Boltzmann constant. When the kinetic temperature is high the OPR is expected to reach a value of 3, while in cold neutral media its value is expected to be below 1 ~\citep{Srianand2005b}. Hence, the OPR, in agreement with the excitation temperature $T_{01}$, also points toward a cold neutral medium.

\subsubsection{Excitation diagram}
\begin{figure}[t]
\centering
\includegraphics[scale=0.22,natwidth=6cm,natheight=6cm]{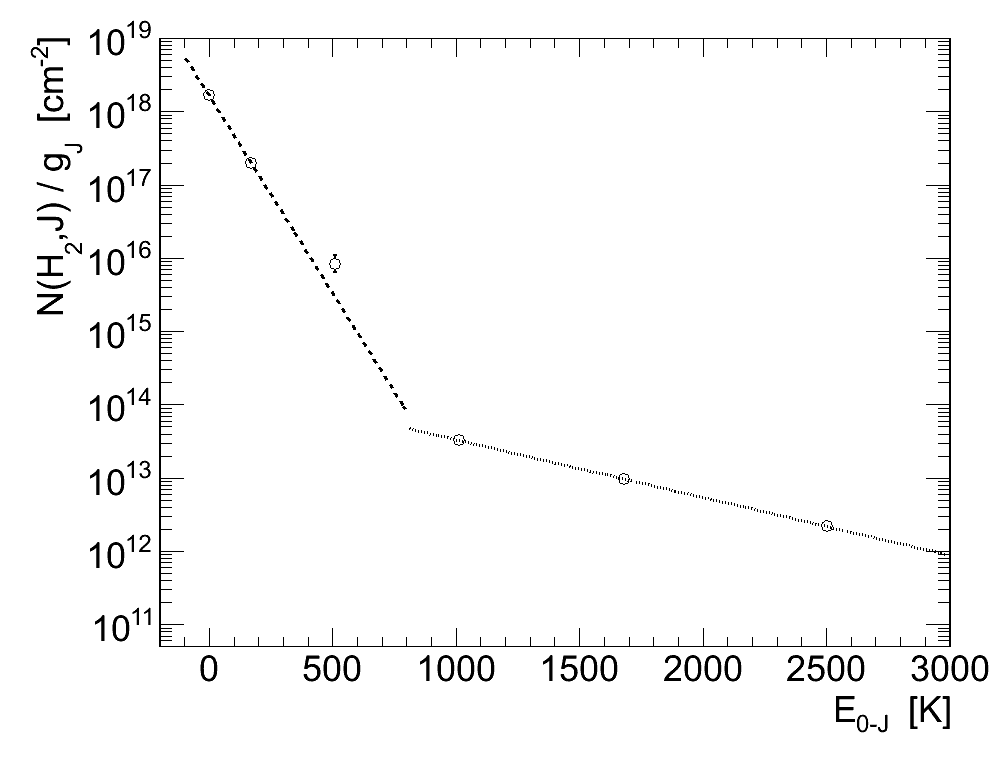}
\caption{Excitation diagram of \htwo. Two different behaviors are characterised by two different excitation temperatures, one for $J=0,\,1,$~and~$2$ and one for $J=3,\,4,$~and~$5$. The following two levels are somehow incompatible with both of these characteristic temperatures, however the uncertainty on their column densities is too large to elaborate further.}
\label{fig:H2Exc}
\end{figure}

The excitation diagram, i.e the ratio of the column density in a level to its statistical weight versus the energy difference between this level and $J=0$, is presented in Fig.~\ref{fig:H2Exc}. As it is shown by the two fitting lines, there seems to be two different excitation regimes, namely one for levels $J=0\,...\,2$ with $T_{0J}=(80.7\pm0.8)\,$K, and another for levels $J=3\,...\,5$ with $T_{3J}=(551^{+42}_{-37})\,$K (we limit the analysis to $J=5$ since the uncertainty on the column densities of $J=6$ and $J=7$ is large). We can readily see that $T_{0J,\,J=1,\,2}$ is compatible with $T_{01}$. If $T_{ij}\simeq T(\rm{OPR})\,(\simeq T_{01})$, then the excitation is dominated by collisions~\citep{Srianand2005b}, and this seems to be the case below $J=3$. From this level and above, other mechanisms intervene, such as radiation pumping (excitation of lower levels to higher levels by UV radiation) or formation pumping (preferred formation of \htwo\ in excited states, which then de-excite to the lower levels). From Table~\ref{tab:Texc} we see that $T_{3J}$ is of the same order than $T_{34}$ and $T_{35}$. 

The Doppler parameter increases with increasing rotational level. This could be the result of higher rotational levels arising mostly from the outer parts of the cloud, that are heated by photoelectric effect on dust grains by the surrounding UV flux~\citep{Ledoux2003}.

\subsection{Ambient UV flux}

\subsubsection{UV flux from heating/cooling equilibrium}
We proceed to estimate the ambient UV flux using the same procedure used by~\citet{Wolfe2003}. Assuming equilibrium between the photo-electric heating of the gas and the cooling by the \ciie $\left( ^2\rm{P}_{3/2} \rightarrow ^2\rm{P}_{1/2} \right)$ emission, together with similar interstellar medium conditions between the galaxy and DLAs (same type of dust grains, temperature and density,~\citep[][]{Weingartner2001}), we can write $\frac{l_c}{l_c^{\rm{MW}}}\simeq \kappa\frac{\rm{F}_{\rm{UV}}}{\rm{F}_{\rm{UV}}^{\rm{MW}}}$, where $\kappa=k_{\rm{DLA}}/k_{\rm{MW}}$ is the relative dust to gas ratio, and ${k}_{\rm{DLA}}$ and ${k}_{\rm{MW}}$ are the local dust to gas ratio at the DLA and the MW and $l_c$ the cooling rate.

The relative dust to gas ratio has been estimated to be $\kappa~\simeq~0.08$ in the DLA we study. Then, the average energy loss per hydrogen atom determines the cooling rate
\begin{equation}
l_c~=~\frac{\rm{h} \nu \left( ^2\rm{P}_{3/2} \rightarrow ^2{\rm{P}_{1/2} }\right) N \left( \rm{C\,\mathsc{II}^\star}~^2 \rm{P}_{3/2} \right) \rm{A}_{21}}{ N \left( \rm{H} \right)} \rm{erg}~\rm{s}^{-1},
\end{equation}
with $\lambda \left( ^2\rm{P}_{3/2} \rightarrow ^2\rm{P}_{1/2} \right)\simeq 158\mu$m and A$_{21}\simeq2.4\times10^{-6}$s$^{-1}$, the wavelength and decay rate of the spontaneous photon emitting $^2\rm{P}_{3/2} \rightarrow ^2\rm{P}_{1/2}$ transition~\citep{Pottasch1979,Wolfe2003}. We estimate $N$(\ciie)~$\simeq10^{14.1}$cm$^{-2}$ (see Table~\ref{tab:H2CI}) and $N$(\hi)~$\simeq10^{21.0}$cm$^{-2}$. This leads to $l_c\simeq10^{-26.4}\rm{erg}~\rm{s}^{-1}$ per hydrogen atom. From this we derive ${F}_{\rm{UV}}\sim 0.82\times {F}_{\rm{UV}}^{\rm{MW}}$.

\subsubsection{Rotational excitation and radiation field}
We can estimate the ambient UV radiation using the excitation of the high J levels since we have seen that for $J\geq3$ the excitation temperature as deduced from the level populations is substantially larger than $T_{kin}$, hence these levels are excited by fluorescence but also formation of \htwo\ in high rotational levels. Following~\citet{Noterdaeme2007a}, formation pumping is related --at equilibrium-- to the photodissociation rate, such that $R_{\rm{form}} \,n_{\rm{p}} \, n($H$ )= R_{\rm{diss}} \,n($\htwo$)$, where $R_{\rm{form}}$, $n_{\rm{p}}$ and $R_{\rm{diss}}$ stand for the \htwo\ formation rate, the proton density and the \htwo\ photodissociation rate. The latter can be related to the photoabsorption rate $\beta$ --which can bring us to the ambient UV flux-- by $R_{\rm{diss}}=\zeta\,\beta$, where $\zeta=0.11$ is the fraction of photodissociation to photoabsorption. With this we come to $R_{\rm{form}} \,n_{\rm{p}} \, n($H$ )/n($\htwo$) = R_{\rm{diss}} = \, \zeta\,\beta$. Hence, we can write the equilibrium between spontaneous decay of $J=4,\,5$ levels (with transition probabilities $A_{4\rightarrow2}=2.8\times10^{-9}\,$s$^{-1}$ and $A_{5\rightarrow3}=9.9\times10^{-9}\,$s$^{-1}$), and UV pumping from the most populated lower levels $J=0,\,1$ (with efficiencies $p_{4,\,0}=0.26$ and $p_{5,\,1}=0.12$ and absorption rates $\beta_0$ and $\beta_1$) and formation pumping into $J=4,\,5$ levels (with $f_4\simeq0.19$ and $f_5\simeq0.44$ fractions) as
\begin{equation}
p_{4,\,0}\,\beta_0 \, n \left( \rm{H}_2 , \, J=0 \right) + f_4\, R_{\rm{form}} \,n_{\rm{p}} \, n\left(\rm{H} \right) = A_{4\rightarrow2} \, n \left( \rm{H}_2, \, J=4 \right)
\end{equation}
and
\begin{equation}
p_{5,\,1}\,\beta_1 \, n \left( \rm{H}_2 , \, J=1 \right) + f_5\, R_{\rm{form}} \,n_{\rm{p}} \, n\left(\rm{H} \right) = A_{5\rightarrow3} \, n \left( \rm{H}_2, \, J=5 \right).
\end{equation}
Dividing by the \htwo\ density, assuming the homogeneity of the gas cloud (and therefore $n($\htwo$,\,J)/n($\htwo$)=N($\htwo$,\,J)/N($\htwo$)$) and rearranging these equations we get
\begin{equation}
\beta_0 \, \left( p_{4,\,0}\,\frac{N\left( \rm{H}_2 , \, J=0 \right)}{N\left( \rm{H}_2 \right)} + \zeta\,f_4 \right) = A_{4\rightarrow2} \frac{N\left( \rm{H}_2 , \, J=4 \right)}{N\left( \rm{H}_2 \right)}
\end{equation}
and
\begin{equation}
\beta_1 \, \left( p_{5,\,1}\,\frac{N\left( \rm{H}_2 , \, J=1 \right)}{N\left( \rm{H}_2 \right)} + \zeta\,f_5 \right) = A_{5\rightarrow3} \frac{N\left( \rm{H}_2 , \, J=5 \right)}{N\left( \rm{H}_2 \right)}.
\end{equation}
Using these relations we get $\beta_0\simeq(4.9\pm0.2)\times10^{-13}\,$s$^{-1}$ and $\beta_1\simeq(18.8\pm1.9)\times10^{-13}\,$s$^{-1}$. These are extremely low values that can be explained by shielding of the inner part of the cloud by outer layers. The large $\log N($\htwo$)/N($\ci$)=5.97\pm0.09$ ratio measured in the cloud confirms this as it has already been pointed out that large \htwo\ column densities at high redshift, and in particular large $\log N($\htwo$)/N($\ci$)$, are associated with low photoabsorption rates~\citep[see][Fig.~7]{Noterdaeme2007a}.

The total shielding $S$ is the product of the molecular hydrogen self-shielding, estimated by $S_{\rm{H}_2}\simeq(N($\htwo$)/10^{14}\,$cm$^{-2})^{-0.75}$ by the dust extinction $S_{\rm{dust}}=\exp(-\tau_{\rm{UV}})$, and allows us to relate the photodissociation rate to $J_{LW}$, the UV intensity at $h\nu=12.87\,$eV averaged over the solid angle, with the relation $R_{\rm{diss}}=\zeta\beta=4\pi\,1.1\times10^{8}\,J_{LW}\,S$. The dust optical depth can be approximated by $\tau_{\rm{UV}}\simeq0.879\,\kappa\,(N($H$)/10^{21}\,$cm$^{-2})$~\cite[see][]{Noterdaeme2007b}, where $\kappa$ is the dust to gas ratio as discussed in Section~\ref{sec:depdust}. Hence, $\tau_{\rm{UV}}\simeq0.07$ leads to $S\simeq(4\times10^{-4})\times(0.9)\simeq 3.8\times10^{-4}$. Self-shielding is the dominant shielding mechanism as usual. With this and using $\beta=\beta_0$ we come to $J_{\rm{LW}}\simeq8\times10^{-11}\,\beta_0/S\simeq8.5\times10^{-20}\,$erg$\,$s$^{-1}\,$cm$^{-2}\,$Hz$^{-1}\,$sr$^{-1}$, which is a factor $\chi=2.6$ larger than in the solar vicinity. Note that this is larger by about a factor of four compared to the value derived from the \ciie\ absorption line. The two measurements are comparable however and have no a priori reason to match exactly as the \ciie\ measurement corresponds to a mean flux in the host galaxy when the \htwo\ measurement is related to the flux in the vicinity of the \htwo-bearing cloud.

Following~\citet{Hirashita2005} we can relate $\chi$ to the surface star forming rate $\Sigma_{SFR} \simeq \chi \times 1.7 \times10^{-3}$M$_{\odot}\,$yr$^{-1}\,$kpc$^{-2} \simeq 4.4 \times10^{-3}$M$_{\odot}\,$yr$^{-1}\,$kpc$^{-2}$.

\subsection{Neutral carbon fine-structure and the extension of the molecular cloud}\label{sec:CIfine}
The relative populations of the fine structure levels of neutral carbon ground state is determined by the ambient conditions through collisions and excitation by the UV flux and the CMB radiation. We can estimate the density of the gas because for densities $\geq10\,$cm$^{-3}$ and temperatures $T_{kin}\sim$100$\,$K, these populations depend primarily on the hydrogen density ~\citep[see][Fig.~2]{Silva2002}.

We measure $N($\cie$) / N($\ci$)=10^{-0.10\pm0.11}$ and estimate the hydrogen density to be in the range $n($H$)\simeq(40-140)\,$cm$\,^{-3}$ with a central value of 80$\,$cm$\,^{-3}$~\citep[from][Fig.~8]{Noterdaeme2007a}.

We can thus estimate the characteristic length of the cloud $l\simeq N($\hi$)/n($H$)$: $2.3\,$pc$\, \lesssim l \lesssim 7.9\,$pc.  The \hi\ column density in the main component, which bears molecules, is certainly overestimated, since species such as \feii, \siii, \oi, \cii\ or \Niii\ are found in many components spanning over 200~\kms. Therefore we can safely conclude that $l < 8\,$pc. Assuming for simplicity a spherical shape, this translates into an angular size of $\theta_{DLA} < 0.9$~mas, however there is no reason for the length scale along the line-of-sight to be the same as the transverse extension. Moreover, the average density at the molecular cloud could be slightly different, since there is a small 250~\ms\ displacement in the centroid of the neutral carbon and \htwo\ absorption features.

Bearing this in mind, for comparison, the angular size of a typical BLR (of extension $l_{BLR}\sim1$~pc) amounts to $\theta_{BLR}\sim0.1$~mas. This is fully consistent with full coverage of the BLR by the DLA. Nonetheless, the size of the molecular cloud could be much smaller (\citet{Balashev2011} estimated that the molecular cloud at $z=2.34$ toward Q1232+082 has an extension of ($0.15~\pm~0.05$)~pc), hence it could well be two orders of magnitude smaller than 0.9~mas. With this it is impossible to conclude of the actuality of the partial coverage of the BLR by the \htwo-bearing cloud.


\section{Constraints to the cosmic variation of the proton-to-electron mass ratio with \qso}\label{sec:VoC}
\begin{figure*} 
\centering
\includegraphics[width=0.75\hsize]{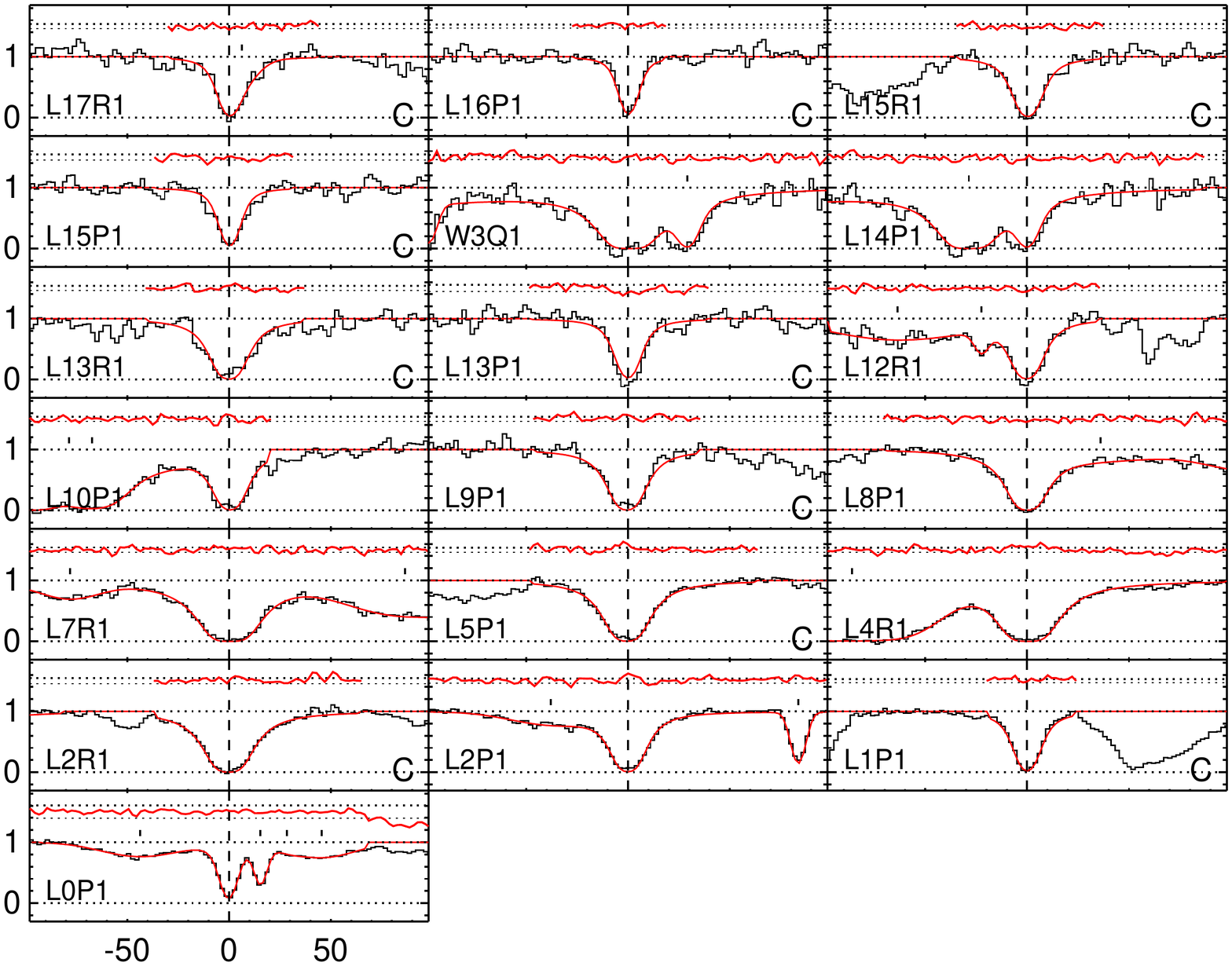}
\caption{Absorption profile of \htwo\ transitions from the $J$ = 1 level and the best-fitting Voigt profile. The normalized residual (i.e.([data]$-$[model])/[error]) for each fit is also shown in the top of each panel along with the 1$\sigma$ line. We present the clean absorption lines by putting a letter "C" in the right bottom of these transitions.}
\vskip -14.0cm
\begin{picture}(400,400)(0,0)
\put(170, 42){\large Velocity (\kms)}
\end{picture}
\label{fig_J1}
\end{figure*}

In the case of \htwo\ molecule, the energy difference between rotational states and vibrational states is proportional to the reduced mass of the system and its square root respectively. Using intervening molecular absorption lines seen in the high-$z$ quasar spectra for measuring \dmm\ (i.e. \dmm~$\equiv (\mu_z- \mu_0)/\mu_0$ where $\mu_z$ and $\mu_0$ are the values of proton-to-electron mass ratio at redshift $z$ and today) in the distant universe was first proposed by \citet{Thompson1975}.

The sensitivity of the wavelength of the i'th \htwo\ transition to the variation of $\mu$ is generally parametrised as
\begin{equation}
\lambda_i = \lambda_i^0 (1+z_{\rm abs})\big{(}1+K_i\frac{\Delta\mu}{\mu}\big{)},
\label{eq_dm}
\end{equation}
where $\lambda_i^0$ is the rest frame wavelength of the transition, $\lambda_i$ is the observed wavelength, $K_i$ is the sensitivity coefficient of i'th transition, given by $K_i=\rm{d}\,\rm{ln} \lambda_i^0 / \rm{d}\,\rm{ln} \mu$ and \zabs\ is the redshift of the \htwo\ system. Here we use the most recent data on $K_i$ given by~\citet{Meshkov2006,Ubachs2007} and the rest wavelengths and oscillator strengths from \citet{Malec2010}. Eq. \ref{eq_dm} can be rearranged as 
\begin{equation}
z_i = z_{\rm abs} + C K_i, ~~~~~ C = (1+z_{\rm abs})\frac{\Delta\mu}{\mu}
\label{eq_dm_a}
\end{equation}
which clearly shows that \zabs\ is only the mean redshift of transitions with $K_i$ = 0. $z_i$ is the redshift of the i'th \htwo\ transition. Eq.~\ref{eq_dm_a} is sometimes presented as
\begin{equation}
z_{red} \equiv \frac{(z_i - z_{abs})}{(1 + z_{abs})} = K_i \frac{\Delta\mu}{\mu}
\label{eq_dm_b}
\end{equation}
which shows the value of \dmm\ can be determined using the reduced redshift ($z_{red}$) versus $K_i$. At present, measurements of \dmm\ using \htwo\ are limited to 6 \htwo-bearing DLAs at $z \ge$ 2 \citep[see][]{Varshalovich1993,Cowie1995,Levshakov2002,Ivanchik2005,Reinhold2006,Ubachs2007,Thompson2009,Wendt2011,Wendt2012,Rahmani2013}. All these measurements are consistent with \dmm\ being zero at the level of 10 ppm. Here we present a new \dmm\ constraint using \zabs = 2.6586 system towards \qso. 

Like in \citet{Rahmani2013} we use two approaches to measure \dmm. In the first approach we fit the \htwo\ lines using single velocity component and estimate the redshift for each \htwo\ transitions, and measure \dmm\ using Eq.~\ref{eq_dm_b}. In the second approach we explicitly use \dmm\ as one of the fitting parameters in addition to N, $b$ and $z$ in VPFIT. This approach allows for a multi-component fit of the \htwo\ lines. The results of \dmm\ for various approaches are given in Table~\ref{tab:dmumu_estimations}.

To carry out a measurement of $\mu$ we need to choose a suitable set of \htwo\ lines. Although \htwo\ absorption from high J-levels (J = 4 -- 7) are detected toward \qso, they are too weak to lead to very accurate redshift measurements, as required for this study. Therefore, we reject \htwo\ lines from these high-J levels while measuring \dmm, and only use \htwo\ absorption features from J = 0 -- 3. By carefully inspecting the combined spectrum we identified 81 lines suitable for \dmm\ measurements. A list of \htwo\ transitions we used is tabulated in Table~\ref{fitting_res1}, where clean lines are highlighted. 38 out of 81 lines are mildly blended with the intervening \Lya\ absorption of the intergalactic medium. We accurately model surrounding contaminations using multi-component Voigt-profile fitting, while simultaneously fitting the \htwo\ lines (see Fig.~\ref{fig_J1}).

\subsection{Systematic wavelength shifts: cross-correlation analysis}
\citet{D'Odorico2000} have shown that the resetting of the grating between an object exposure and the ThAr calibration lamp exposure can result in an error of the order of a few hundred meters per second in the wavelength calibration. To minimize the errors introduced via such systematics in our \dmm\ measurements we do not use those exposures without attached mode ThAr calibration lamps. We further exclude the exposure with 1389 seconds of EXPTIME (10th row of Table~\ref{tab:Exposures}) as the quality of this spectrum is very poor. Therefore, the combined spectrum to be used for measuring \dmm\ is made of 14 exposures with SNR of between 11 -- 31 over our wavelength range of interest.

\begin{figure} 
\centering
\includegraphics[width=0.98\hsize]{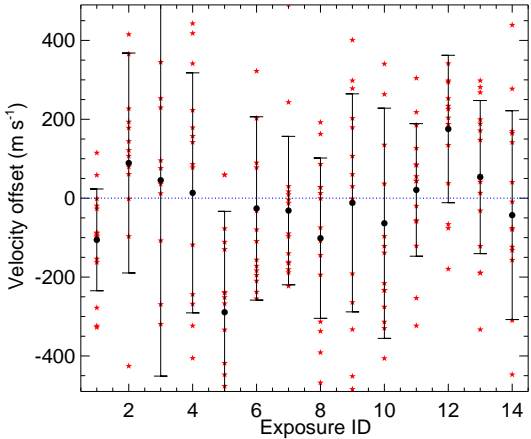}
\caption{Results of cross-correlation analysis between individual spectra and the combined one for each order. Bars present the standard deviations of the shifts measured for individual spectra.}
\label{fig_crcor}
\end{figure}

The shortcomings of the ThAr calibration of VLT/UVES spectra have been shown by a number of authors~\citep{Chand2006,Levshakov2006,Molaro2008,Thompson2009,Whitmore2010,Agafonova2011,Wendt2011,Rahmani2012,Agafonova2013,Rahmani2013}. Here we carry out a cross-correlation analysis between the combined spectrum and the individual exposures to estimate the offset between them over the wavelength range of echelle orders of the blue arm. To do so we rebin each pixel of size 2.0 \kms\ into 20 sub-pixels of size 100 \ms\ and measure the offset as corresponding to the minimum value of the $\chi^2$ estimator of the flux differences in each window~\citep[see][for more detail]{Rahmani2013}. Each cross-correlating window spans an echelle order. The accuracy of this method is well demonstrated via a Monte Carlo simulation analysis in~\citet{Rahmani2013}. Fig.~\ref{fig_crcor} shows the results of such cross-correlation analysis.

\begin{figure} 
\centering
\includegraphics[width=0.98\hsize]{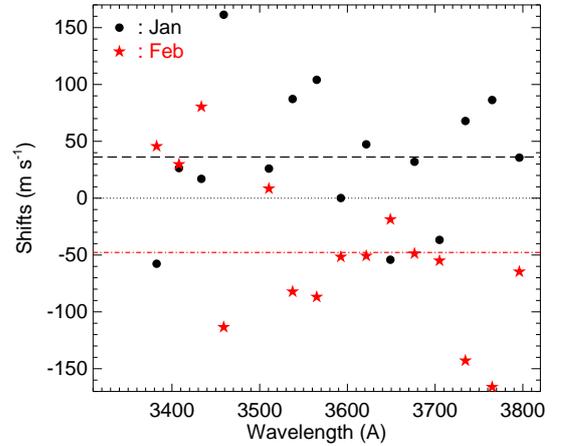}
\caption{Cross-correlation analysis between the combined spectrum of the exposures taken in early January and combined spectrum of those in February. }
\label{fig_cr_jan_feb}
\end{figure}

\begin{figure*} 
\centering
\includegraphics[width=0.75\hsize]{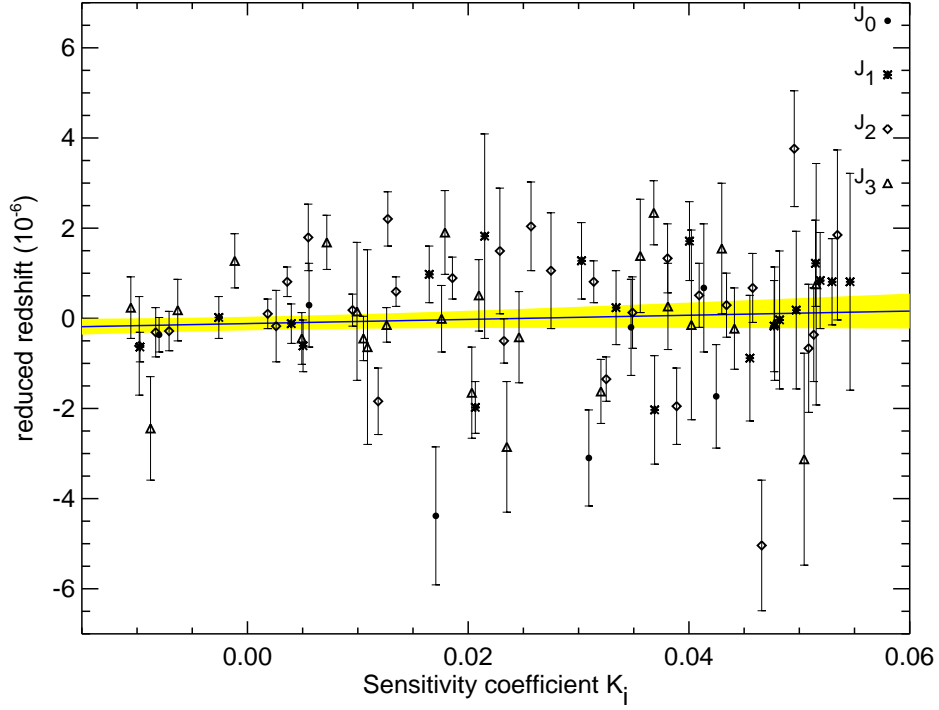}
\caption{Reduced redshift vs the $K_i$ for the fit with different $b$ for different $J$-levels. Different $J$-levels are plotted with different symbols. The best-fitting linear line is also shown. The yellow region shows the 1$\sigma$ error in the fitted line.} 
\label{zvskplot}
\end{figure*}

We observe that EXP5 has the maximum constant offset of $-$287 \ms, while other exposures show constant offsets consistent with zero shift. We correct the measured shifts for individual orders of each exposure and make a new combined spectrum, and subsequently analyse its impact on \dmm.

In order to check the stability of the spectra over a one-month period of observations, we make two combined spectra of EXP7 -- EXP14 and EXP17 -- EXP20. Fig.~\ref{fig_cr_jan_feb} shows the results of cross-correlation of these two spectra with the combined spectrum. The first three circle-asterisks pairs at $\lambda <$ 3450 \AA\ and the pair at 3650 \AA\ present a opposite trend in comparison to the rest of pairs. However, apart from a  constant shift of $\sim$ 80 \ms, these is not any strong evidence for the existence of possible wavelength dependent systematics.

Furthermore, \citet{Rahmani2013} compared the asteroids spectra from UVES and the solar spectrum to discover a wavelength dependent systematic error in UVES observations. Moreover, they showed that such systematics can mimic a \dmm\ in the range of 2.5 -- 13.7 ppm which is changing in different epochs from 2010 to 2012. Such a systematic cannot be revealed from the cross-correlation analysis between the individual exposures and the combined spectrum of this quasar, i.e. without a reference spectrum. Unfortunately we do not have asteroids observations with the same settings of and close in time to our science observations, hence we are not in a position to check whether such a drift is present in our data. Ceres has been observed with 346 setting of UVES on 05-12-2007 and without attached mode ThAr lamp. However, such an observation is not an appropriate reference to study the effect of a wavelength drift in 390 setting observations, hence we do not correct the data using this drift measurement.

\subsection{Limits on \dmm\ at $z=2.6586$ -- First approach}\label{sec:mudan}
Most measurements of \dmm\ using \htwo\ use the measured slope between the reduced redshifts and $K_i$. The most important step in this approach is to measure the redshifts and associated errors of a set of chosen \htwo\ absorption lines. 

To measure the redshifts of the suitable \htwo\ lines we first choose a model in which all the J-levels have the same $b$ parameter. The best-fitting model in this case has a reduced $\chi^2$ ($\chi^2_{\rm red}$) of 1.57. Inspecting the spectrum and the normalised residual (([data]-[model])/[error]) of the best-fitting model show that the relatively large $\chi^2_{\rm red}$ is mainly due to the underestimation of the flux error and not due to a poor Voigt profile model. A $z$-vs-$K$ analysis of the fitted redshifts based on a linear regression gives \dmm~$=(5.0~\pm~6.1)$~ppm. The quoted error is obtained using the bootstrap technique. Indeed, we generate 2000 random realizations of the measured redshifts and estimate \dmm\ for each realization. We finally quote the 1$\sigma$ scatter of the 2000 \dmm\ thus obtained as the estimated error. 

The physical conditions in this \htwo\ system and also other \htwo\ systems~\citep[see][]{Noterdaeme2007b} shows that different $J$-levels may bear different turbulent broadening parameters. In particular, $J$ = 0 and 1 have smaller $b$-parameters compared to higher $J$-levels. Hence we define a second model in which different $J$-levels are allowed to have different $b$-parameters. The best such fitted model has a $\chi^2_{\rm red}$ = 1.55. The $z$-vs-$K$ analysis out of these fitted redshifts yields \dmm~$=(4.6~\pm~5.9)$~ppm, which is very much consistent with the value obtained with the first model. Fig.~\ref{zvskplot} presents the results of this fit.

\begin{table*}[!ht]
\caption{\dmm\ estimations using \htwo\ lines in the z$_{\rm abs}=2.6586$ \htwo-bearing cloud toward \qso. The quoted errors in \dmm\ come mainly from statistical errors.}
\begin{center}
\begin{tabular}{c c c c c c c c c}
\hline
\multicolumn{9}{c}{\dmm~$\times 10^{6}$} \\
\hline
\hline
&&& \multicolumn{3}{c}{1 component} & \multicolumn{3}{c}{shift-corrected} \\
\hline
Fit & Error & $b$ & \dmm\ & $\chi^2_{red}$ & AICC & \dmm\ & $\chi^2_{red}$ & AICC\\
\hline
l-b-l & bootstrap & tied & 5.0~$\pm$~6.1 & 1.57 & ... & 8.1~$\pm$~6.6 & 1.53 & ... \\
l-b-l & bootstrap & free & 4.6~$\pm$~5.9 & 1.55 & ... & 7.6~$\pm$~6.4 & 1.51 & ... \\
VPFIT & weighted & tied & 4.8~$\pm$~4.2 & 1.58 & 7318 & 5.9~$\pm$~4.2 & 1.57 & 7299 \\
VPFIT & weighted & free & 5.5~$\pm$~4.3 & 1.58 & 7301 & 6.7~$\pm$~4.2 & 1.55 & 7204 \\
\hline
&&&\multicolumn{3}{c}{2 components}&\multicolumn{3}{c}{shift-corrected} \\
VPFIT & weighted & free & 6.5~$\pm$~4.3 & 1.57 & 7253& 7.4~$\pm$~4.3 & 1.53& 7150 \\
\hline
\end{tabular}
\end{center}
\label{tab:dmumu_estimations}
\end{table*}

Applying the above discussed models on the shift-corrected combined spectrum we find \dmm\ of (8.1$\pm$6.6)~ppm and (7.6$\pm$6.4)~ppm for tied and untied $b$, respectively. While being consistent both values are $\sim$~3~ppm larger in comparison with the uncorrected ones.

\subsection{Limits on \dmm\ at $z=2.6586$ directly from VPFIT}
Next, we include \dmm\ as a parameter of the fits. Using a single component model and assuming the fitted $b$ parameter value to be the same for all J-levels, the best fit converges to \dmm~$=~(4.8~\pm~4.2)$~ppm with an overall reduced $\chi^2$ of 1.60. The quoted error of 4.2~ppm in \dmm\ is already scaled with $\sqrt{\chi^2_{red}}$. Similar scaling of the statistical errors will be implemented whenever we find $\chi^2_{red} > 1.0$. This result is very much consistent with the previous findings. We apply the same analysis to the combined spectrum made of CPL generated 1-d spectra. For this model we find \dmm~$=~(9.1~\pm~4.7)$~ppm for such a combined spectrum. The two values differ by $\sim$ 1$\sigma$. Therefore systematic errors as large as 4.3 ppm can be produced if using the final 1-d spectra generated by CPL.

In the last column of Table~\ref{tab:dmumu_estimations} we provide the Akaike information criteria~\citep[AIC;][]{Akaike1974} corrected for the finite sample size~\citep[AICC;][]{Sugiura1978} as given in Eq.~4 of~\citet{King2011}. While the \dmm\ error from the bootstrap is sensitive to the redshift distribution of \htwo\ lines the VPFIT errors purely reflect the statistical errors. Hence, the larger error from the bootstrap method can be used to quantify the associated systematic errors. If we quadratically add 4.4 ppm to the VPFIT error (which is 4.2 ppm) we get the bootstrap error. Therefore, we can associate a systematic error to each VPFIT measurement  by comparing the VPFIT error with the bootstrap error.

In the fourth line of Table~\ref{tab:dmumu_estimations} we give the results of the fit when $b$ parameters in different $J$-levels are allowed to be different. While the best-fitting \dmm\ of (5.5~$\pm$~4.3)~ppm is consistent with the case when we use a common $b$ value for all $J$-levels, the AICC value is slightly better. In this case the comparison of the estimated error from the bootstrap method (second line in Table~\ref{tab:dmumu_estimations}) suggests a systematic error of 4.0~ppm.

We also consider two velocity component models. The two-component fit with common $b$ for different $J$ levels systematically drops one of the components while minimizing the $\chi^2$. However, when the Doppler parameter is different for different $J$-levels we are able to obtain a consistent two-component fit. The two components are separated by (0.19~$\pm$~0.11)~\kms. We find \dmm\ = $(6.5~\pm~4.3)$~ppm for such a fit. The results are summarized in the last row of Table~\ref{tab:dmumu_estimations}. We also present the 2 component fit results using shift-corrected data in Table~\ref{tab:dmumu_estimations}. The values of \dmm\ are larger by $\sim$ 2--3 ppm for all the models after correcting the spectra for the shifts. It is clear that while there is a marginal improvement in the $\chi^2_{red}$ and AICC the final results are very much consistent with one another. Moreover, from Table~\ref{tab:dmumu_estimations} we see that the best model is the two-component model for both shift-corrected and not corrected spectra. Therefore, we choose the two-component \htwo\ fit as the best model of the data. As in~\citet{Rahmani2013} we quote the final error in \dmm\ including the systematic error obtained above and the statistical error given by VPFIT. Therefore, we consider the best-fitting measurement to be \dmm$~=~(7.4~\pm~4.3_{\rm stat}~\pm~5.1_{\rm sys})$~ppm. 

\begin{figure}[t]
\centering
\hspace*{-0.5cm}	 
\includegraphics[scale=0.38,natwidth=6cm,natheight=6cm]{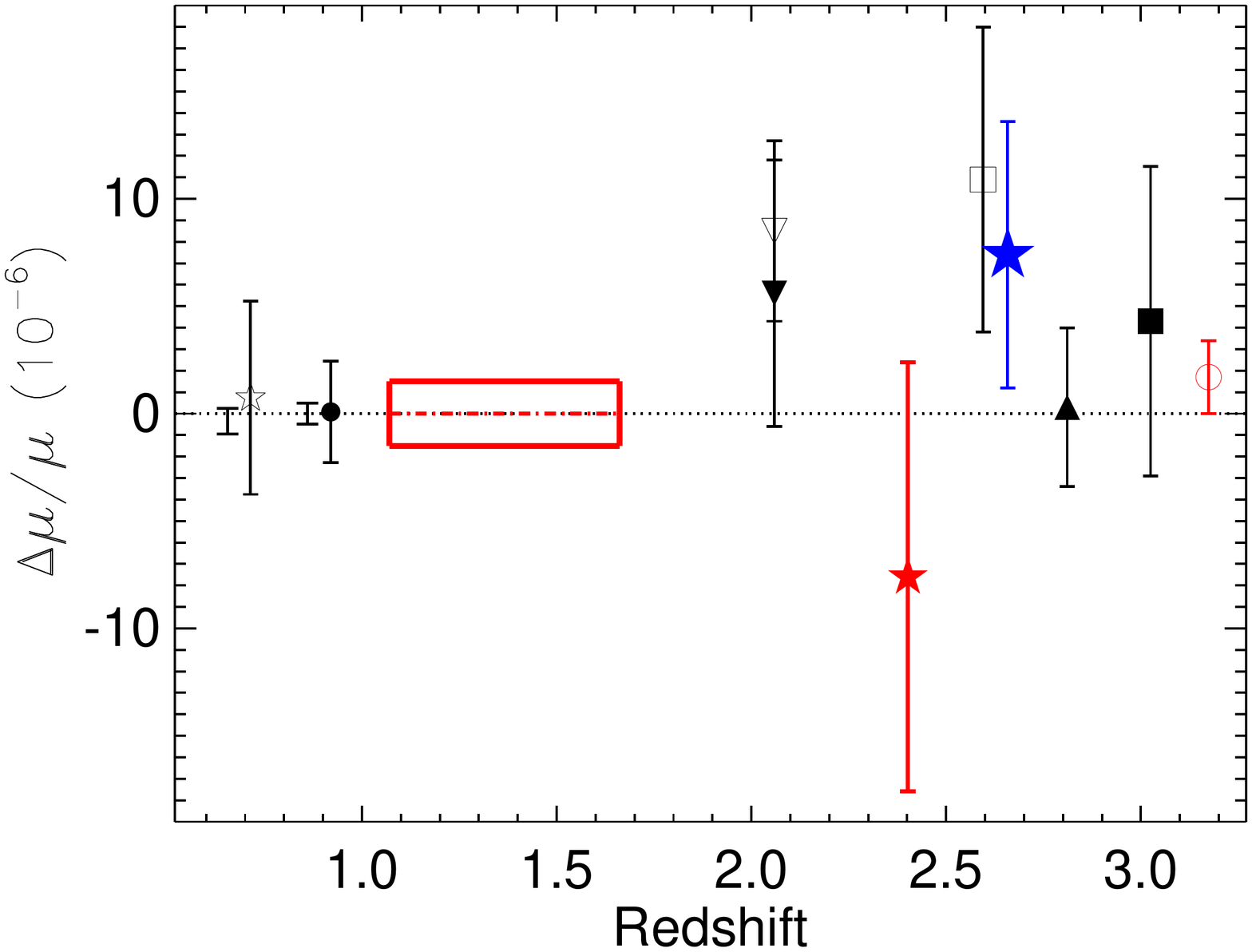}
\caption{Comparing \dmm\ measurement in this work and those in the literature. All measurements at 2.0 $< z < $ 3.1 are based on the analysis of \htwo\ absorption. The filled larger blue star shows our result and the smaller red star shows the result 
from~\citet{Rahmani2013}. The downwards empty and filled triangles are the \dmm\ measurements from~\citet{vanWeerdenburg2011} and~\citet{Malec2010}. The filled upward triangle and the empty and filled squares are respectively from~\citet{King2011}, \citet{King2008}, and~\citet{Wendt2012}. The solid box and the open circle present the constraint obtained respectively by~\citet{Rahmani2012} and~\citet{Srianand2010} based on the comparison between 21-cm and metal lines in \MgII\ absorbers under the assumption that $\alpha$ and $g_p$ have not varied. The \dmm\ at $z < 1$ are based on ammonia and methanol inversion transitions that their 5$\sigma$ errors are shown. The two measurements at $z \sim$ 0.89 with larger and smaller errors are respectively from~\citet[][]{Henkel2009} and~\citet[][]{Bagdonaite2013} based on the same system. The two \dmm\ at $z \sim$ 0.684 with larger and smaller errors are respectively from~\citet{Murphy2008} and~\citet{Kanekar2011} based on the same system.}
\label{fig:s_dmm}
\end{figure}


\section{Conclusions and discussion}
We have studied the physical properties of the molecular hydrogen gas associated with the Damped \Lya\ system at $z_{\rm abs}$=2.6586 towards the high redshift ($z_{\rm em}$~=~3.09) quasar \qso\ using VLT-UVES data with a total integration time of 23~hours.

We find that the DLA (log~$N$(\hi)(cm$^{-2}$)~=~21.03$\pm$0.08) has typical characteristics of the high-redshift DLA population associated to molecular clouds, with metallicity [Zn/H]~=~$-0.91\pm 0.09$ and depletion of iron relative to zinc [Zn/Fe]~=~0.45$\pm$0.06, and a hydrogen molecular fraction of log~$f$~=~-2.19$^{+0.07}_{-0.08}$. Molecular hydrogen is detected up to $J=7$ and the excitation diagram exhibits two temperatures, $T=80.7\pm 0.8$~ and $551\pm40$~K for J~$<$~3 and $>$~3 respectively. The ambient UV radiation field, derived from the \cii$\lambda$156$\mu$ radiation and from the analysis of the \htwo\ UV pumping, is of the order of the MW field.

We study the possibility that the \htwo\ bearing cloud does not cover the background source completely. For this we analyse the distribution of residual flux observed at the bottom of saturated \htwo\ lines and estimate the relative contributions from the BLR and AGN continuum to the emitted flux. We find that there is weak evidence for a possible excess of residual light from the BLR. Given the density derived from the \ci\ absorption lines ($n_{\rm H}$ in the range 40$-$140~cm$^{-3}$) and the \hi\ column density, we derive a dimension of the cloud of the order of 2.5$-$8~pc. This is consistent with the fact that the residual flux, if real, is small and the cloud covers most of the BLR.

We have attempted to measure \hd\ at the same redshift as \htwo. By stacking the best defined features we have been able to set an upper limit of $N($\hd$)\lesssim10^{13.65~\pm~0.07}$cm$^{-2}$. Deuterium is formed in the primordial universe, and is subsequently progressively destroyed in stars. Therefore one expects the abundance of deuterium to be at most equal to the primordial value (stemming from primordial nucleosynthesis) in most metal poor gas, and smaller if its local destruction is already onset. In molecular clouds, deuterated molecular hydrogen is expected to form along with \htwo\ in shielded environments. In high redshift quasar absorption lines, only 6 detections have been reported so far: toward Q1232+082~\citep{Varshalovich2001,Ivanchik2010}; toward Q143912.04+111740.4~\citep{Srianand2008,Noterdaeme2008b}; toward J0812+3208 and Q1331+170~\citep{Balashev2010}; toward J123714.60+064759.5~\citep{Noterdaeme2010}; and toward J2123–0500~\citep{Tumlinson2010}. When \hd\ is detected, it is possible to probe the deuterium abundance by studying the $N($\hd$)/2N($\htwo$)$ ratio which should be a lower limit for D/H (a smaller deuterium fraction in molecules would be the consequence of differences in chemical molecular formation processes and of differential photo-dissociation~\citep[see e.g.][]{Tumlinson2010}). All observations up to now seem to show that the measured \hd/2\htwo\ ratio is surprisingly high~\citep{Balashev2010,Ivanchik2010}, with typical values of $f_D$ between -4.8 and -4.4~\citep{Tumlinson2010}. The \deut\ fraction, $f_{D}\simeq N($\hd$)/2N($\htwo$)$, in this system is then $\lesssim (6.50\pm1.07)\times10^{-6} = 10^{-5.19\pm0.07}$, to be compared to the primordial abundance of deuterium which is of $\log D/H= -4.60$. Hence in the cloud at \zabs$=2.6586$ toward \qso\ the relative deuterium abundance is half an order of magnitude lower than the primordial value. This value can also be compared to what is seen in the Milky Way. \citet{Lacour2005} have found $f_{D}$ to increase with increasing molecular fraction. Extrapolating toward the molecular fraction of $f=-2.19$ one would have $f_{D}\simeq-7.5$ which is two orders of magnitude below the value we would observe. Large \hd/2\htwo\ ratios seem to be common at high redshift and remain unexplained~\citep[see][]{Tumlinson2010}.

Fourteen $\sim$~1~hour exposures have been taken with attached ThAr calibrations in order to reduce systematics in the wavelength calibration. We used this data to constrain the variation of the $\mu$ at the redshift of the DLA. We selected 81 \htwo\ lines suitable for \dmm\ measurements. We find that the two velocity component model with untied Doppler parameters between different rotational levels can best model the absorption profiles of \htwo\ lines of this system. We measure \dmm$~=(7.4~\pm~4.3_{\rm stat}\pm5.1_{\rm sys})$~ppm for such a model. Our result is consistent with no variation of $\mu$ over the last 11.2 Gyr. If we quadratically add the systematic and statistical errors we get the total error of 6.7 ppm in our \dmm\ measurement. This is $\sim$ 30\% smaller than the error (10.1 ppm) in \dmm\ we have obtained in a recent study of an \htwo\ system towards HE 0027$-$1836 at $z_{abs} = 2.4018$ \citep{Rahmani2013}. The main reason for the smaller error of the current study is the wider $K_i$ range covered by the \htwo\ lines towards Q J 0643$-$5041 compared to those of HE 0027$-$1836. This is an important consideration that should be taken into account in selection of systems for the study of \dmm.

Fig.~\ref{fig:s_dmm} summarizes \dmm\ measurements based on different approaches at different redshifts. Our new measurement is consistent with all the other measurements based on \htwo\ lines. However, we also note that like the majority of other studies our \dmm\ is positive although consistent with zero.   

\citet{King2011} and \citet{vanWeerdenburg2011} used \htwo\ and \hd\ absorbers at respectively $z$~=~2.811 and 2.059 towards Q0528$-$250 and J2123$-$005 to find \dmm~=~$(0.3\pm3.2_{\rm stat}\pm1.9_{\rm sys})\times10^{-6}$ and \dmm~=~$(8.5\pm4.2)\times10^{-6}$. \citet{Wendt2012} and \citet{Rahmani2013} found \dmm~=~$(4.3\pm7.2)\times10^{-6}$ and \dmm~=~$-(7.6\pm10.2)\times10^{-6}$ using the \htwo\ absorber at $z$~=~3.025 and 2.4018 towards Q0347$-$383 and HE~0027$-$1836. 

\citet{King2008} have found  \dmm~=~$(10.9\pm7.1)\times10^{-6}$  at $z$~=~2.595 towards Q0405$-$443. Using these measurements and ours we find the weighted mean of \dmm~=~$(4.5\pm2.2)\times10^{-6}$. As recently wavelength dependent drifts has been reported in UVES observations this can bias the \dmm\ values to positive values. Therefore, caution should be exercised while interpreting such results as whole.

The best constraints on \dmm\ have been achieved by using NH$_3$ or CH$_3$OH absorption lines~\citep{Murphy2008,Henkel2009,Kanekar2011,Bagdonaite2013}. Very high sensitivity of the inversion transitions associated with these molecules to \dmm\ leads to constraint of the order of 10$^{-7}$. The main drawback of this method is that only two systems at $z < 1$ provide the opportunity to carry out such measurements. Based on 21-cm absorption we found \dmm\ = 0.0 $\pm$ 1.5 ppm (at $z \sim$ 1.3 by \citet{Rahmani2012}) and $-(1.7\pm1.7)$ ppm (at $z \sim 3.2$ by \citet{Srianand2010}). While these measurements are more stringent than those based on \htwo\ one has to assume no variation of $\alpha$ and $g_p$ to get constraint on \dmm, which is not the case in high redshift \htwo\ absorption systems.

A further cumulation of data on the line of sight of \qso\ would allow to establish the actuality of partial coverage of the BLR by the \htwo-bearing cloud as well as the possible detection of \hd. However, it seems rather difficult to go much beyond our analysis on the variation of $\mu$ with UVES data. Indeed, systematic drifts in wavelength calibration over the whole blue arm of UVES can only be unveiled by absolute reference observations (of asteroids or an iodine cell, for example) in the same conditions as science observations to correct for instrumental misbehaviours. Such a systematic effect could be the origin of the preferred positive value of \dmm\ seen in this and other similar works. In order to shed light on the actual value of $\mu$ at high redshift, it is crucial to (re-)observe systems, such as the one presented here, with as much control in wavelength calibration as possible to account for systematics that escape our means in the present state of the art.


\section*{Acknowledgement}
\begin{acknowledgements}
This research was supported by the Agence Nationale pour la Recherche under Prog.~ANR-10-BLAN-510-01. R. S. and P. P. J. gratefully acknowledge support from the Indo-French Centre for the Promotion of Advanced Research (Centre Franco-Indien pour la Promotion de la Recherche Avanc\'ee) under contract No. 4304-2. H. R. would like to thank the Institute for research in Fundamental Sciences (IPM, Tehran) for hospitality and providing facilities during his visit in August 2013. The authors would like to thank the referee, Paolo Molaro, for useful comments and discussion.
\end{acknowledgements}


\bibliographystyle{aa}
\bibliography{Q-J0643}


\appendix


\section{Spectral features}
We present here the spectrum bits containing some of the absorption features that were fitted for the system at \zabs$\simeq2.6586$. 

\begin{figure}[tb]
\centering
\hspace*{-0.2cm}
\includegraphics[scale=0.15,natwidth=6cm,natheight=6cm]{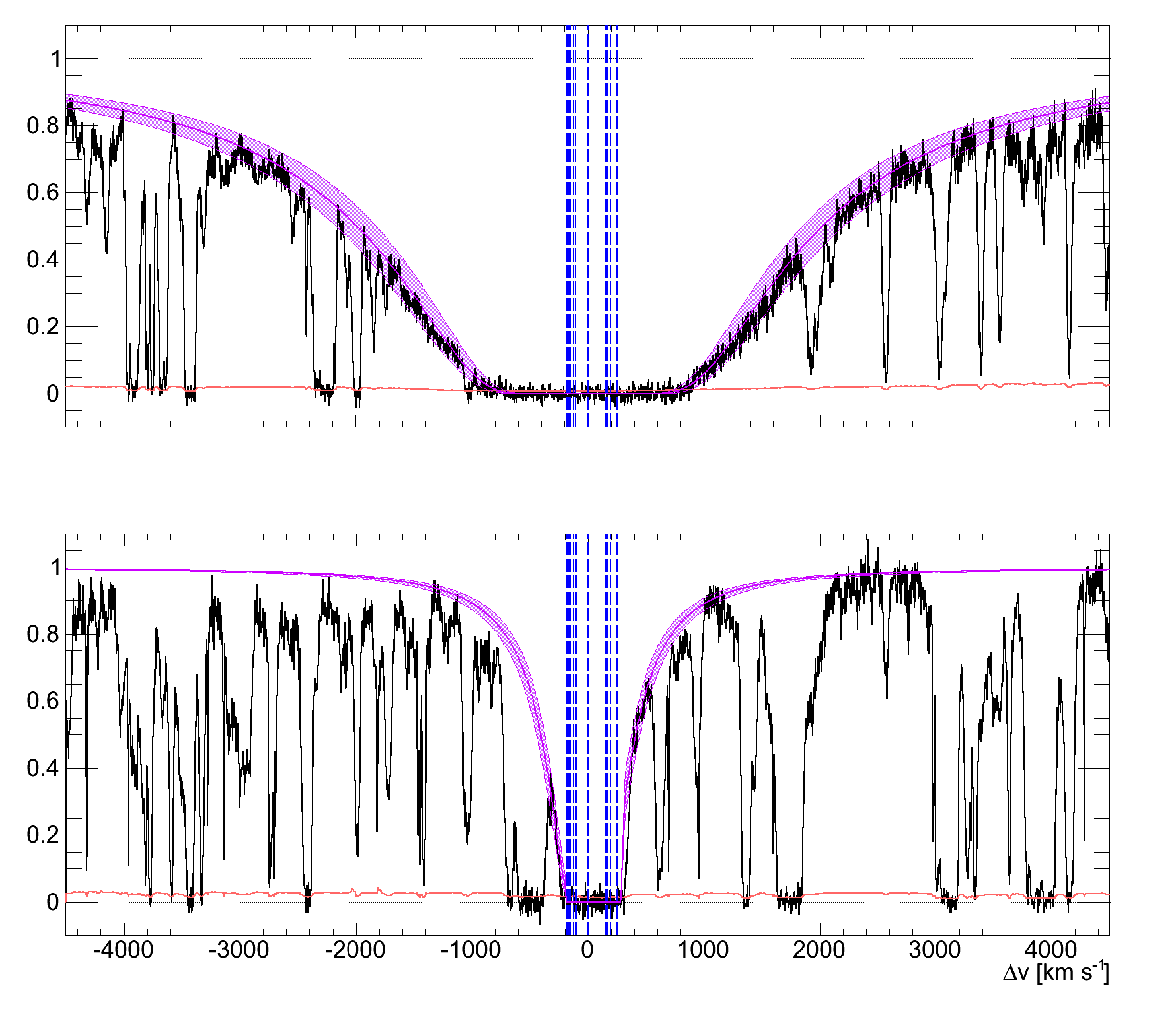}
\caption{Voigt profile fits to the DLA: \Lya\ transition on top, \Lyb\ at lower panel. The fit to the data together with its uncertainty is shown in the shaded area. Vertical dashed lines mark the position of the \hi\ components used for the fit, determined both from the wings of \Lya\ and \Lyb\ and the profile of higher Lyman transitions. The observational error is shown at the bottom for reference.}
\label{fig:DLA}
\end{figure}

Fig.~\ref{fig:DLA} shows the \hi\ absorption together with the statistical error associated to the fit represented by the shaded region.

\begin{figure}[tb]
\centering
\includegraphics[scale=0.122,natwidth=6cm,natheight=6cm]{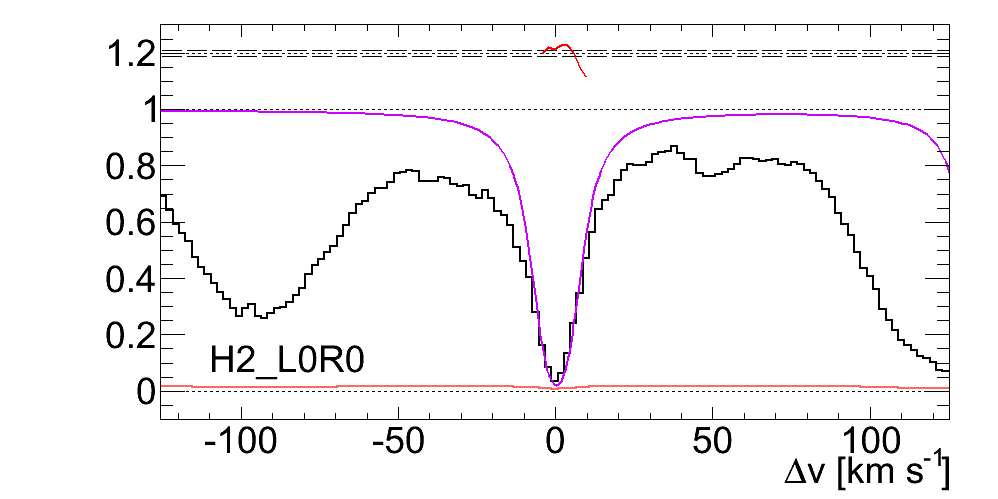}
\includegraphics[scale=0.122,natwidth=6cm,natheight=6cm]{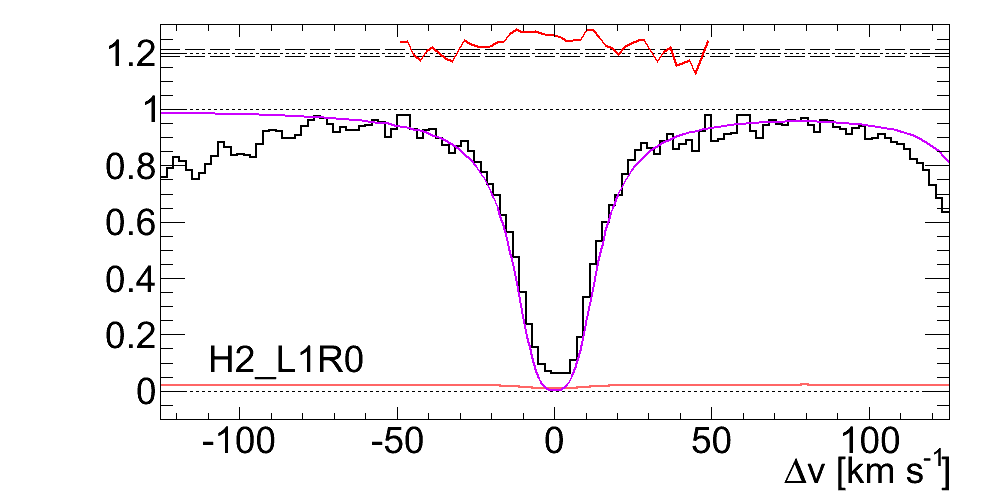}
\includegraphics[scale=0.122,natwidth=6cm,natheight=6cm]{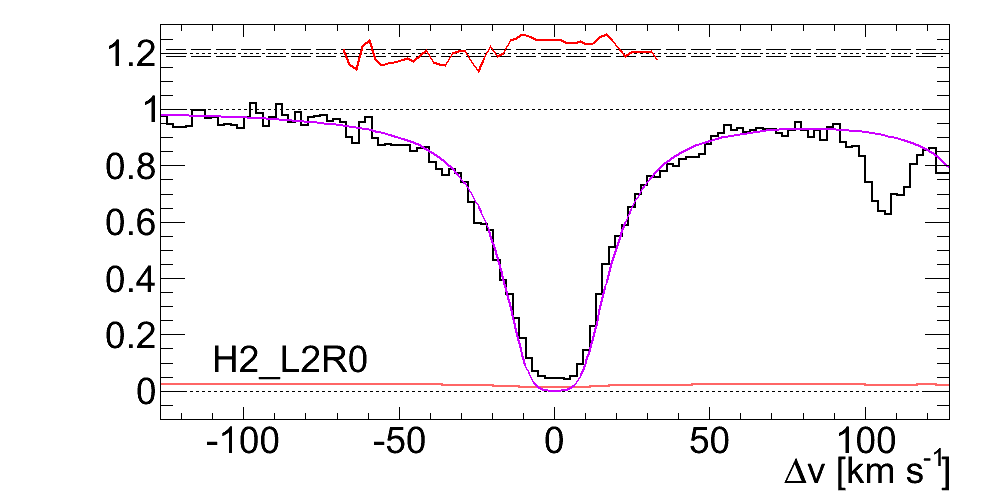}
\includegraphics[scale=0.122,natwidth=6cm,natheight=6cm]{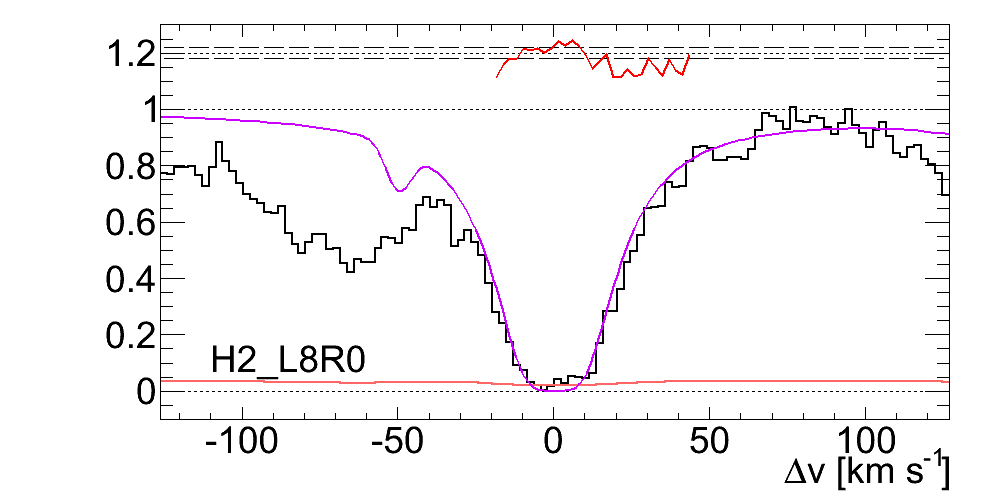}
\includegraphics[scale=0.122,natwidth=6cm,natheight=6cm]{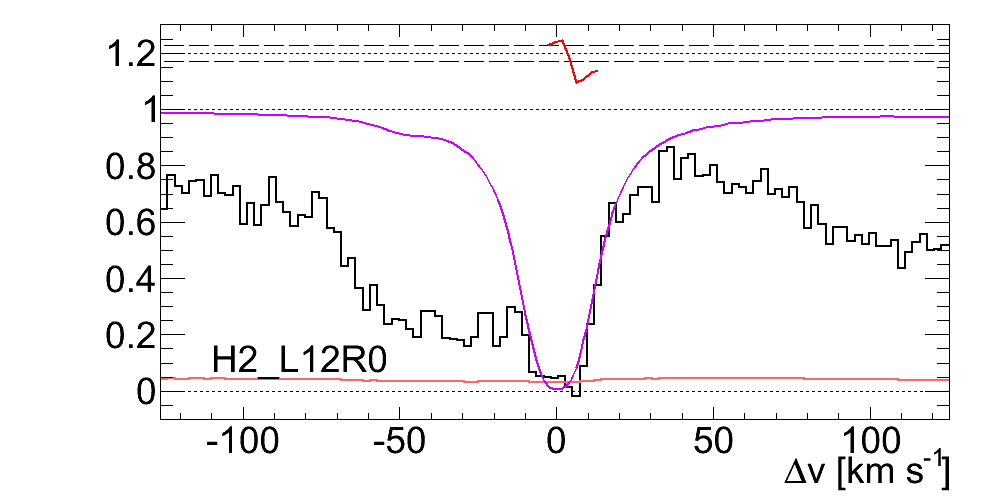}
\includegraphics[scale=0.122,natwidth=6cm,natheight=6cm]{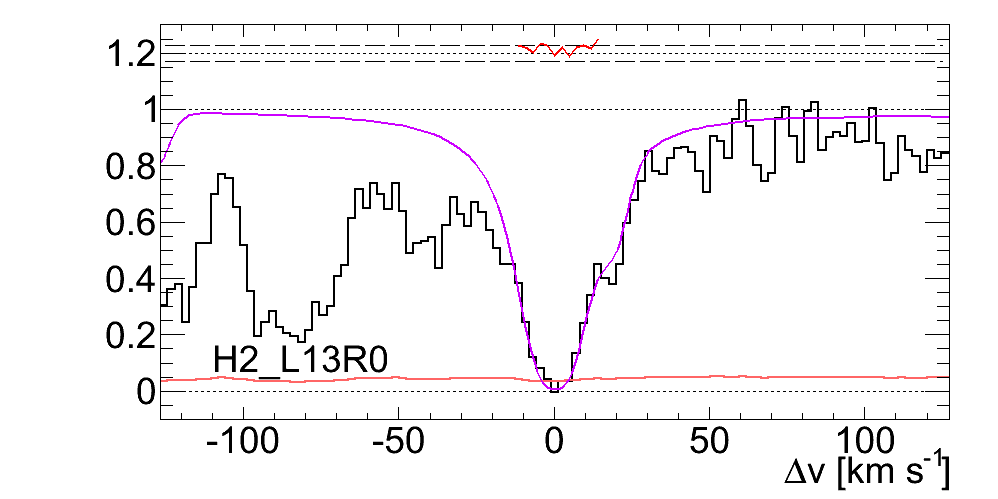}
\caption{Voigt profile fits to \htwo\ J = 0. The fit to the data is represented by a line. Residuals of the fit are shown on top. The observational error is shown at the bottom for reference. Presence of residual flux is obvious in a few cases.}
\label{fig:H2J0}
\end{figure}

\begin{figure}[tb]
\centering
\includegraphics[scale=0.122,natwidth=6cm,natheight=6cm]{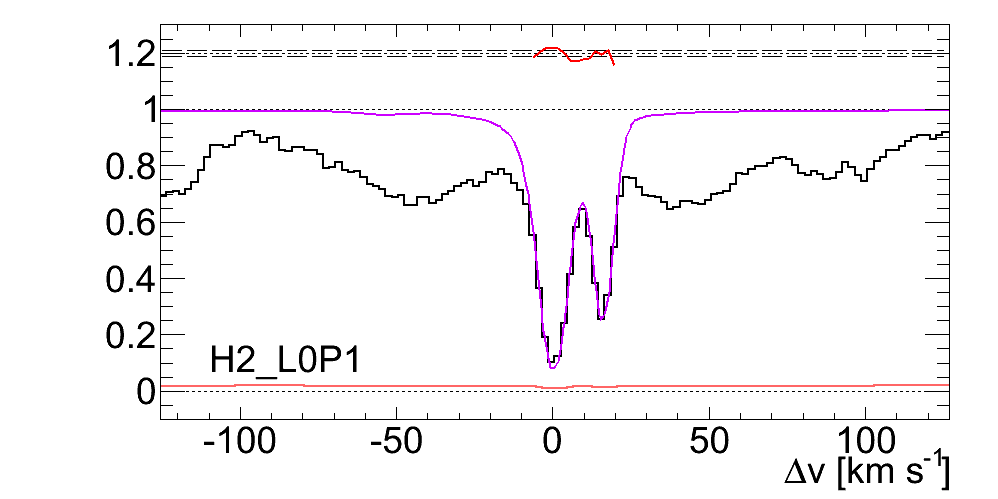}
\includegraphics[scale=0.122,natwidth=6cm,natheight=6cm]{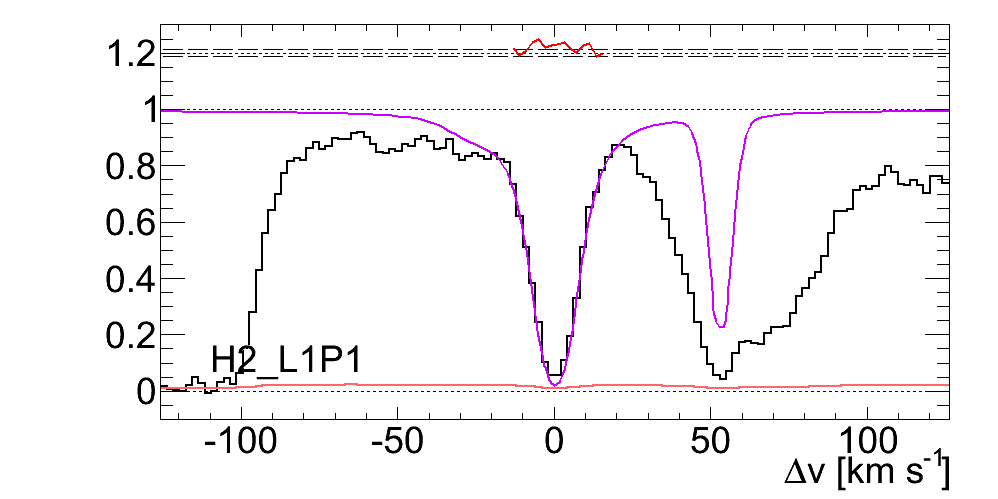}
\includegraphics[scale=0.122,natwidth=6cm,natheight=6cm]{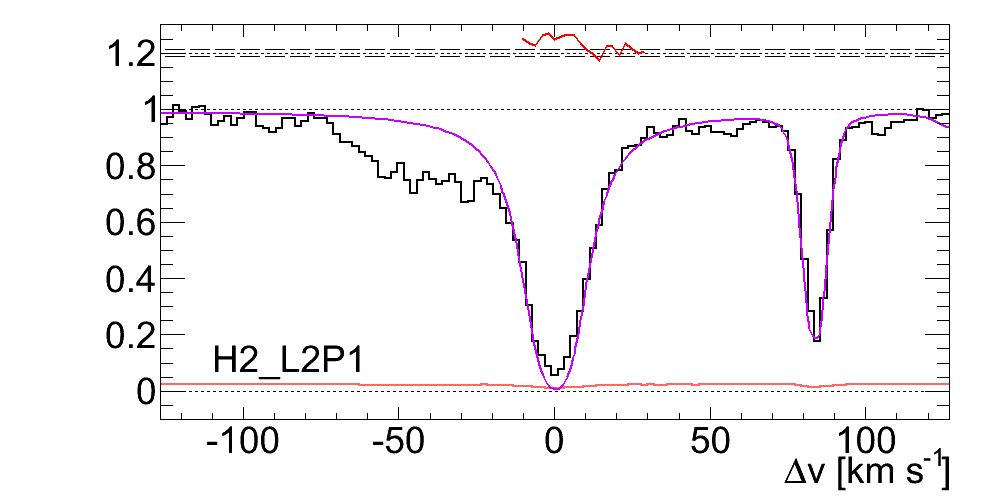}
\includegraphics[scale=0.122,natwidth=6cm,natheight=6cm]{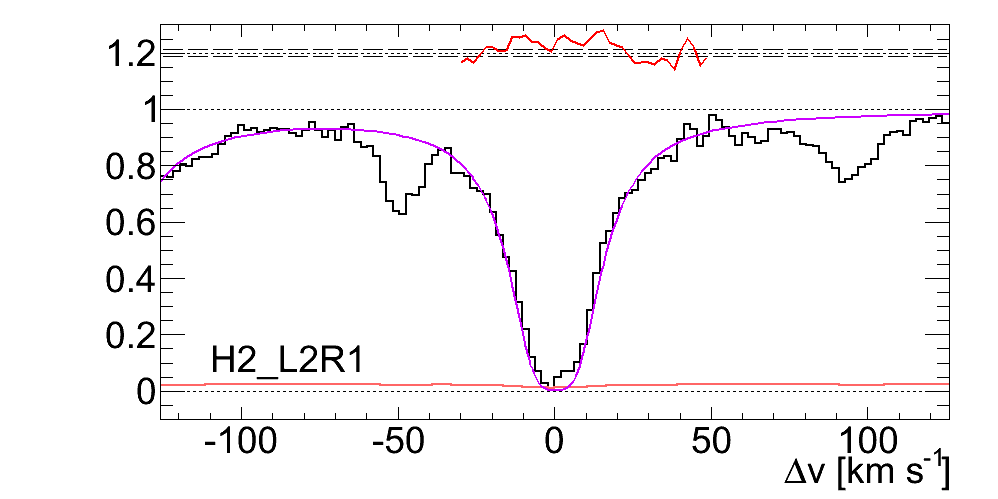}
\includegraphics[scale=0.122,natwidth=6cm,natheight=6cm]{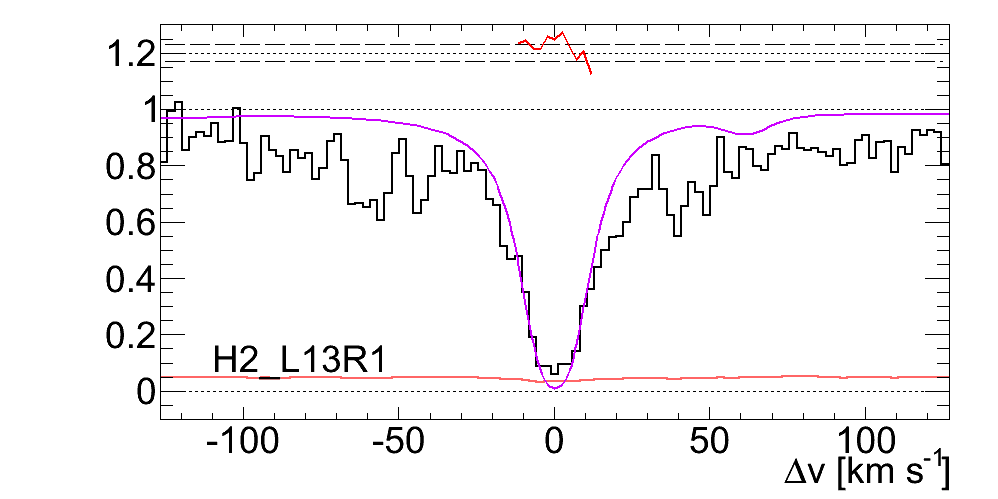}
\includegraphics[scale=0.122,natwidth=6cm,natheight=6cm]{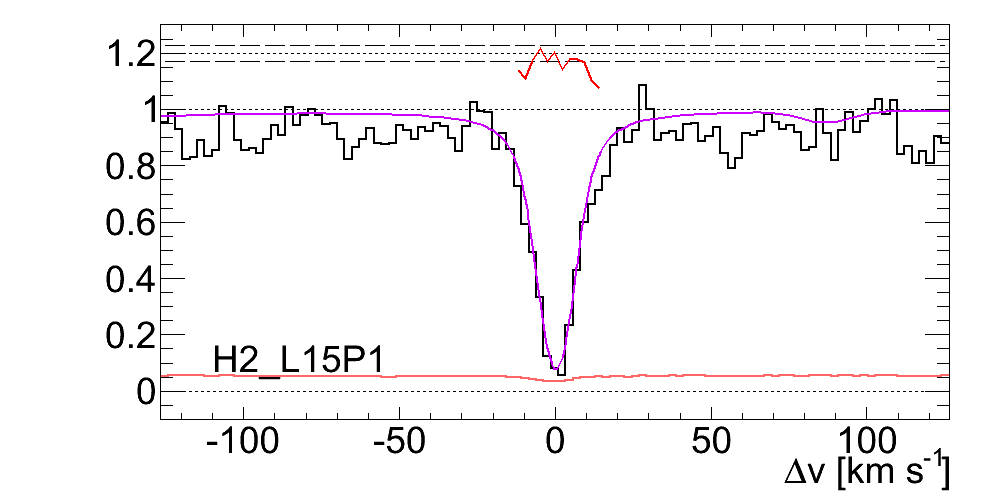}
\includegraphics[scale=0.122,natwidth=6cm,natheight=6cm]{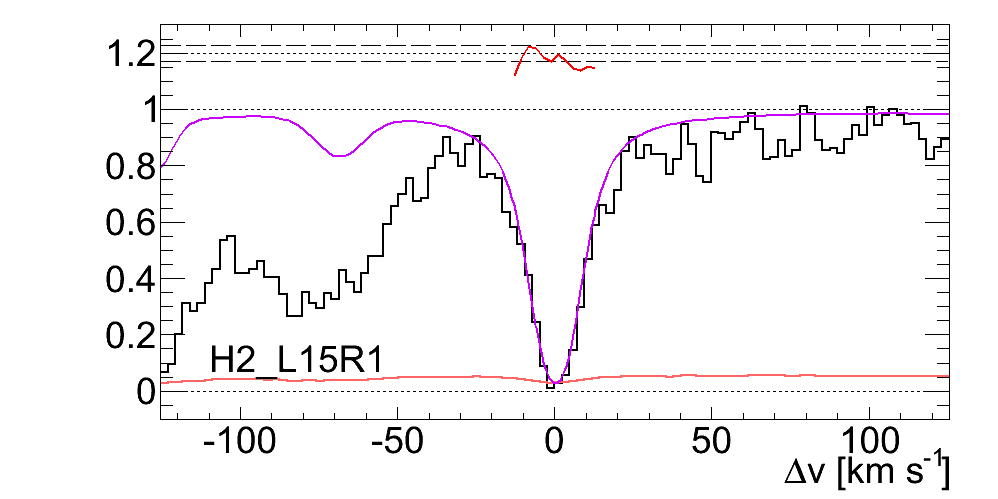}
\includegraphics[scale=0.122,natwidth=6cm,natheight=6cm]{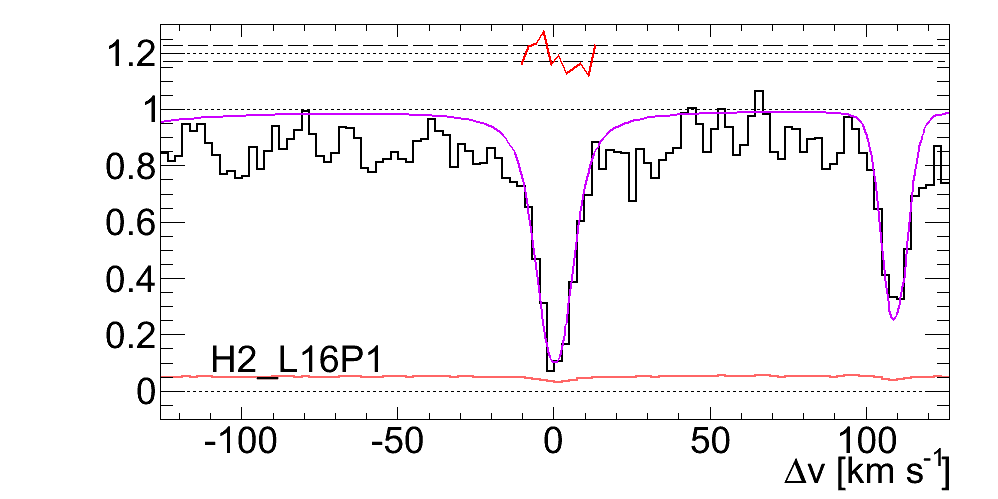}
\includegraphics[scale=0.122,natwidth=6cm,natheight=6cm]{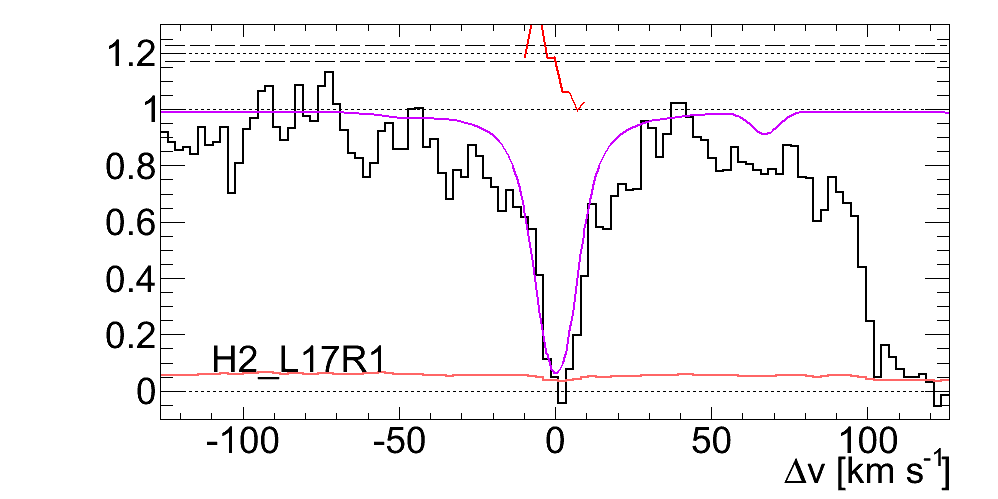}
\caption{Voigt profile fits to \htwo\ J = 1. The fit to the data is represented by a line. Residuals of the fit are shown on top. The observational error is shown at the bottom for reference.}
\label{fig:H2J1}
\end{figure}

\begin{figure}[tb]
\centering
\includegraphics[scale=0.122,natwidth=6cm,natheight=6cm]{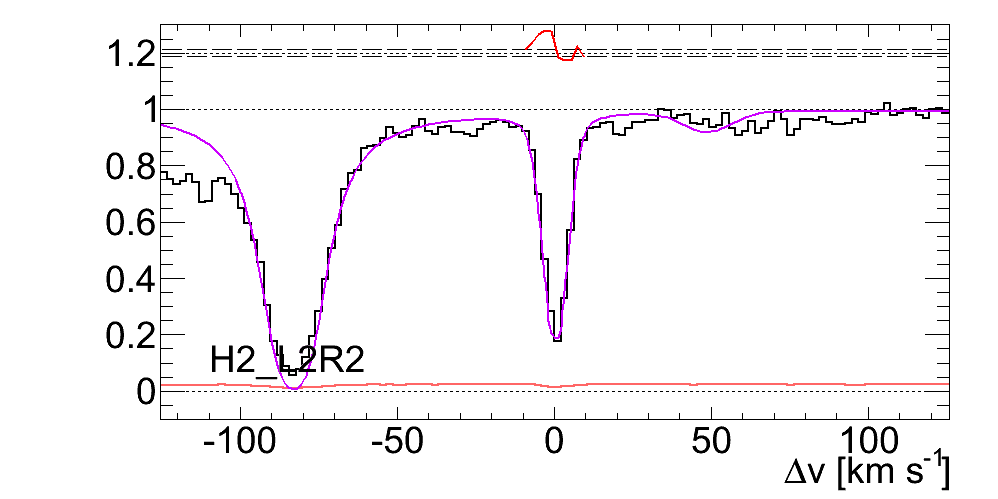}
\includegraphics[scale=0.122,natwidth=6cm,natheight=6cm]{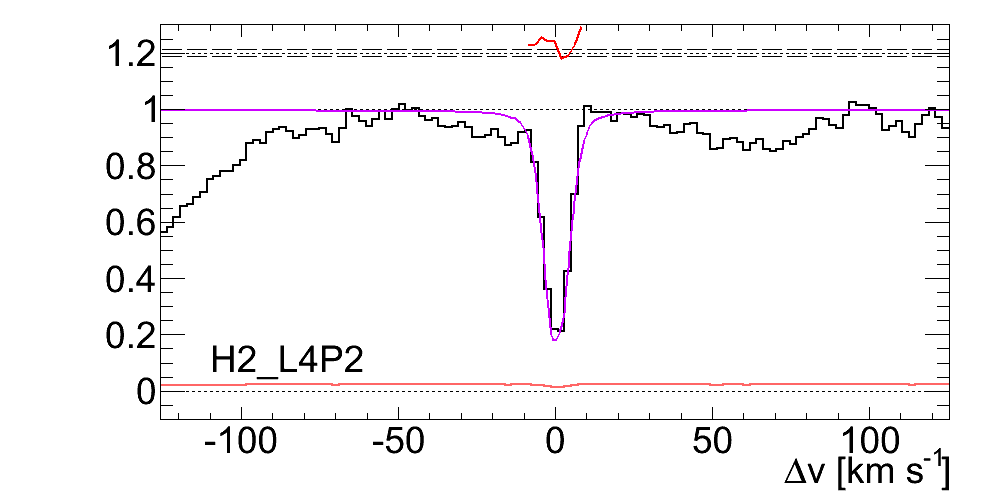}
\includegraphics[scale=0.122,natwidth=6cm,natheight=6cm]{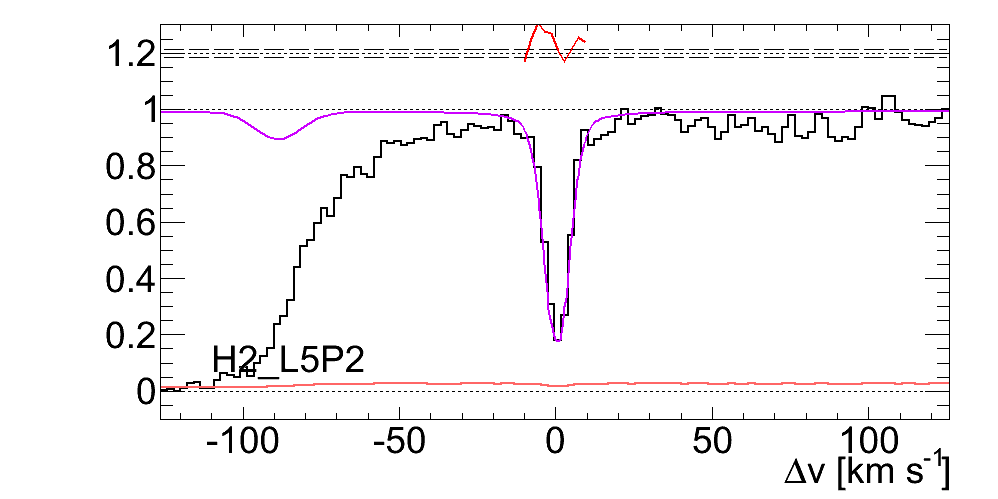}
\includegraphics[scale=0.122,natwidth=6cm,natheight=6cm]{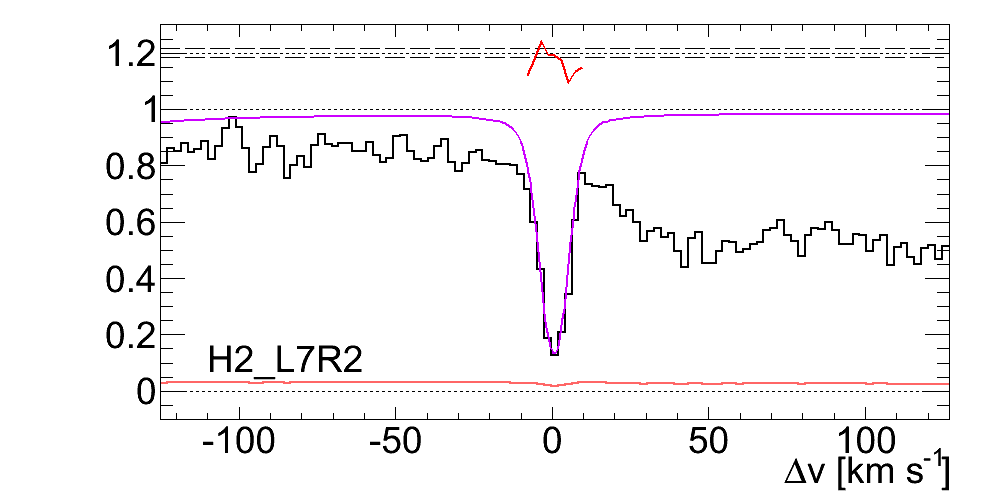}
\includegraphics[scale=0.122,natwidth=6cm,natheight=6cm]{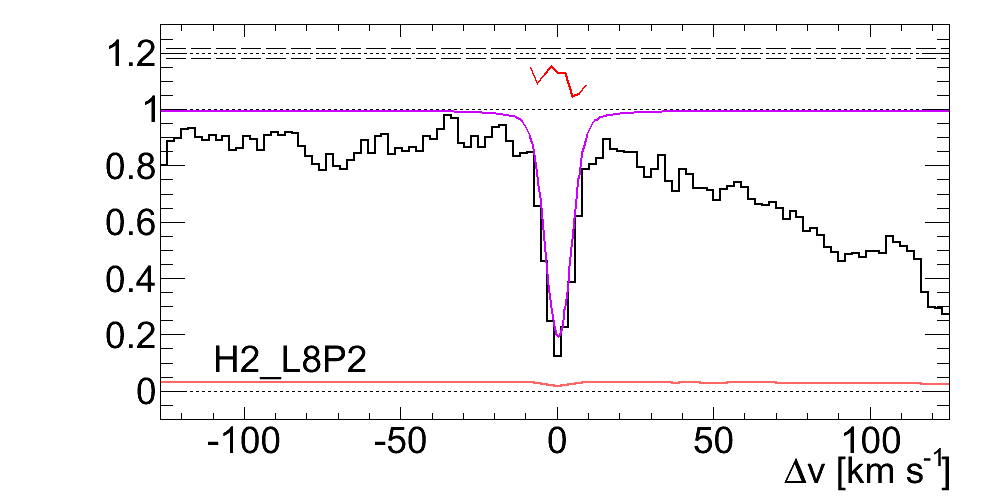}
\includegraphics[scale=0.122,natwidth=6cm,natheight=6cm]{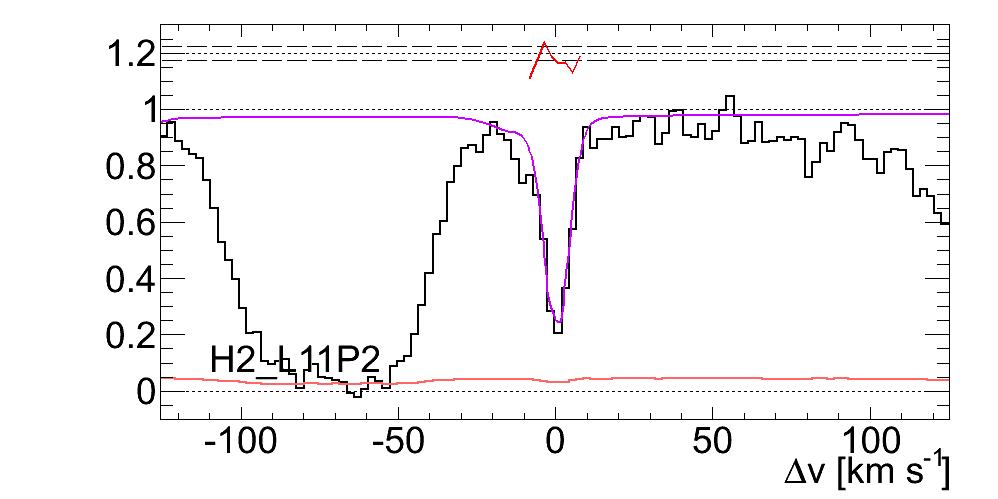}
\includegraphics[scale=0.122,natwidth=6cm,natheight=6cm]{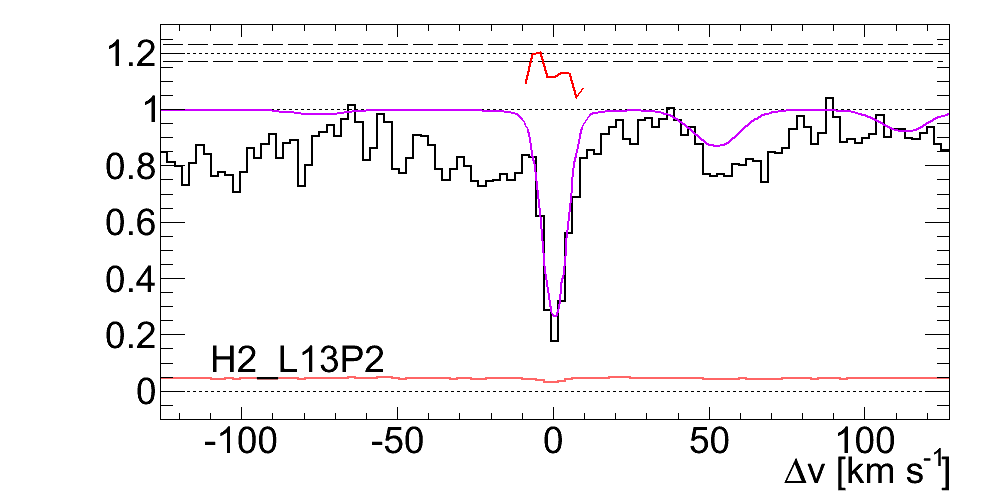}
\includegraphics[scale=0.122,natwidth=6cm,natheight=6cm]{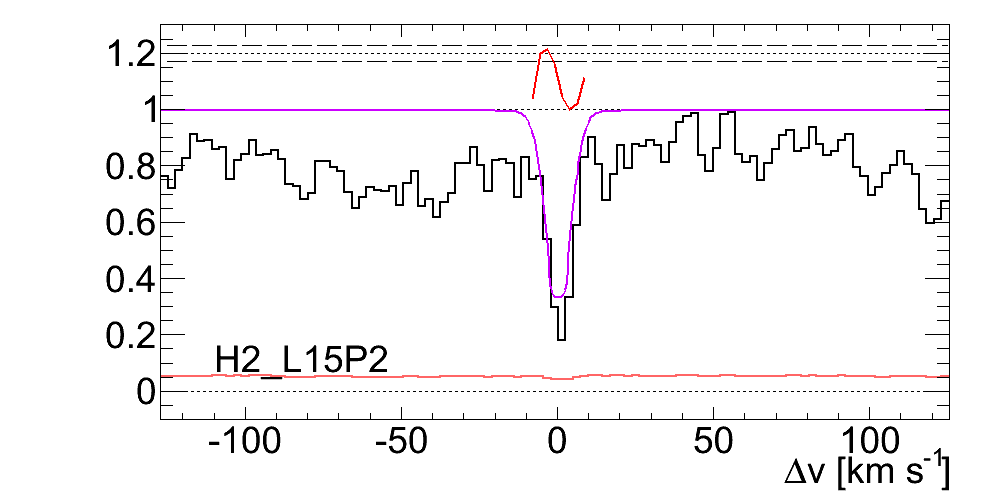}
\includegraphics[scale=0.122,natwidth=6cm,natheight=6cm]{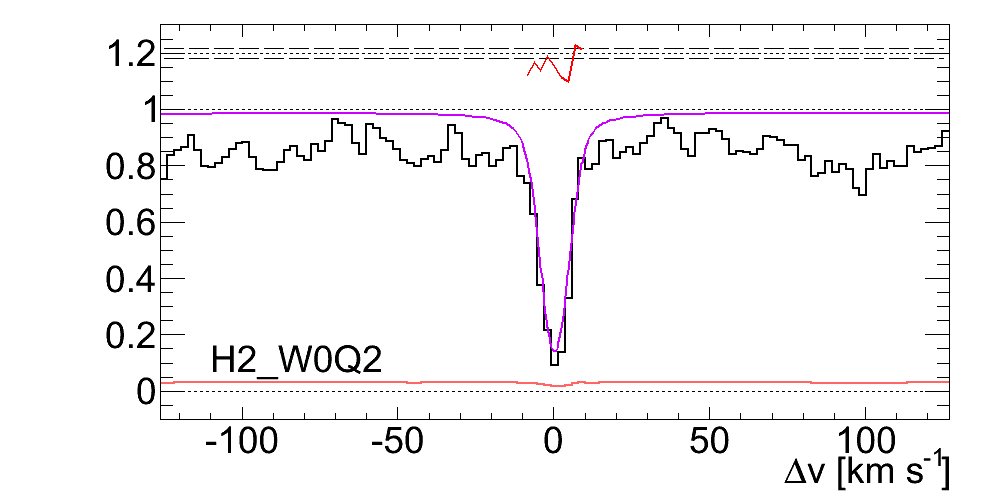}
\includegraphics[scale=0.122,natwidth=6cm,natheight=6cm]{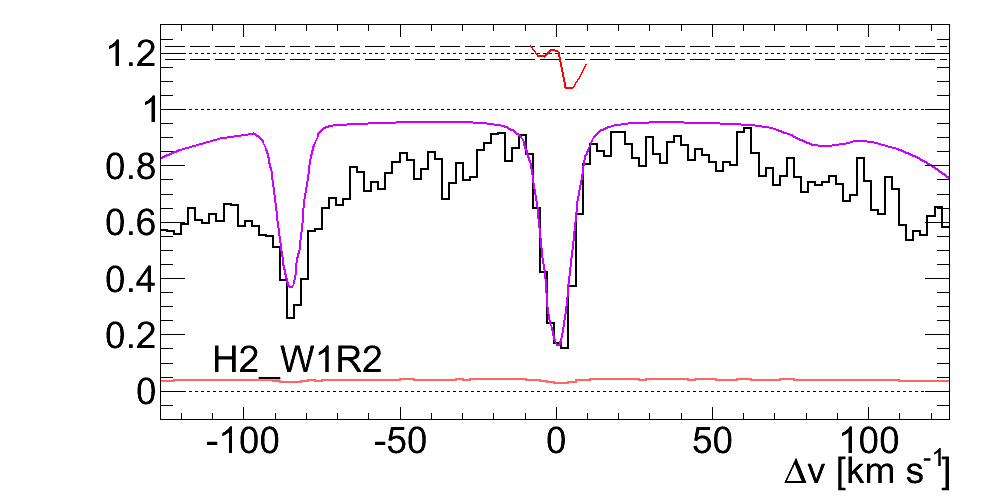}
\includegraphics[scale=0.122,natwidth=6cm,natheight=6cm]{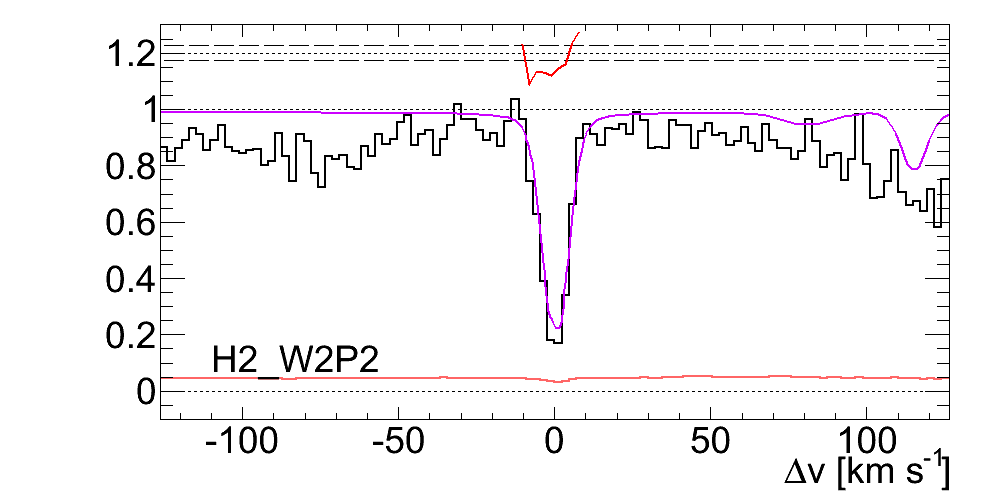}
\includegraphics[scale=0.122,natwidth=6cm,natheight=6cm]{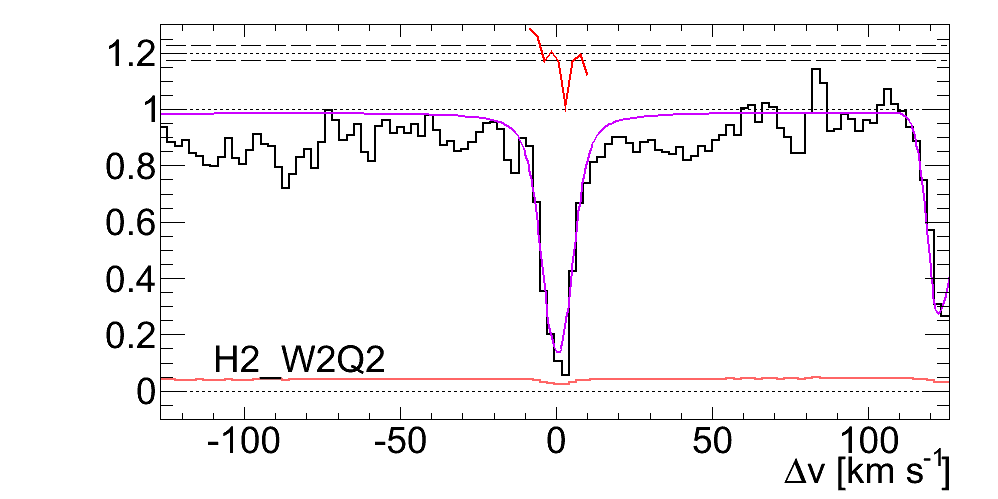}
\caption{Voigt profile fits to \htwo\ J = 2. The fit to the data is represented by a line. Residuals of the fit are shown on top. The observational error is shown at the bottom for reference.}
\label{fig:H2J2}
\end{figure}

\begin{figure}[tb]
\centering
\includegraphics[scale=0.122,natwidth=6cm,natheight=6cm]{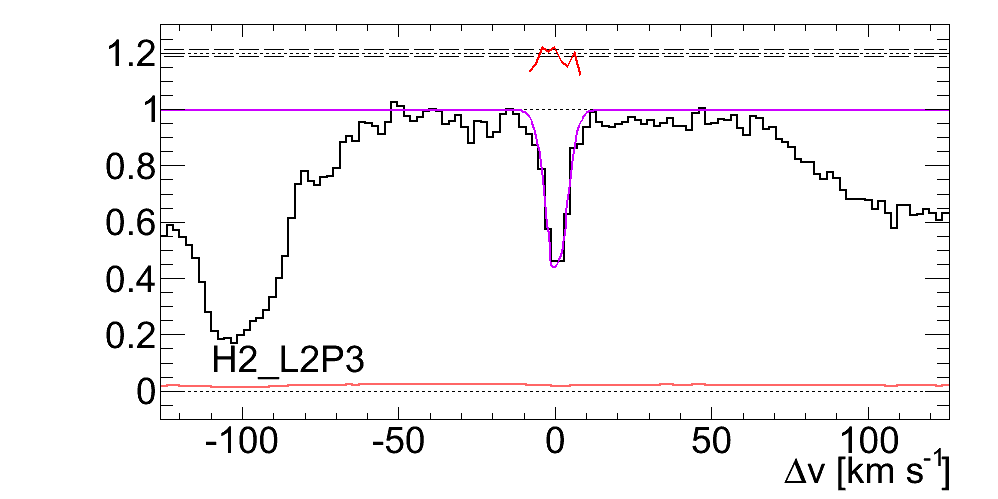}
\includegraphics[scale=0.122,natwidth=6cm,natheight=6cm]{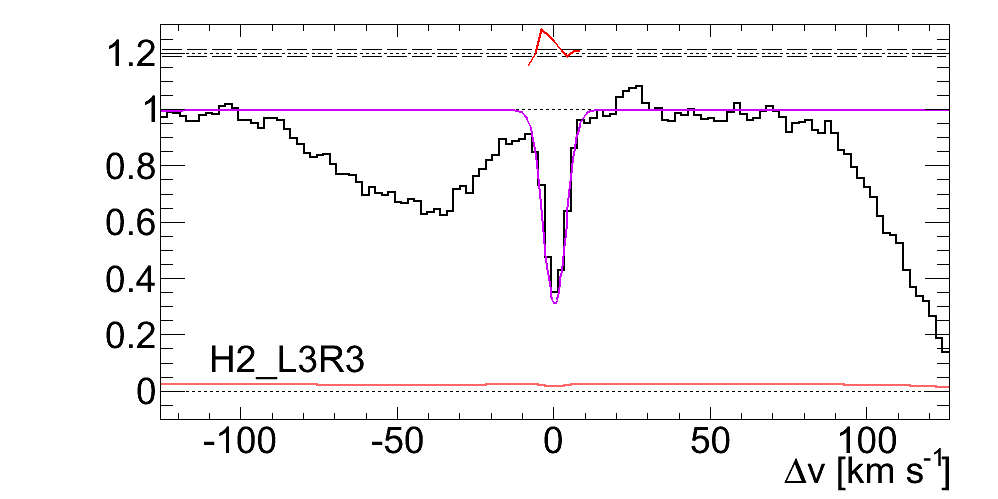}
\includegraphics[scale=0.122,natwidth=6cm,natheight=6cm]{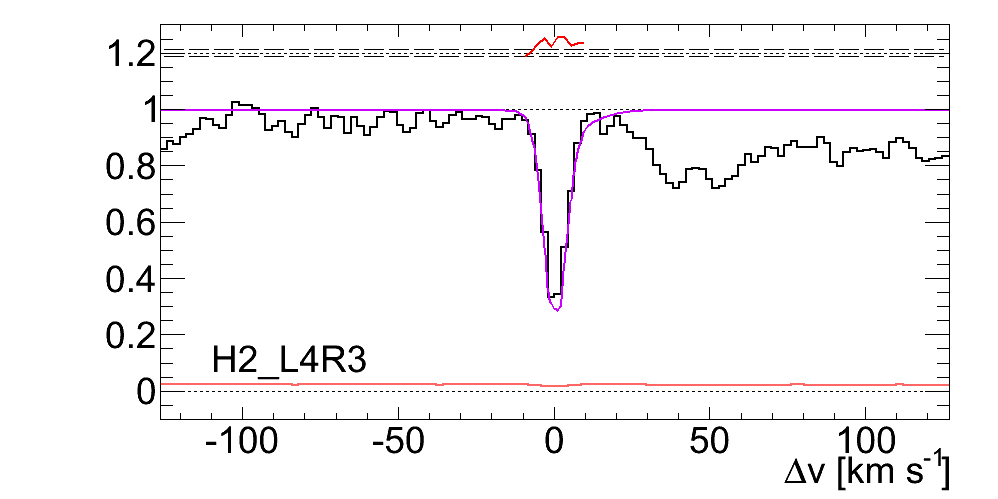}
\includegraphics[scale=0.122,natwidth=6cm,natheight=6cm]{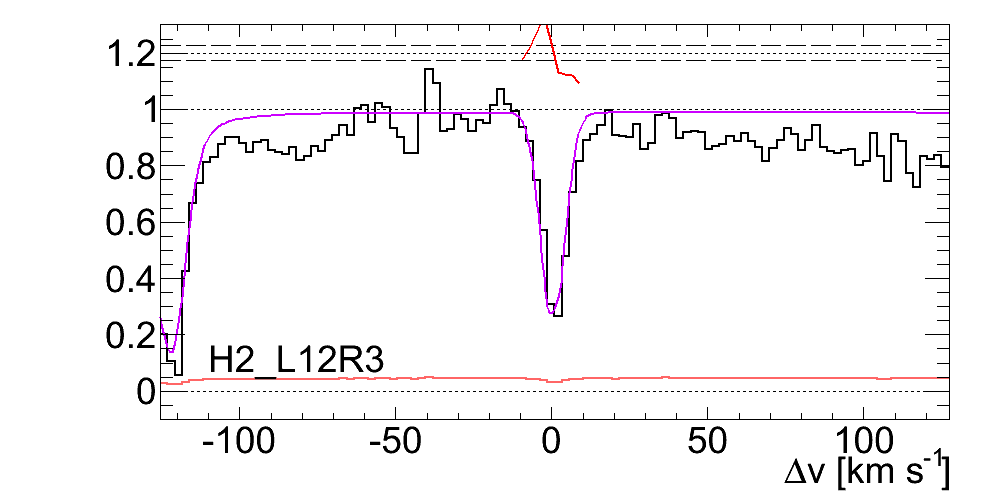}
\includegraphics[scale=0.122,natwidth=6cm,natheight=6cm]{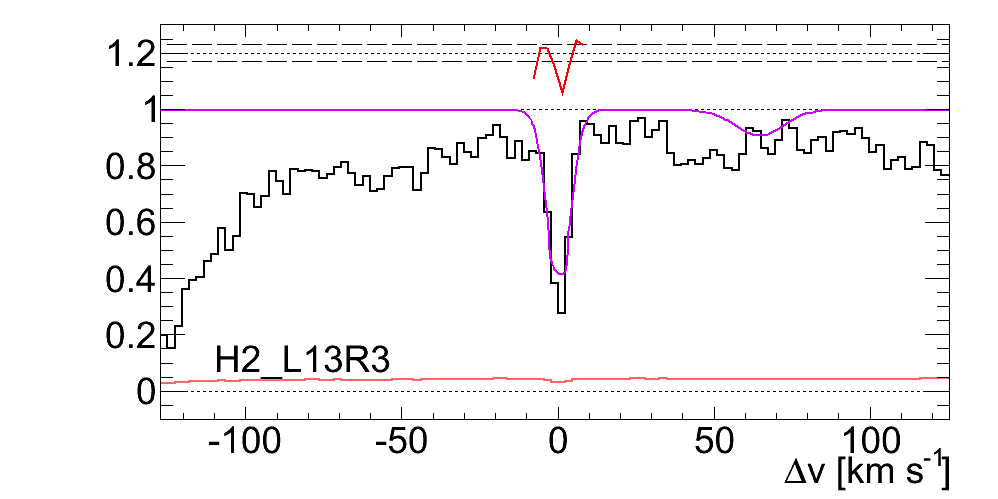}
\includegraphics[scale=0.122,natwidth=6cm,natheight=6cm]{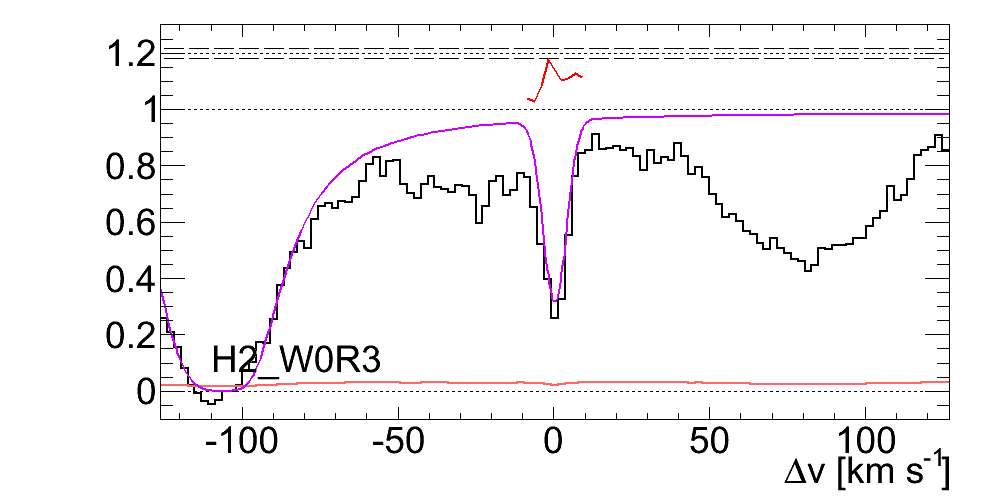}
\includegraphics[scale=0.122,natwidth=6cm,natheight=6cm]{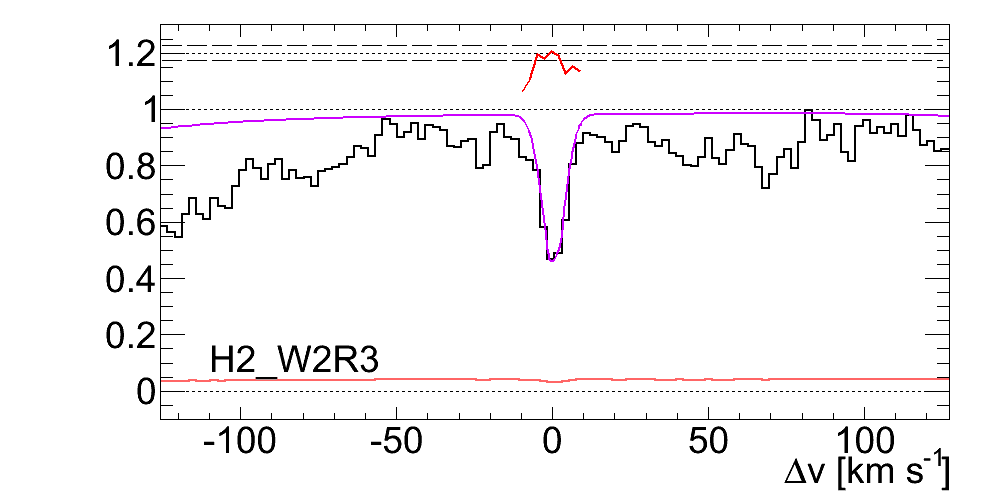}
\includegraphics[scale=0.122,natwidth=6cm,natheight=6cm]{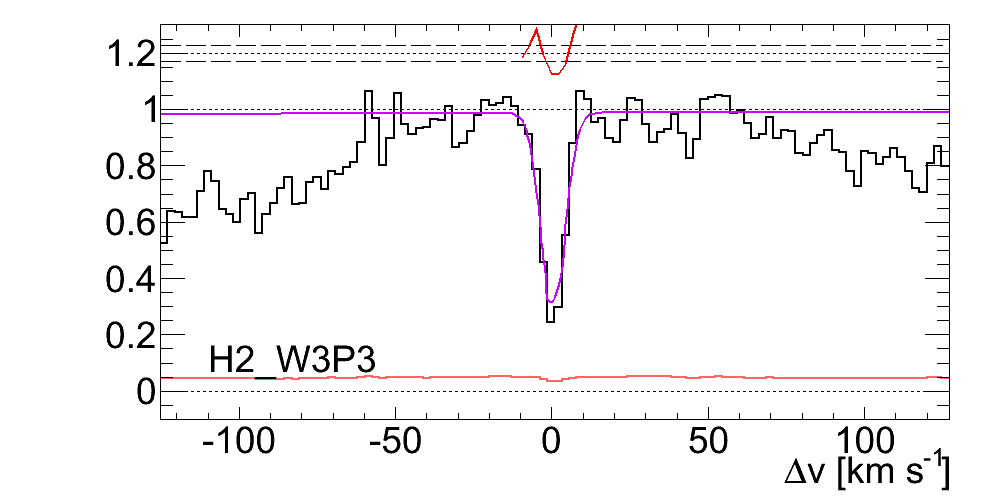}
\caption{Voigt profile fits to \htwo\ J = 3. The fit to the data is represented by a line. Residuals of the fit are shown on top. The observational error is shown at the bottom for reference.}
\label{fig:H2J3}
\end{figure}

\begin{figure}[tb]
\centering
\includegraphics[scale=0.122,natwidth=6cm,natheight=6cm]{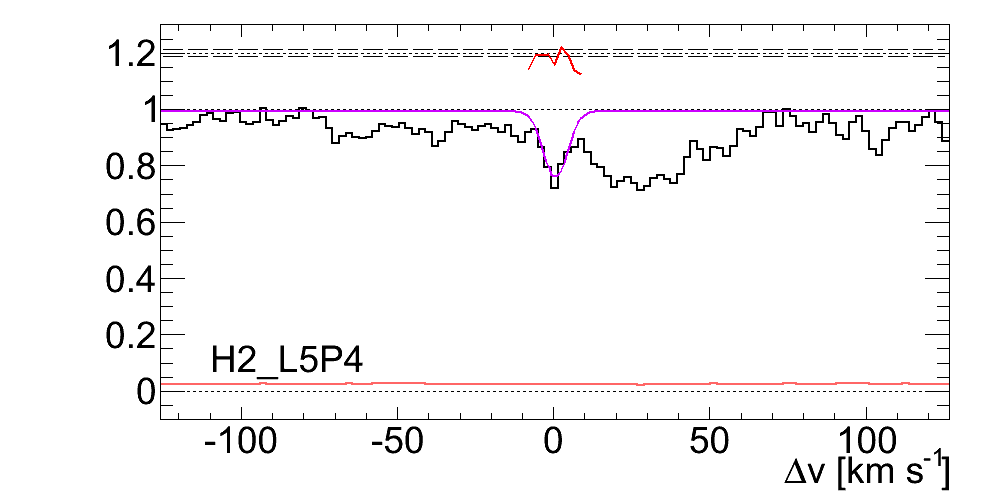}
\includegraphics[scale=0.122,natwidth=6cm,natheight=6cm]{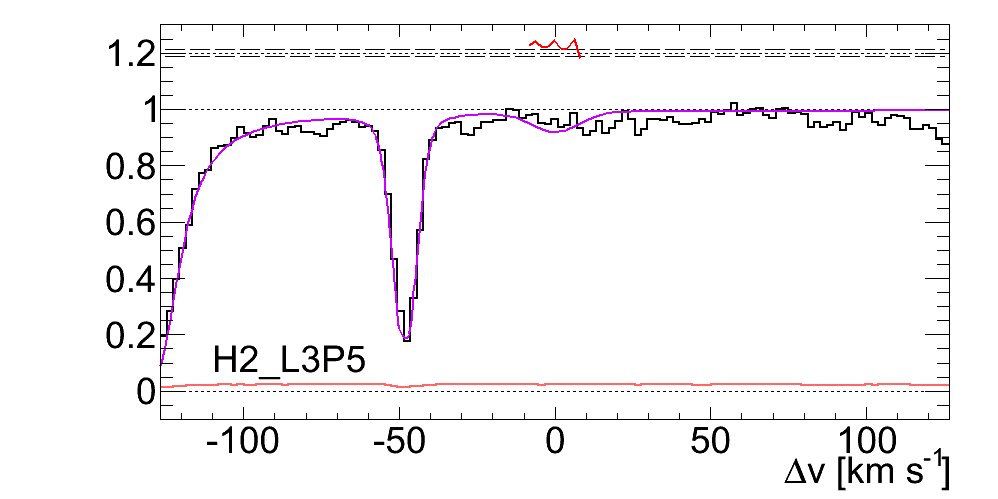}
\includegraphics[scale=0.122,natwidth=6cm,natheight=6cm]{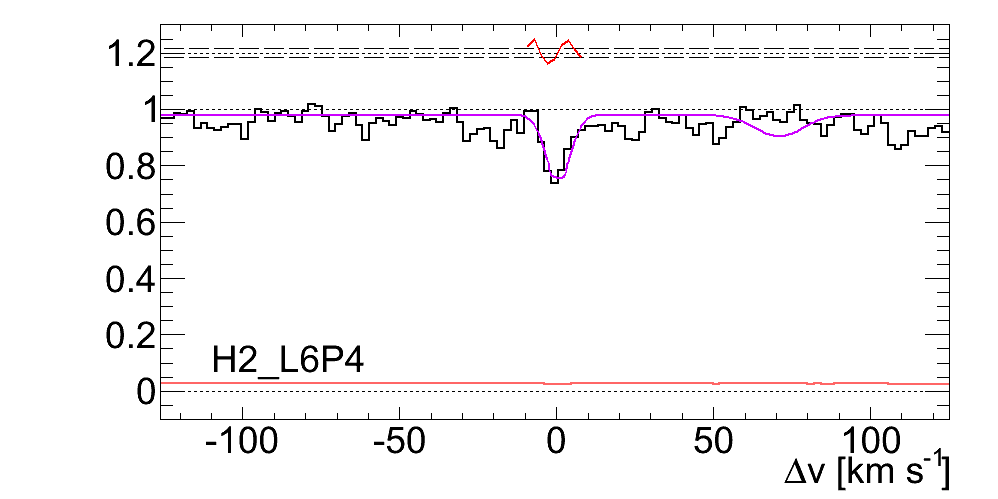}
\includegraphics[scale=0.122,natwidth=6cm,natheight=6cm]{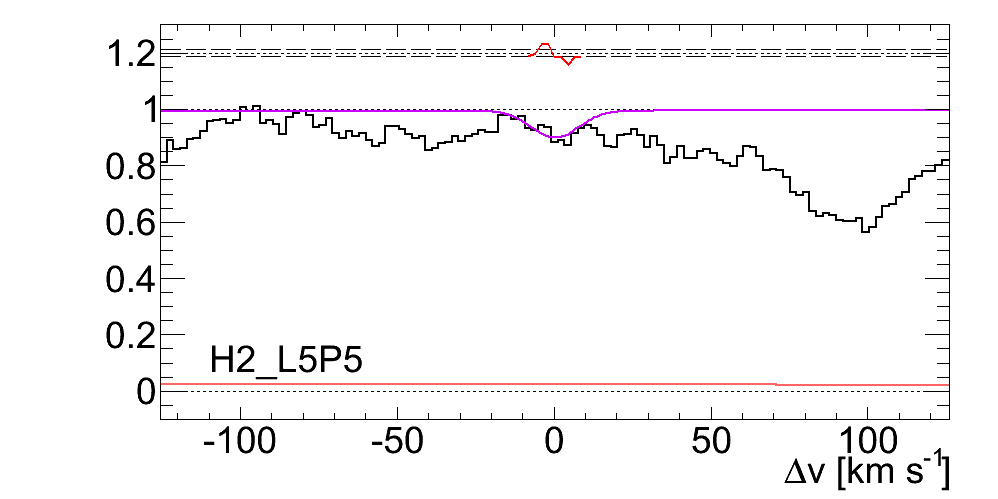}
\includegraphics[scale=0.122,natwidth=6cm,natheight=6cm]{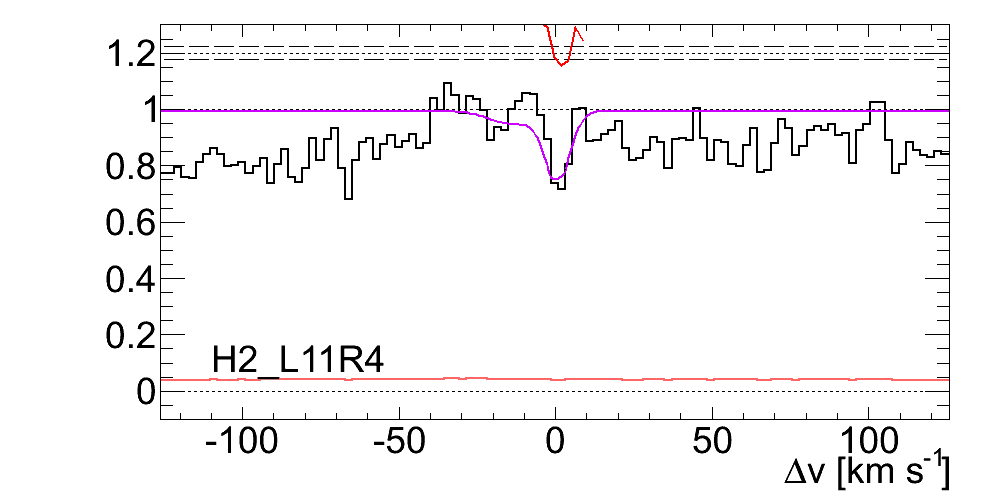}
\includegraphics[scale=0.122,natwidth=6cm,natheight=6cm]{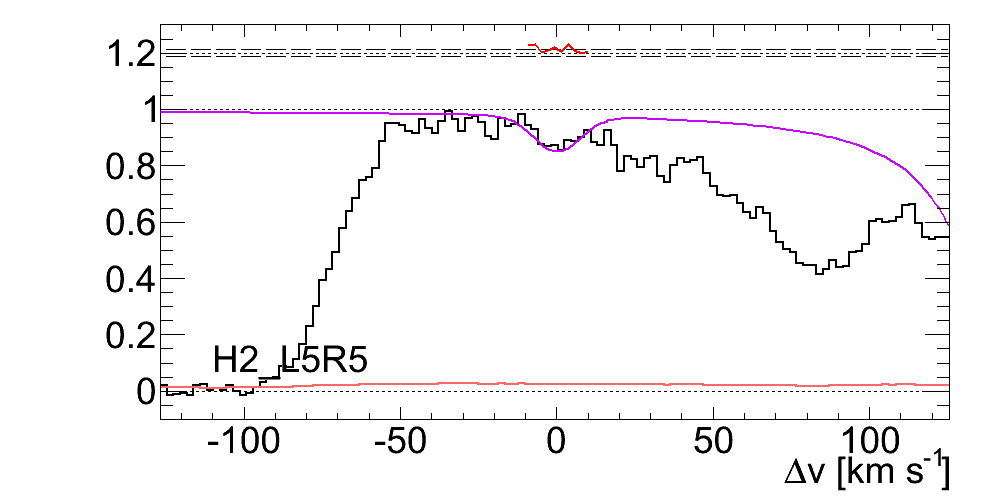}
\caption{Voigt profile fits to \htwo\ J = 4 and 5. The fit to the data is represented by a line. Residuals of the fit are shown on top. The observational error is shown at the bottom for reference.}
\label{fig:H2J45}
\end{figure}

Figs.~\ref{fig:H2J0}, \ref{fig:H2J1}, \ref{fig:H2J2}, \ref{fig:H2J3} and~\ref{fig:H2J45} show the \htwo\ lines ordered by J-level.

\begin{figure}[tb]
\centering
\includegraphics[scale=0.122,natwidth=6cm,natheight=6cm]{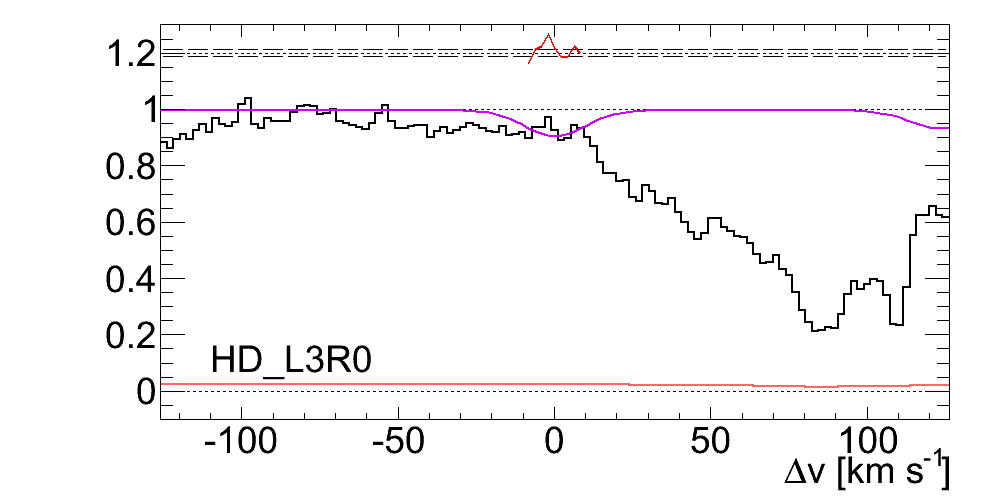}
\includegraphics[scale=0.122,natwidth=6cm,natheight=6cm]{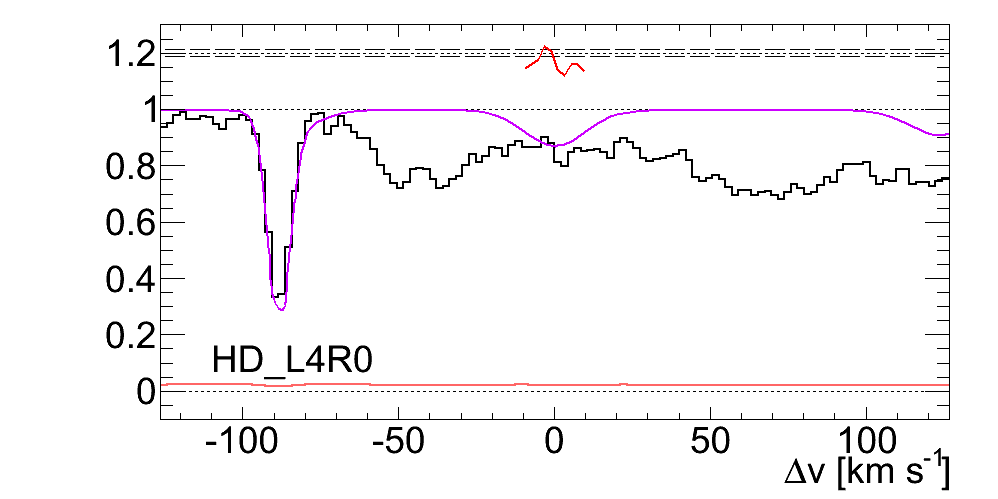}
\includegraphics[scale=0.122,natwidth=6cm,natheight=6cm]{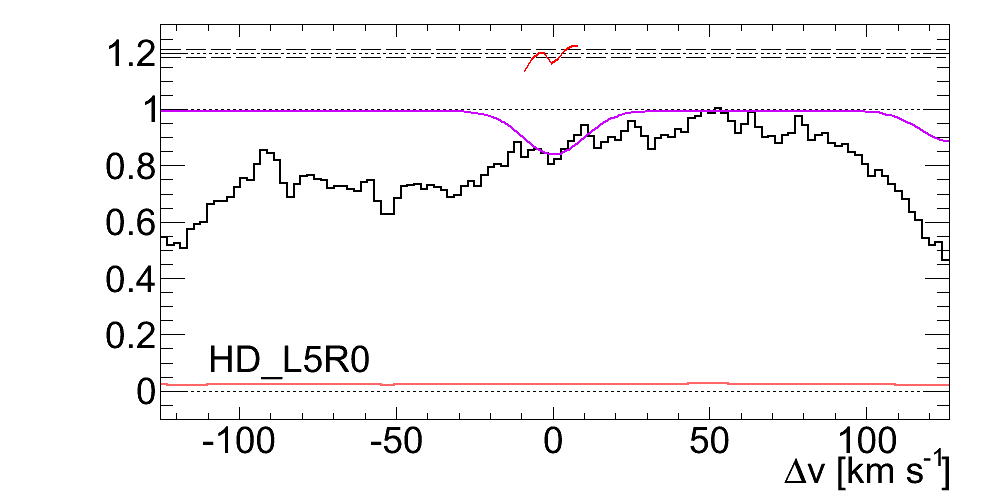}
\includegraphics[scale=0.122,natwidth=6cm,natheight=6cm]{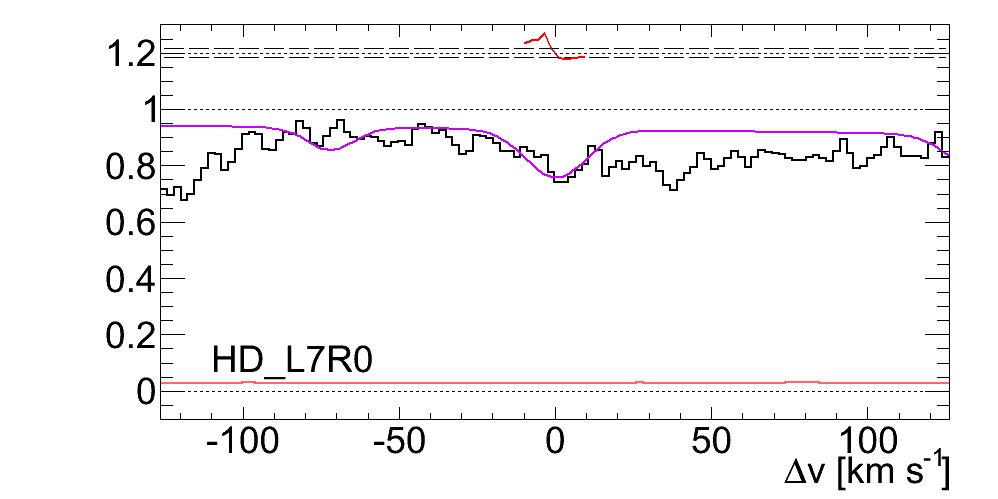}
\includegraphics[scale=0.122,natwidth=6cm,natheight=6cm]{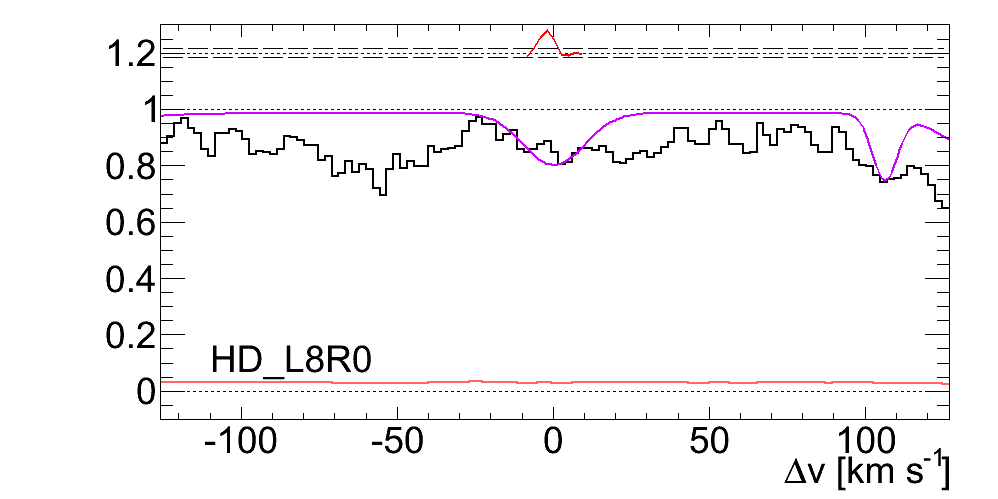}
\includegraphics[scale=0.122,natwidth=6cm,natheight=6cm]{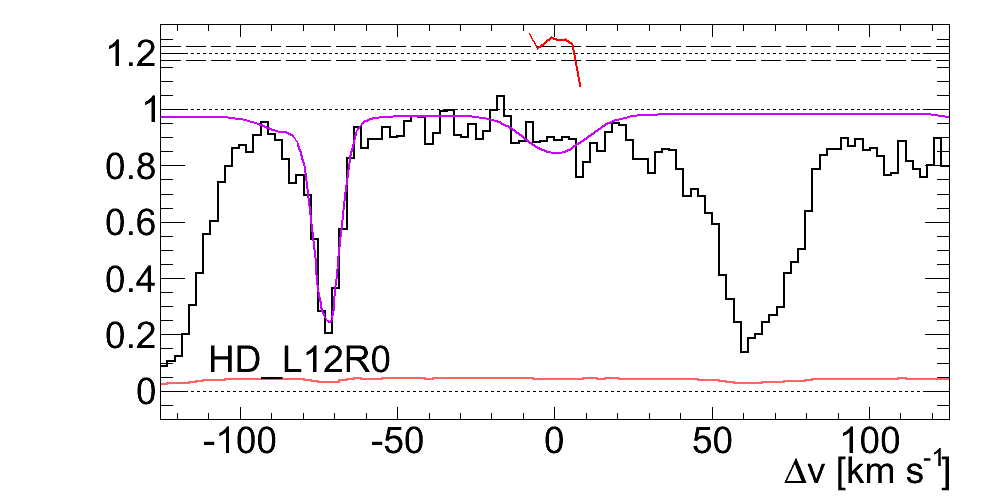}
\includegraphics[scale=0.122,natwidth=6cm,natheight=6cm]{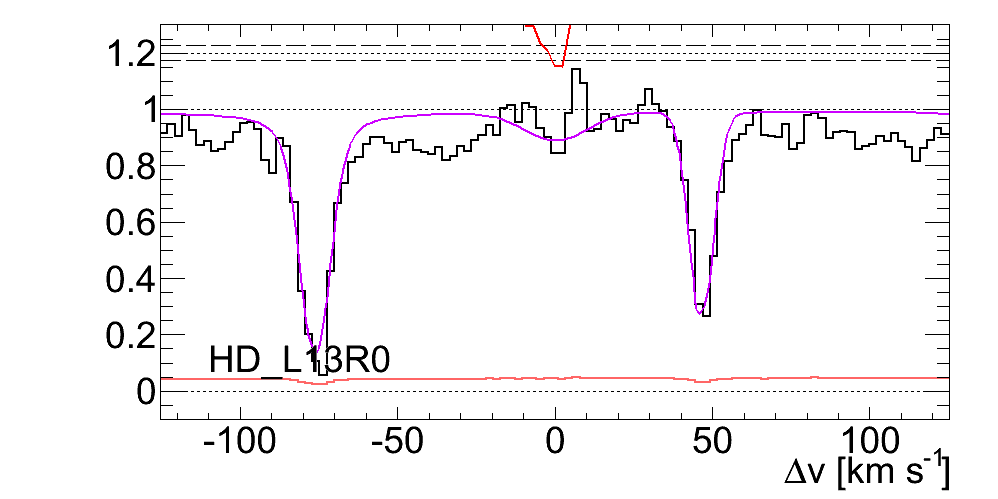}
\includegraphics[scale=0.122,natwidth=6cm,natheight=6cm]{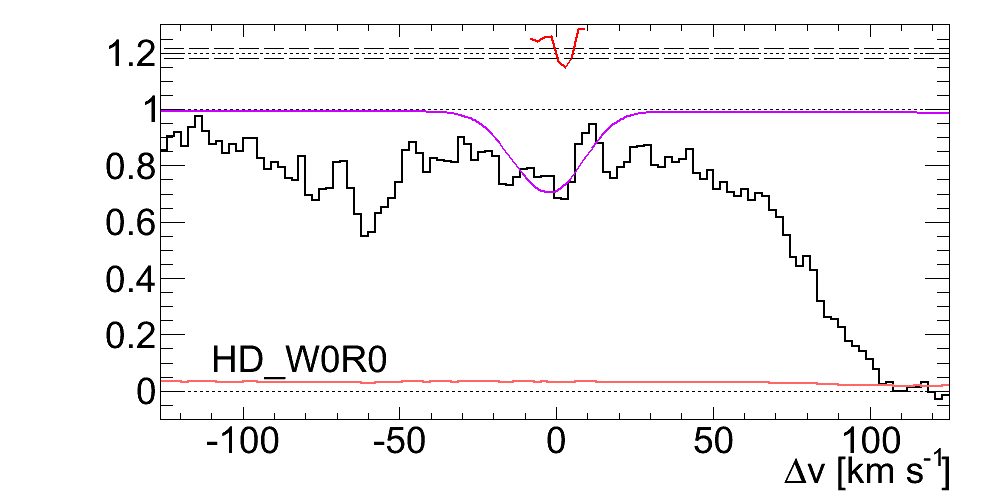}
\includegraphics[scale=0.122,natwidth=6cm,natheight=6cm]{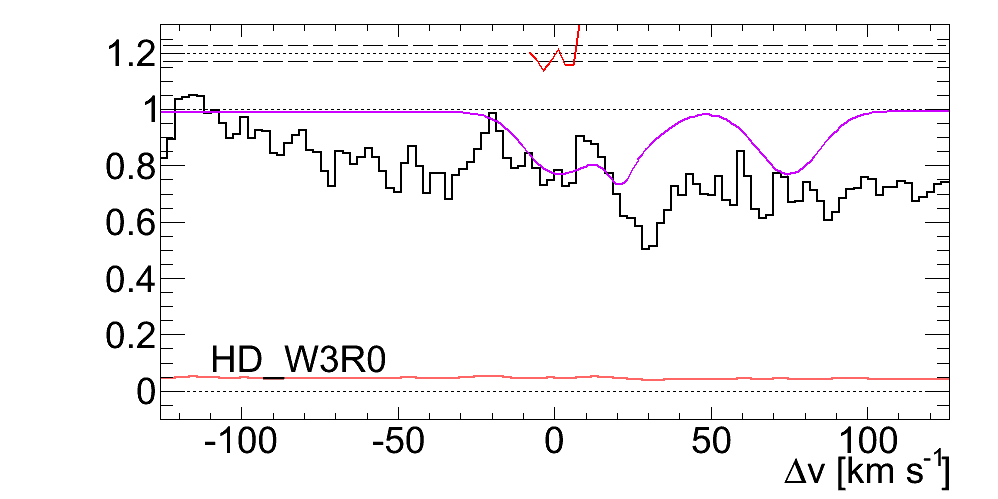}
\caption{Selection of \hd\ J = 0 absorption features. A tentative fit to the data is represented by a line for reference. Residuals of the fit are shown on top. The observational error is shown at the bottom for reference.}
\label{fig:HDJ0}
\end{figure}

\begin{figure}[tb]
\centering
\includegraphics[scale=0.122,natwidth=6cm,natheight=6cm]{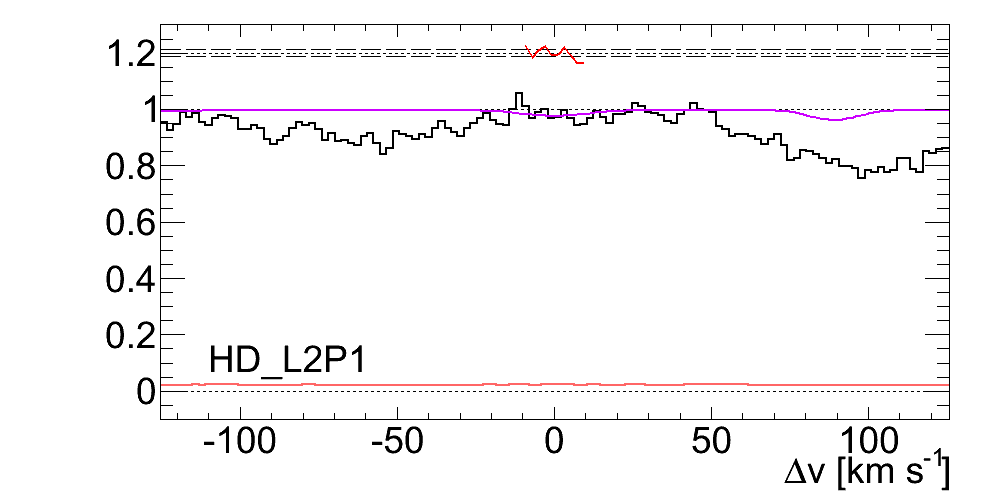}
\includegraphics[scale=0.122,natwidth=6cm,natheight=6cm]{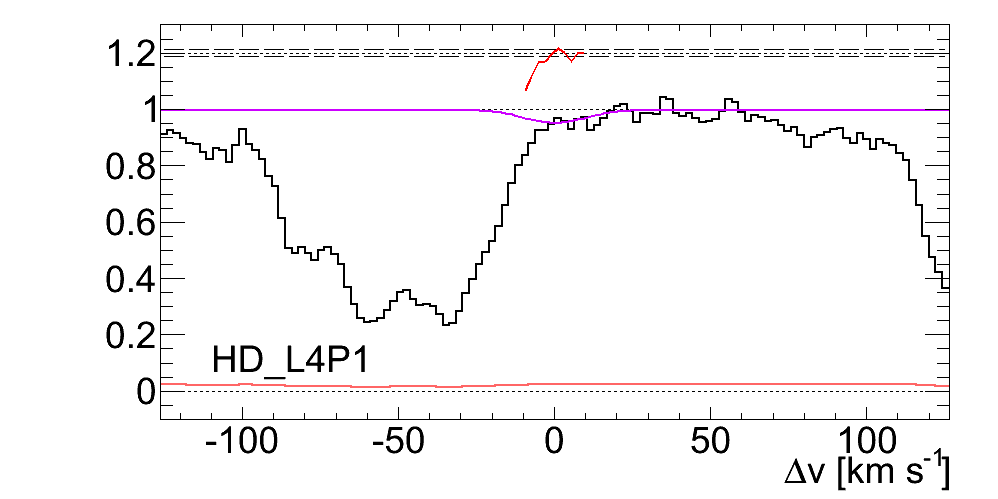}
\includegraphics[scale=0.122,natwidth=6cm,natheight=6cm]{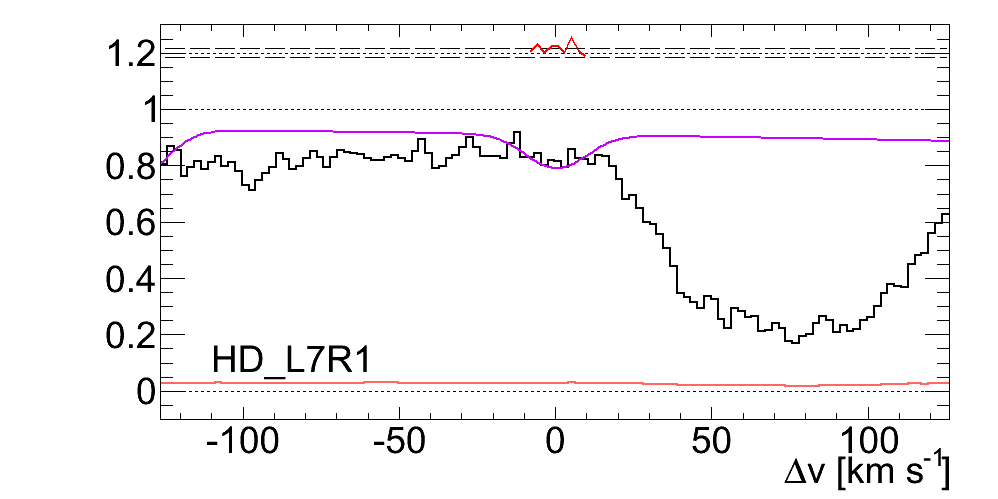}
\includegraphics[scale=0.122,natwidth=6cm,natheight=6cm]{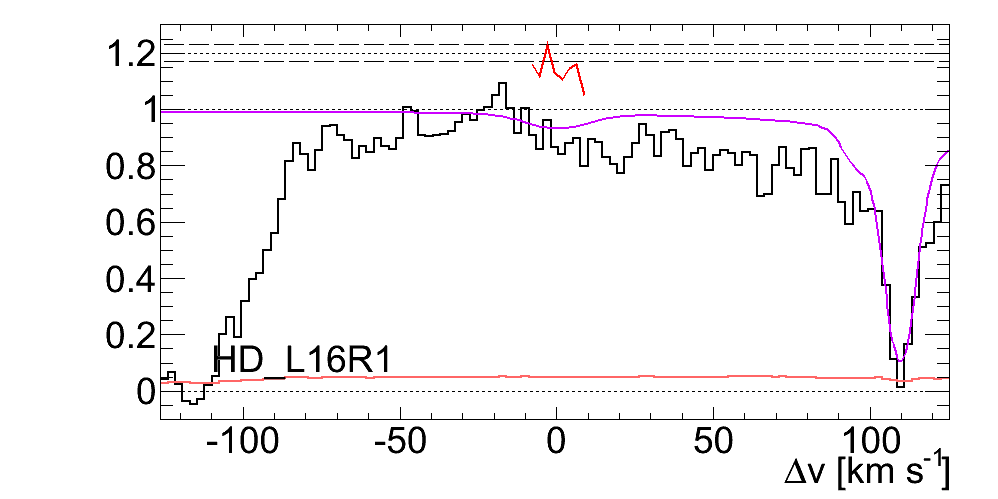}
\includegraphics[scale=0.122,natwidth=6cm,natheight=6cm]{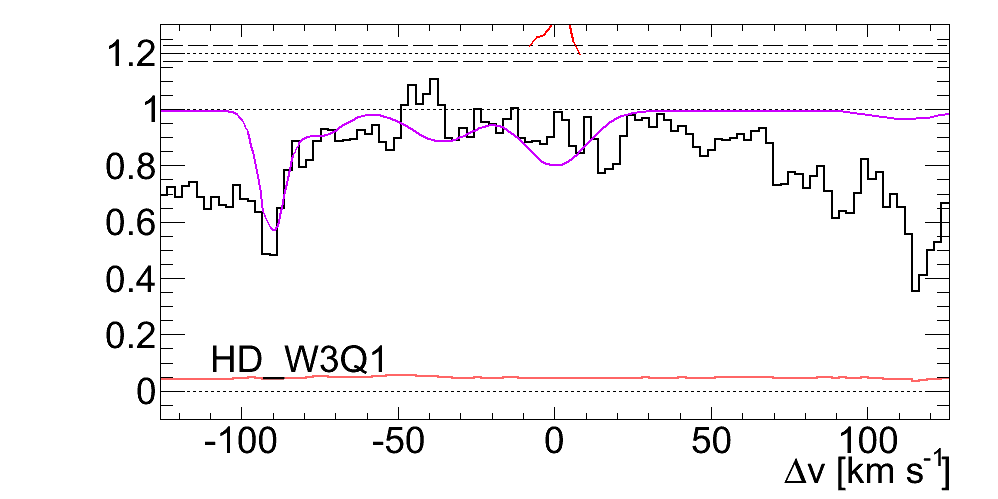}
\caption{Selection of \hd\ J = 1 absorption features. A tentative fit to the data is represented by a line for reference. Residuals of the fit are shown on top. The observational error is shown at the bottom for reference.}
\label{fig:HDJ1}
\end{figure}

Figs.~\ref{fig:HDJ0} and~\ref{fig:HDJ1} show the attempt to fit \hd\ absorptions. The results of this fit are possibly overestimating the column density and Doppler parameter.

\begin{figure}[tb]
\centering
\includegraphics[scale=0.122,natwidth=6cm,natheight=6cm]{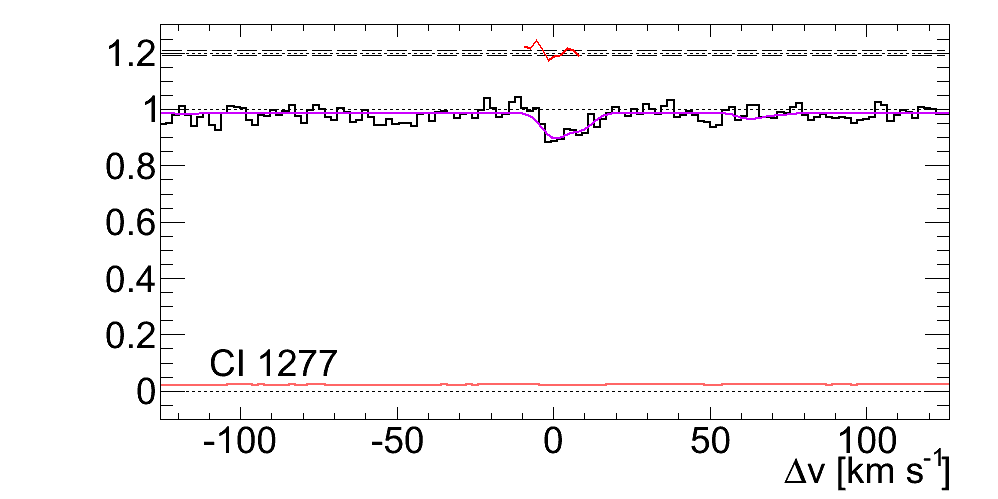}
\includegraphics[scale=0.122,natwidth=6cm,natheight=6cm]{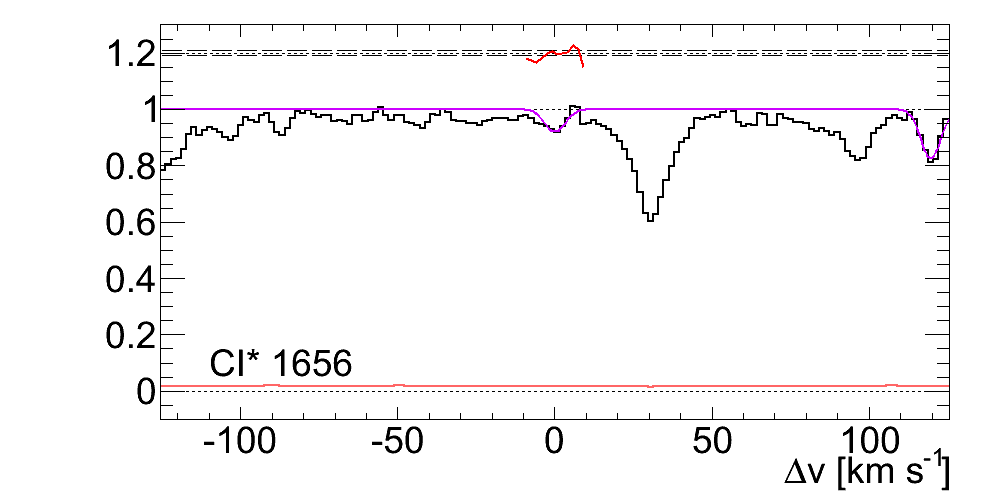}
\includegraphics[scale=0.122,natwidth=6cm,natheight=6cm]{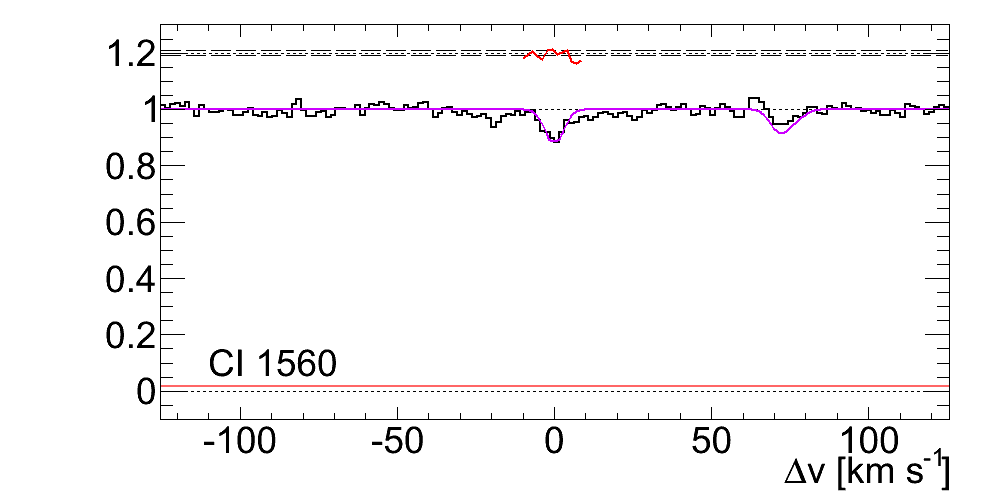}
\includegraphics[scale=0.122,natwidth=6cm,natheight=6cm]{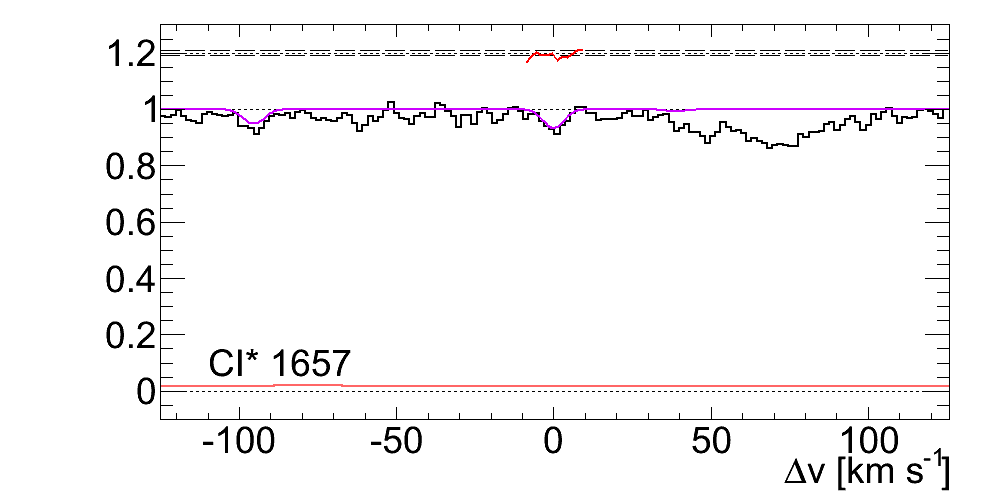}
\includegraphics[scale=0.122,natwidth=6cm,natheight=6cm]{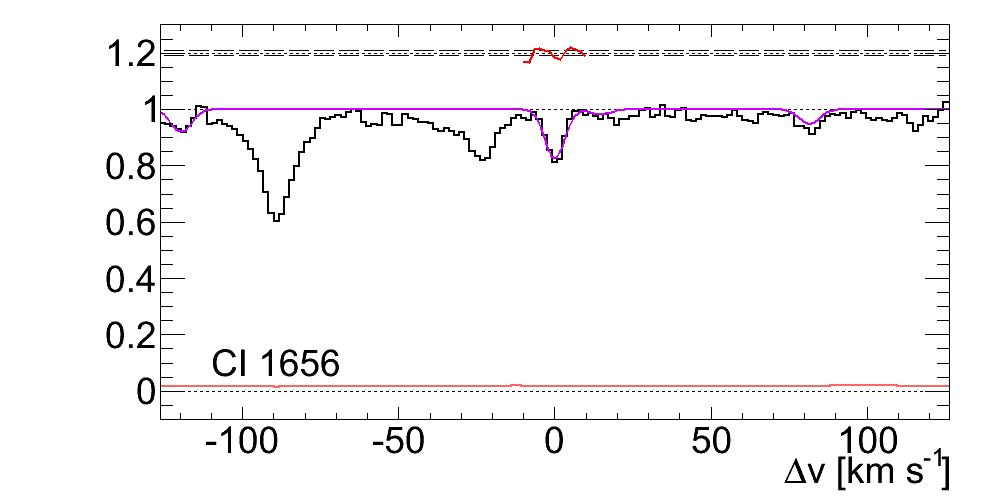}
\includegraphics[scale=0.122,natwidth=6cm,natheight=6cm]{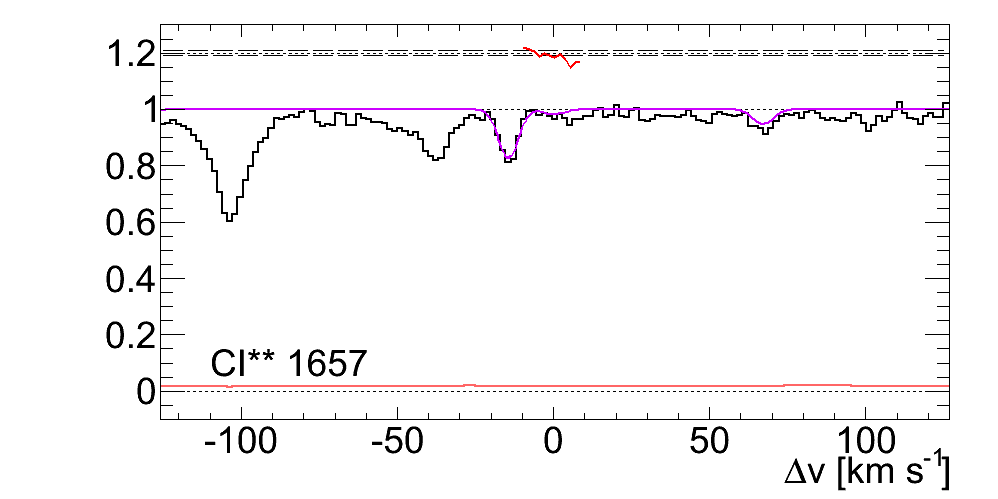}
\includegraphics[scale=0.122,natwidth=6cm,natheight=6cm]{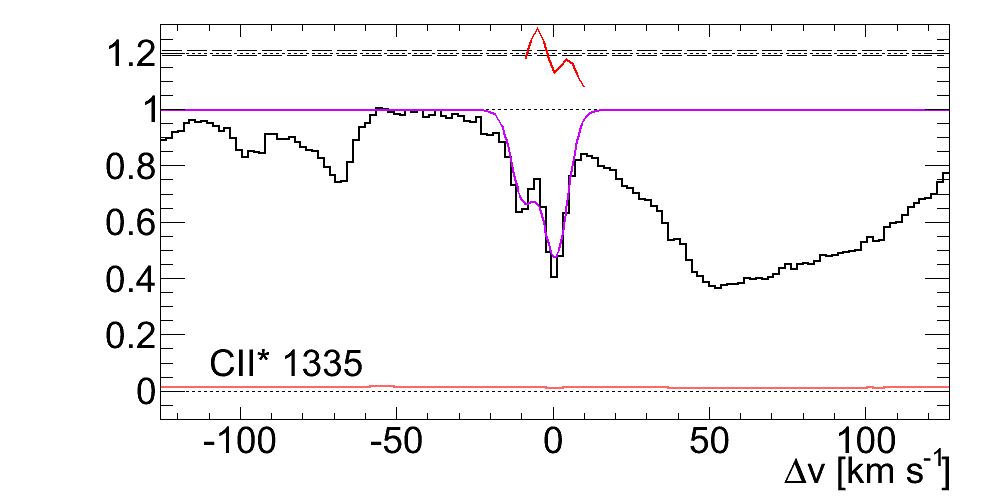}
\caption{Voigt profile fits to carbon features. The fit to the data is represented by a line. Residuals of the fit are shown on top. The observational error is shown at the bottom for reference.}
\label{fig:Carbon}
\end{figure}

Fig.~\ref{fig:Carbon} shows the low ionisation carbon features.

\begin{figure}[tb]
\centering
\includegraphics[scale=0.122,natwidth=6cm,natheight=6cm]{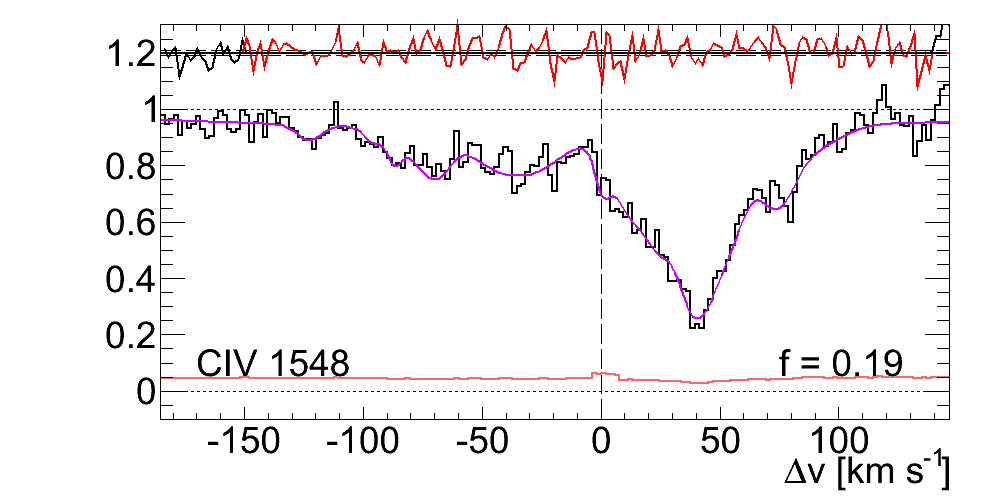}
\includegraphics[scale=0.122,natwidth=6cm,natheight=6cm]{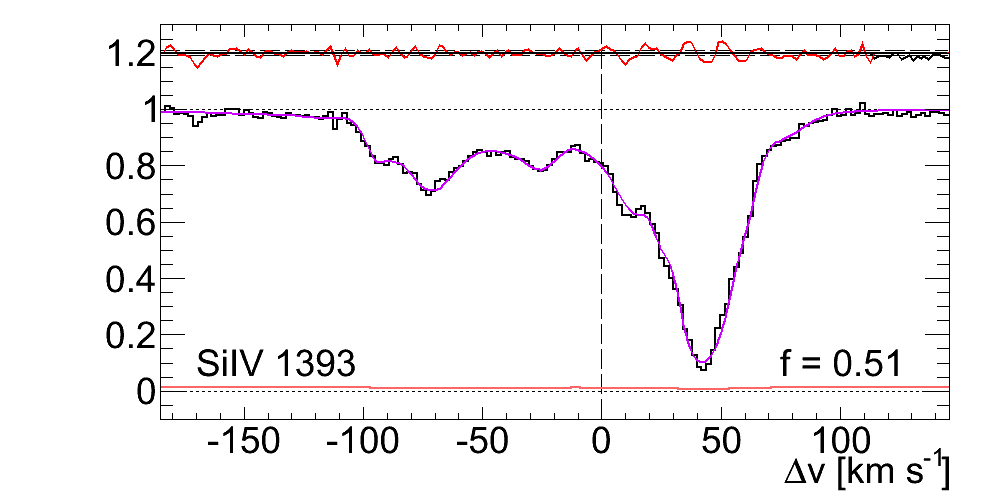}
\includegraphics[scale=0.122,natwidth=6cm,natheight=6cm]{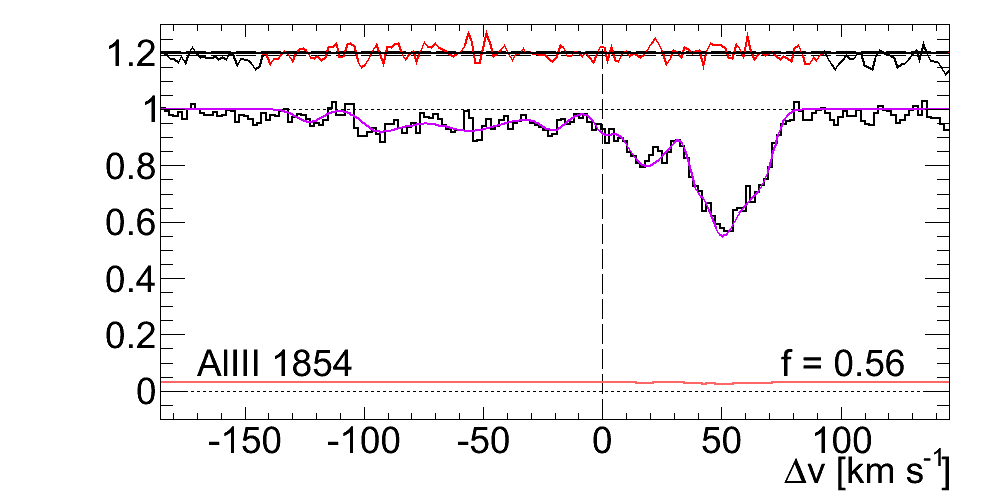}
\includegraphics[scale=0.122,natwidth=6cm,natheight=6cm]{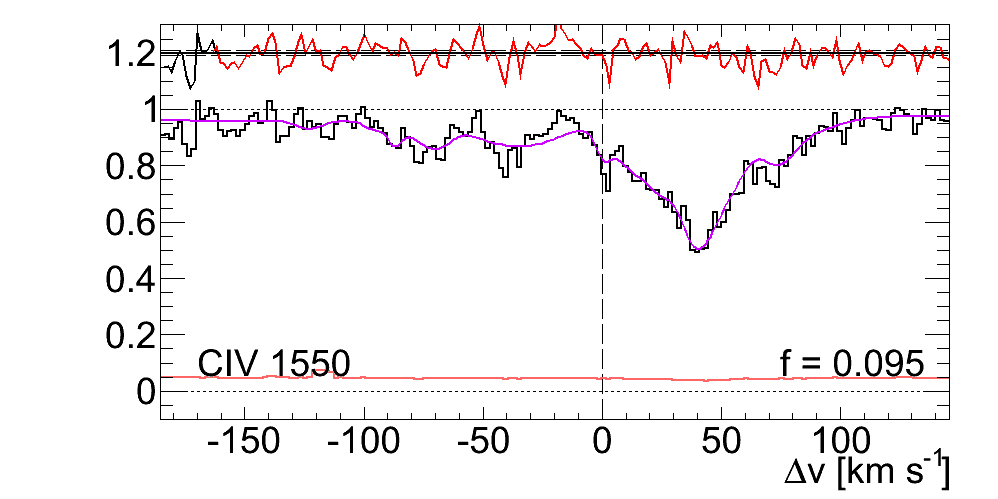}
\includegraphics[scale=0.122,natwidth=6cm,natheight=6cm]{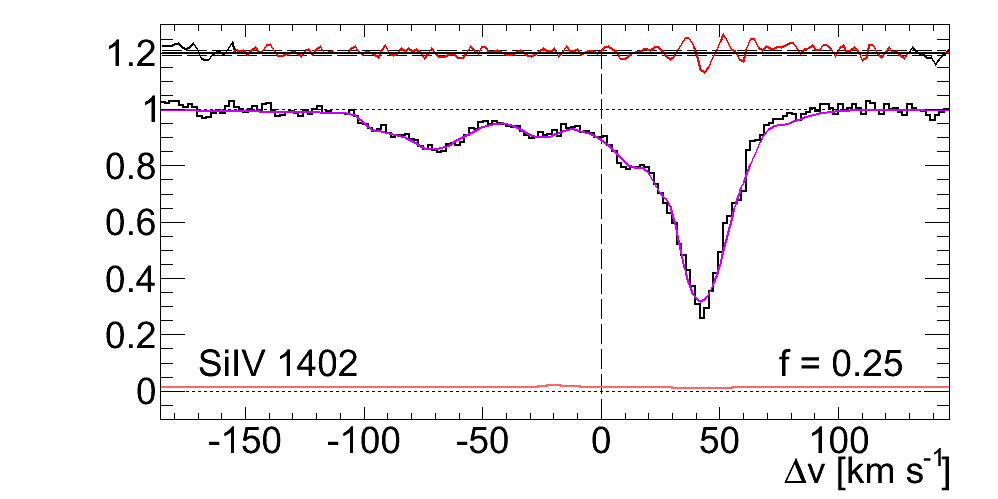}
\includegraphics[scale=0.122,natwidth=6cm,natheight=6cm]{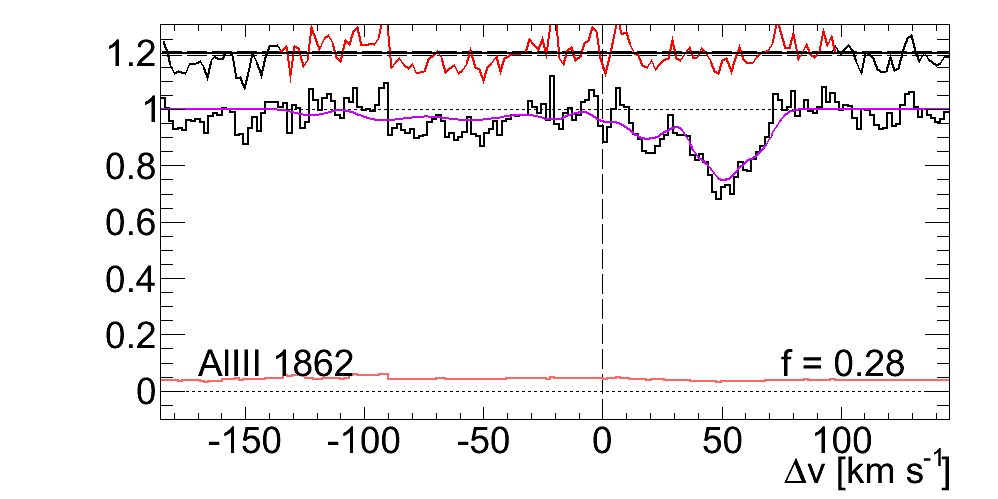}
\caption{Multiple-component Voigt profile fits to high ionisation element profiles. The fit to the data is represented by a line. Residuals of the fit are shown on top, in red when corresponding to the intervals used for the fit, in black otherwise. The observational error is shown at the bottom for reference.}
\label{fig:Metals_h}
\end{figure}

\begin{figure*}[tb]
\centering
\includegraphics[scale=0.122,natwidth=6cm,natheight=6cm]{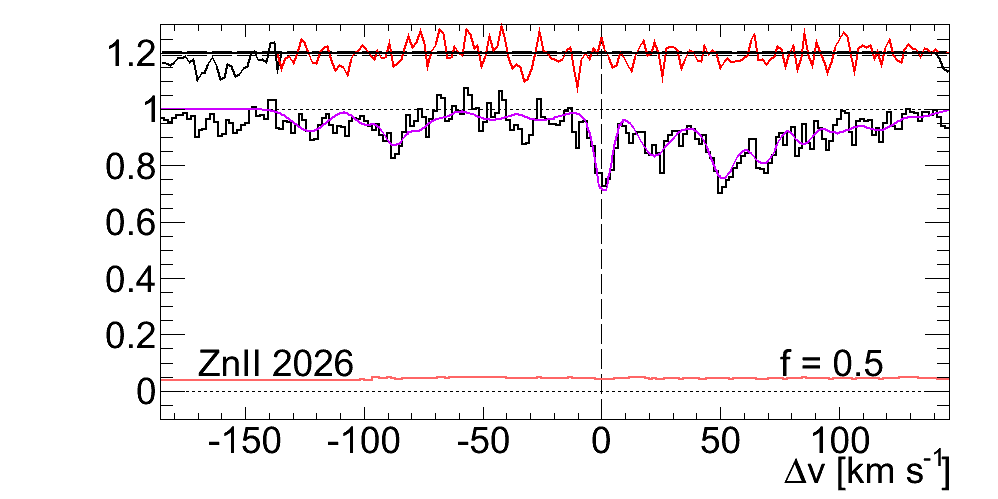}
\includegraphics[scale=0.122,natwidth=6cm,natheight=6cm]{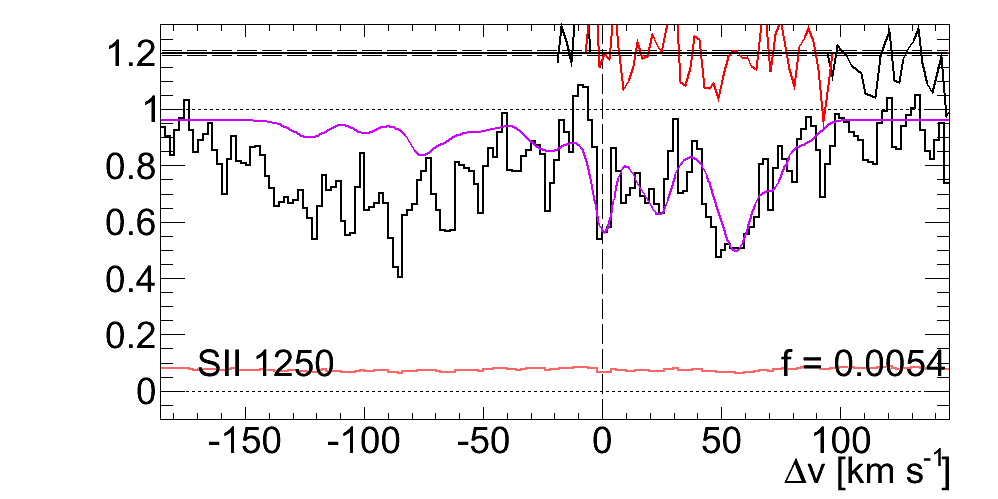}
\includegraphics[scale=0.122,natwidth=6cm,natheight=6cm]{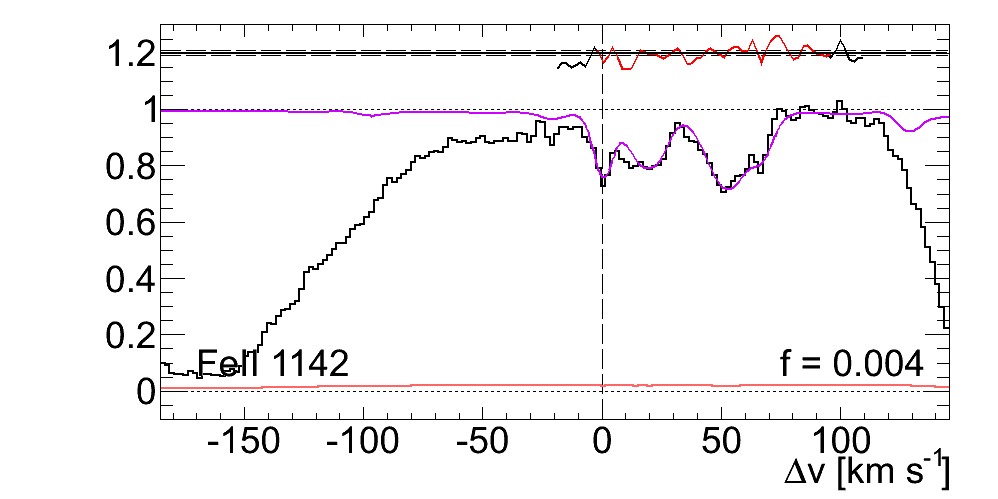}
\includegraphics[scale=0.122,natwidth=6cm,natheight=6cm]{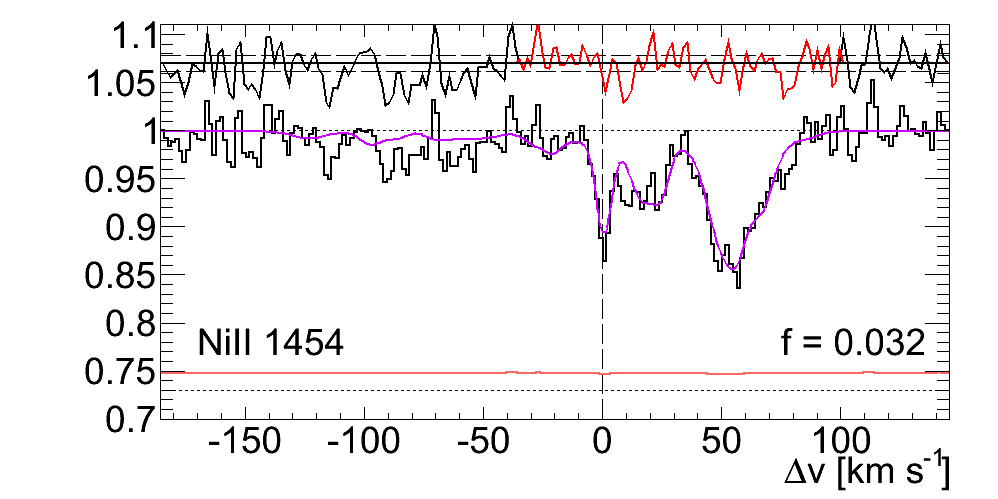}
\includegraphics[scale=0.122,natwidth=6cm,natheight=6cm]{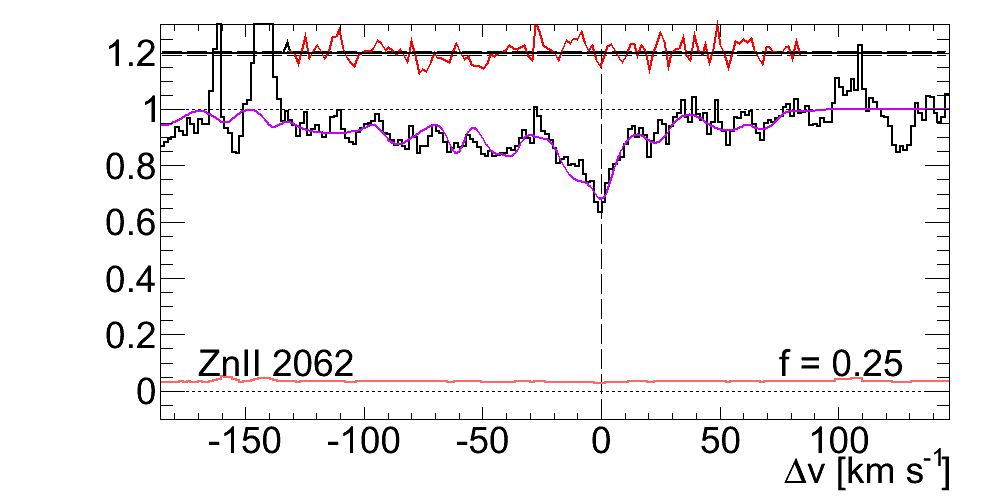}
\includegraphics[scale=0.122,natwidth=6cm,natheight=6cm]{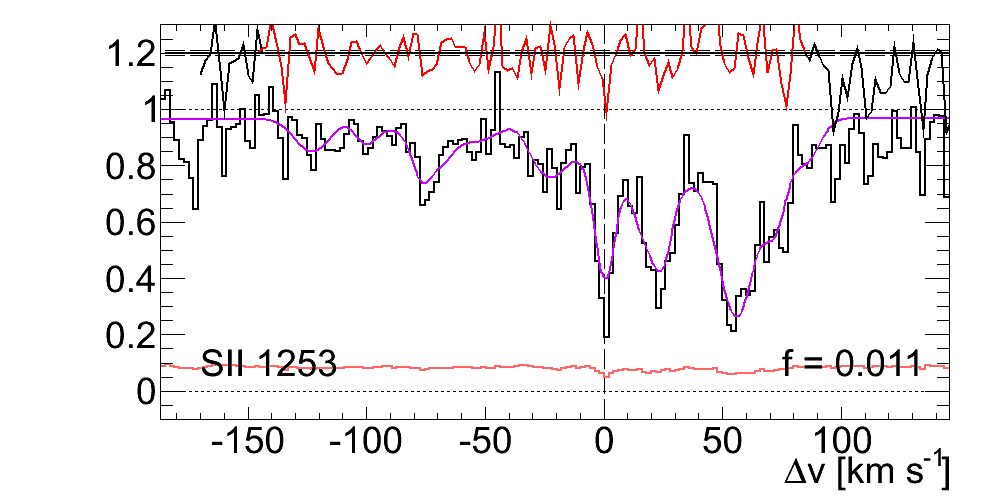}
\includegraphics[scale=0.122,natwidth=6cm,natheight=6cm]{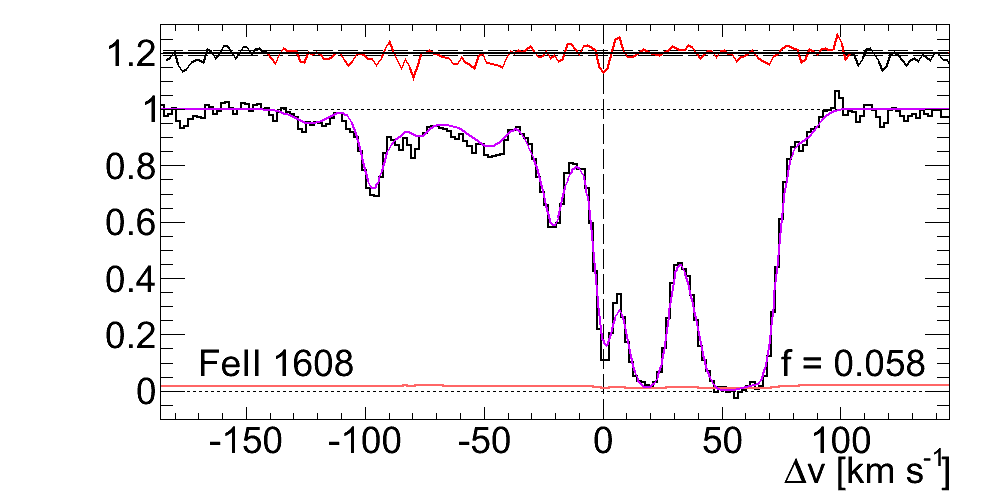}
\includegraphics[scale=0.122,natwidth=6cm,natheight=6cm]{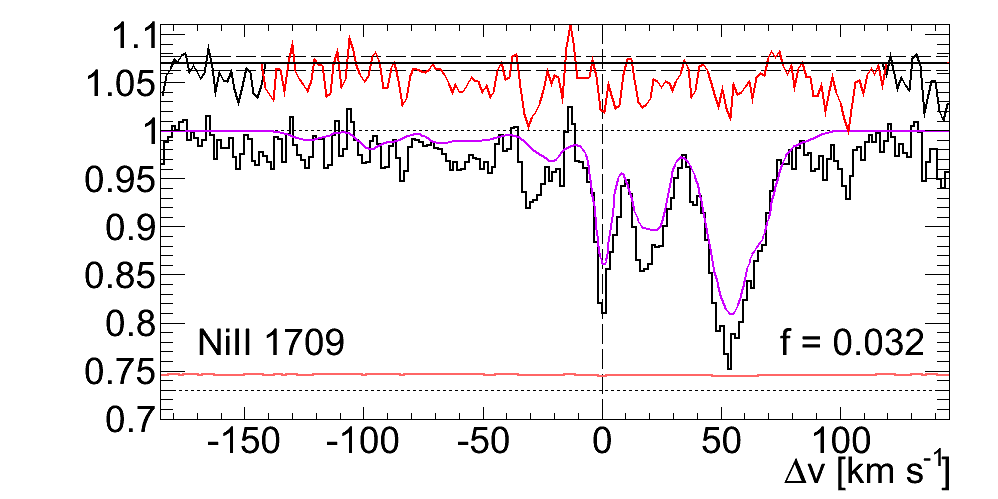}
\includegraphics[scale=0.122,natwidth=6cm,natheight=6cm]{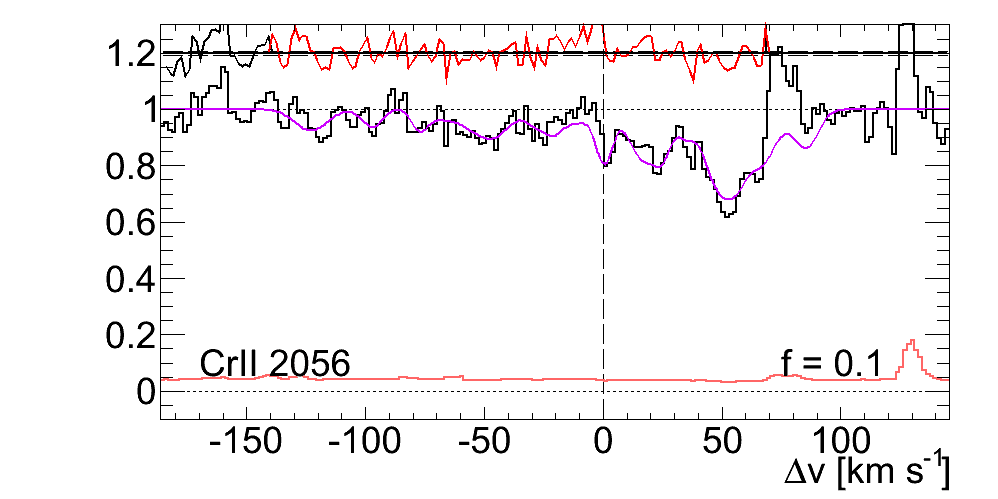}
\includegraphics[scale=0.122,natwidth=6cm,natheight=6cm]{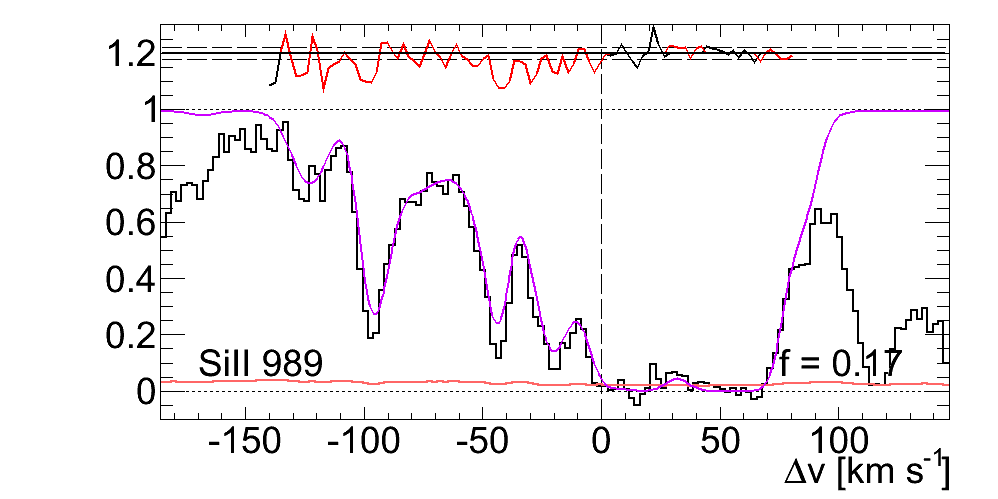}
\includegraphics[scale=0.122,natwidth=6cm,natheight=6cm]{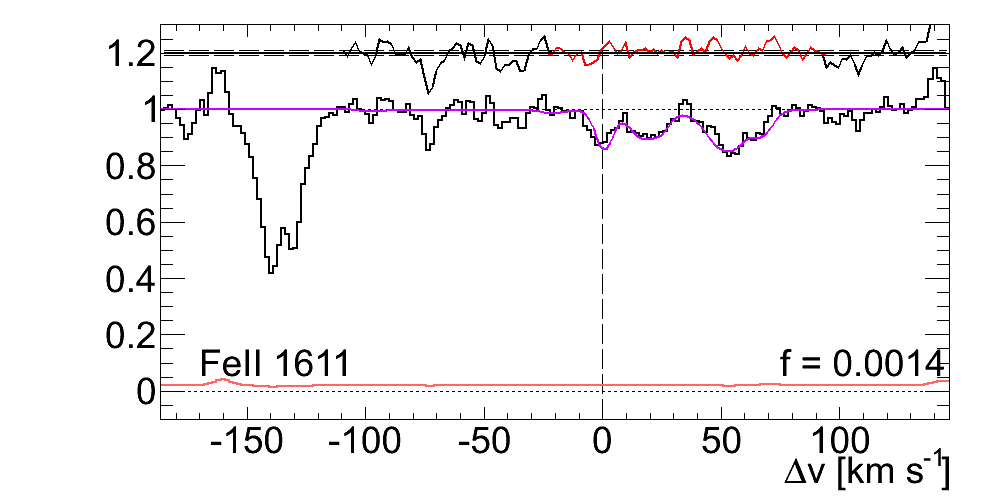}
\includegraphics[scale=0.122,natwidth=6cm,natheight=6cm]{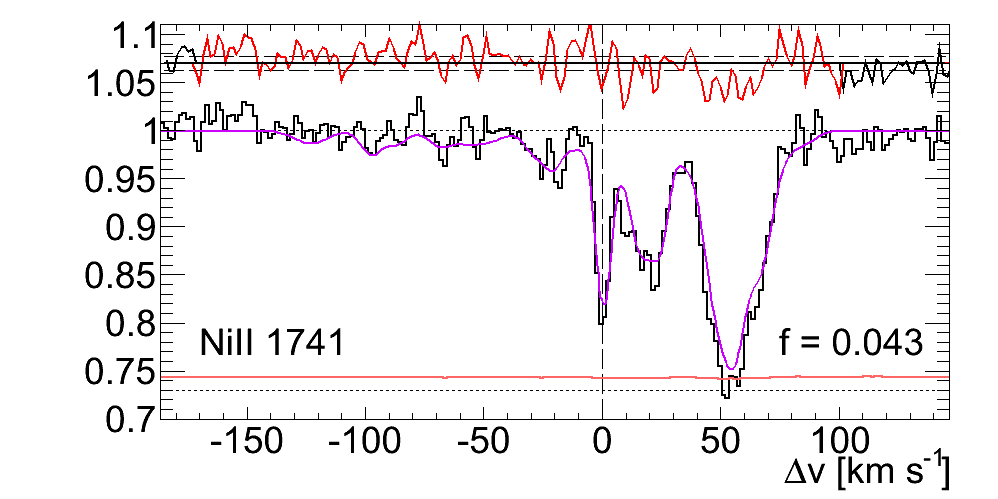}
\includegraphics[scale=0.122,natwidth=6cm,natheight=6cm]{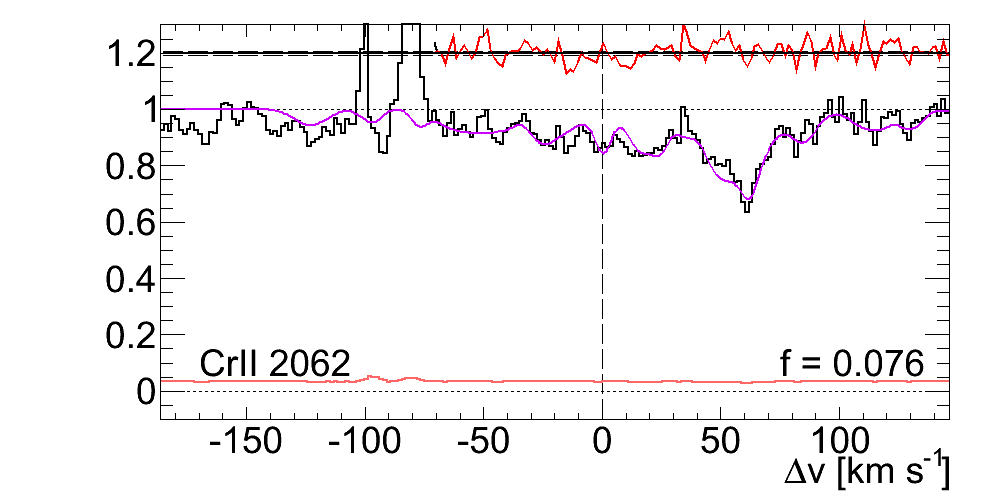}
\includegraphics[scale=0.122,natwidth=6cm,natheight=6cm]{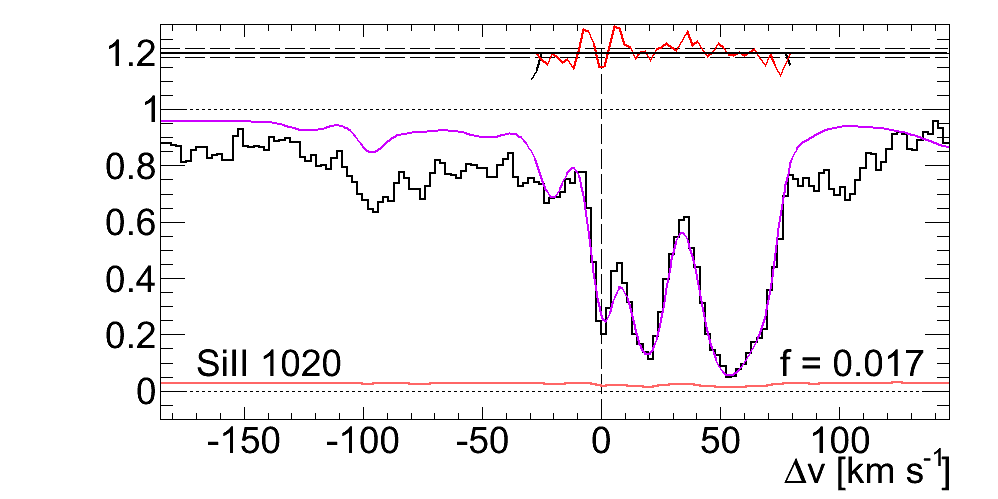}
\includegraphics[scale=0.122,natwidth=6cm,natheight=6cm]{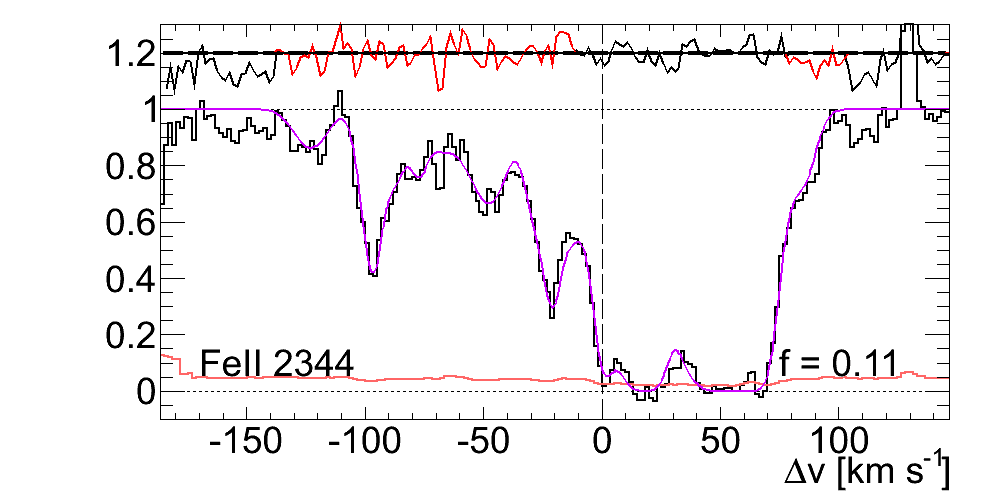}
\includegraphics[scale=0.122,natwidth=6cm,natheight=6cm]{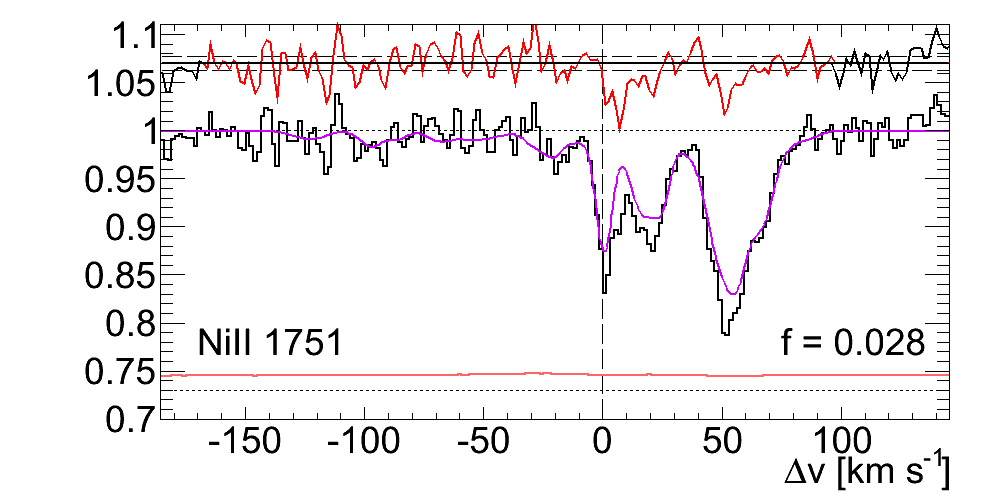}
\includegraphics[scale=0.122,natwidth=6cm,natheight=6cm]{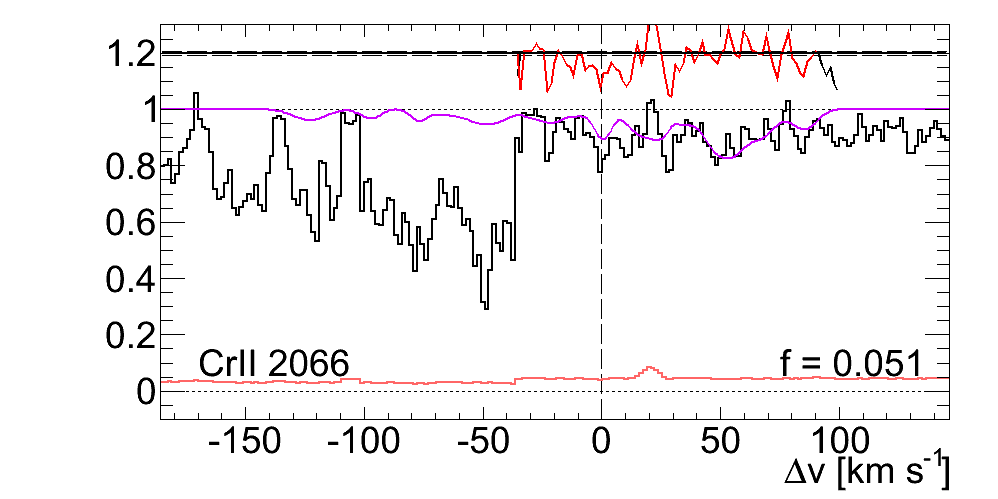}
\includegraphics[scale=0.122,natwidth=6cm,natheight=6cm]{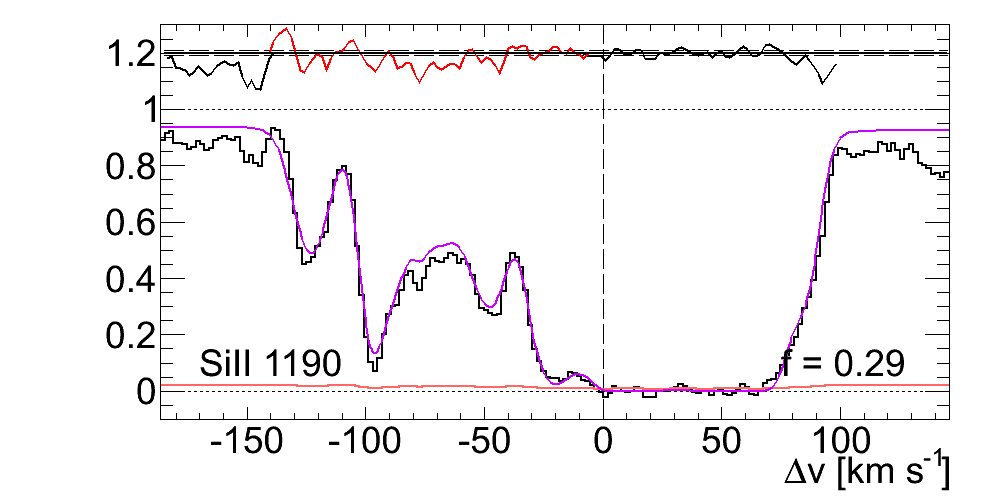}
\includegraphics[scale=0.122,natwidth=6cm,natheight=6cm]{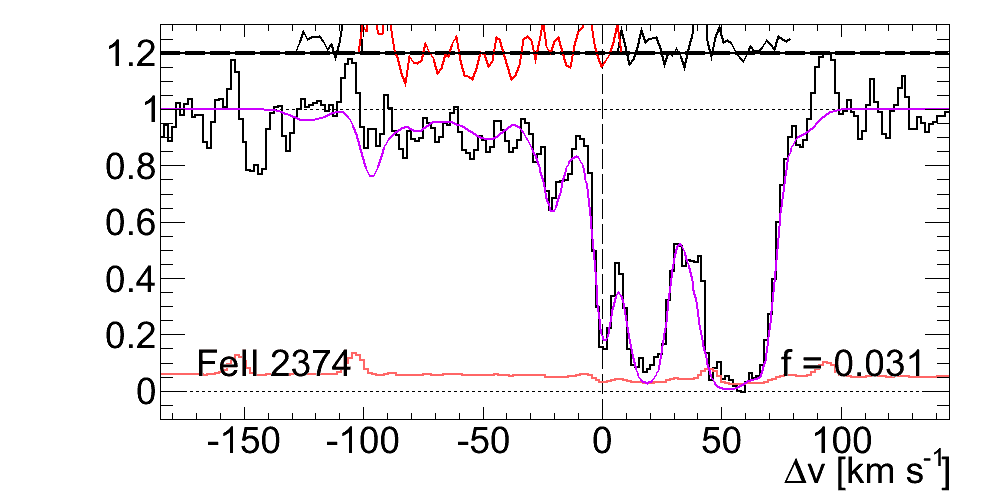}
\includegraphics[scale=0.122,natwidth=6cm,natheight=6cm]{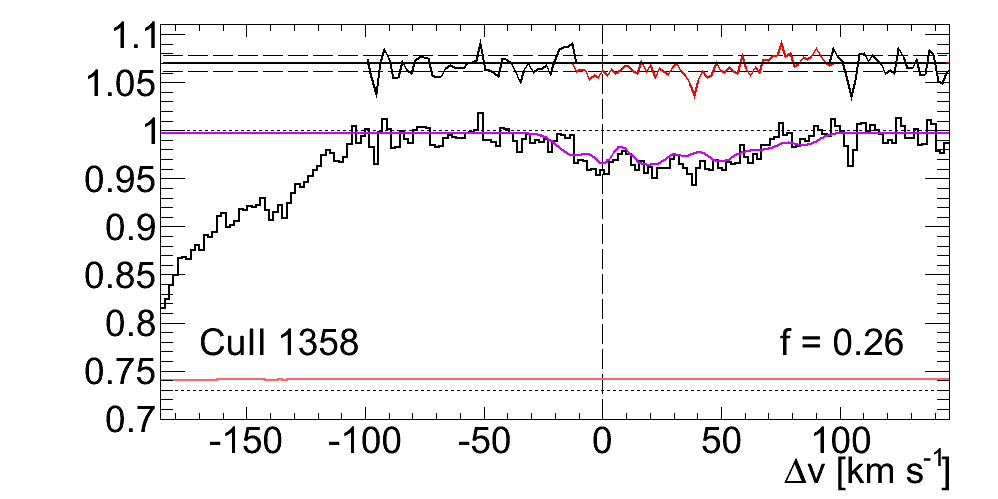}
\includegraphics[scale=0.122,natwidth=6cm,natheight=6cm]{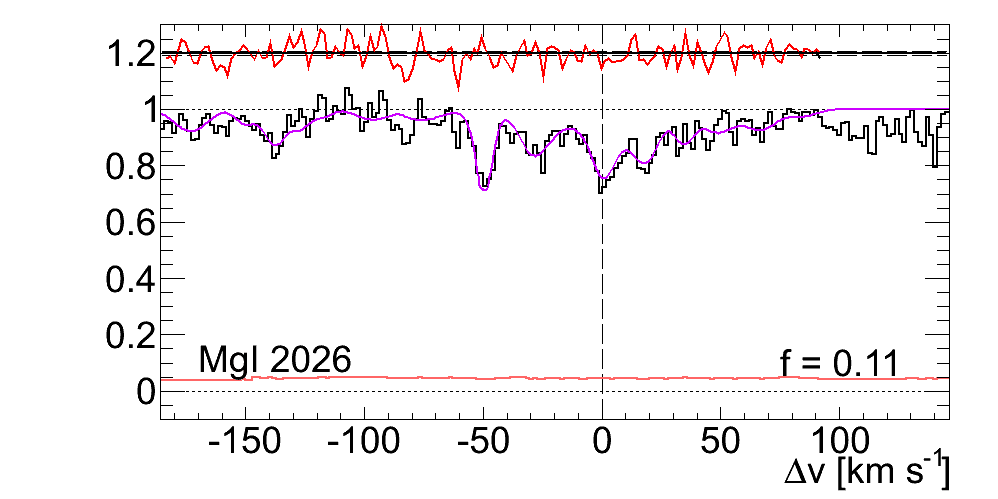}
\includegraphics[scale=0.122,natwidth=6cm,natheight=6cm]{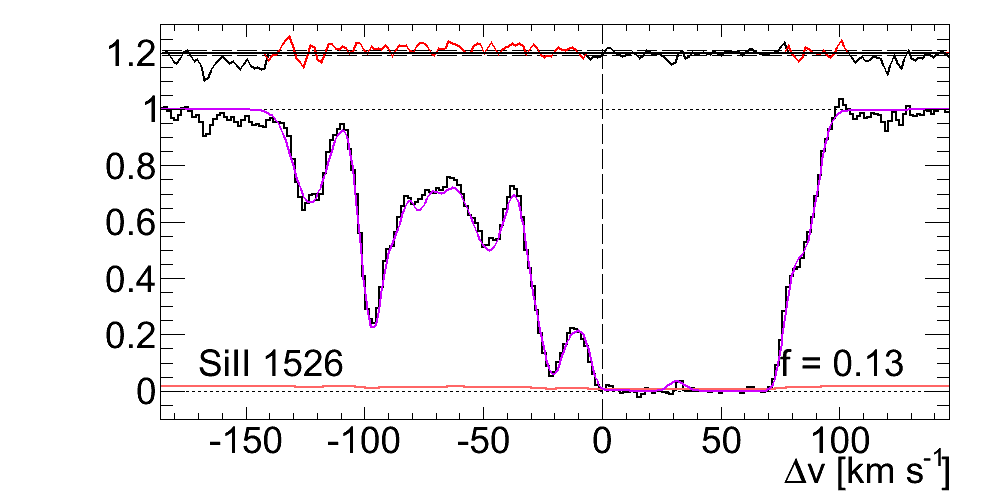}
\includegraphics[scale=0.122,natwidth=6cm,natheight=6cm]{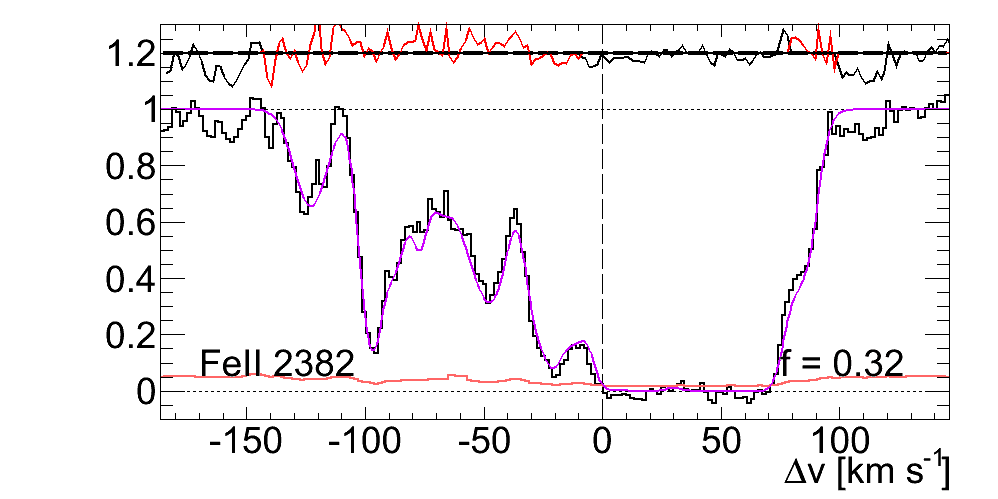}
\includegraphics[scale=0.122,natwidth=6cm,natheight=6cm]{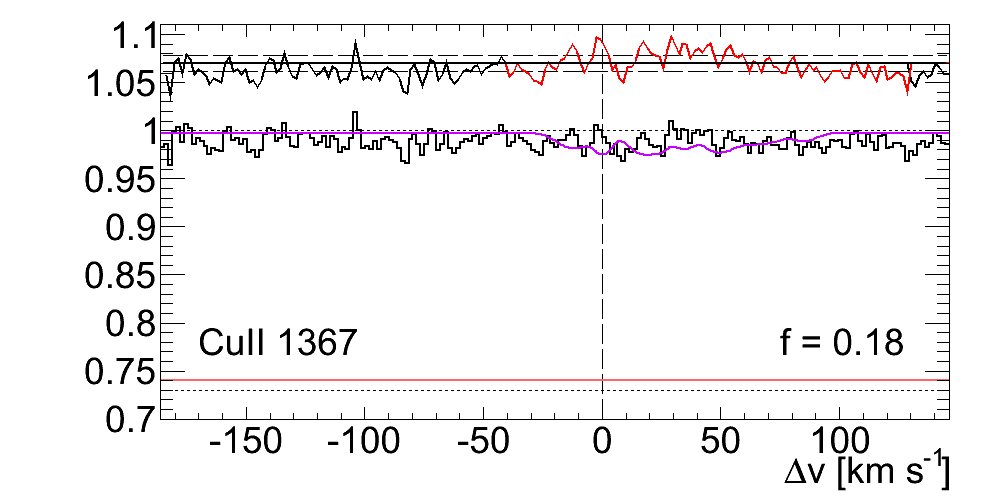}
\includegraphics[scale=0.122,natwidth=6cm,natheight=6cm]{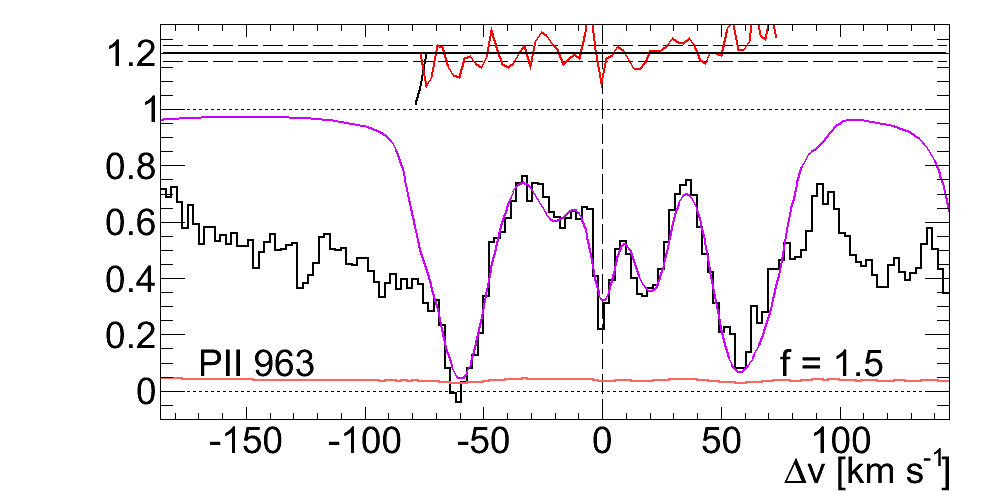}
\includegraphics[scale=0.122,natwidth=6cm,natheight=6cm]{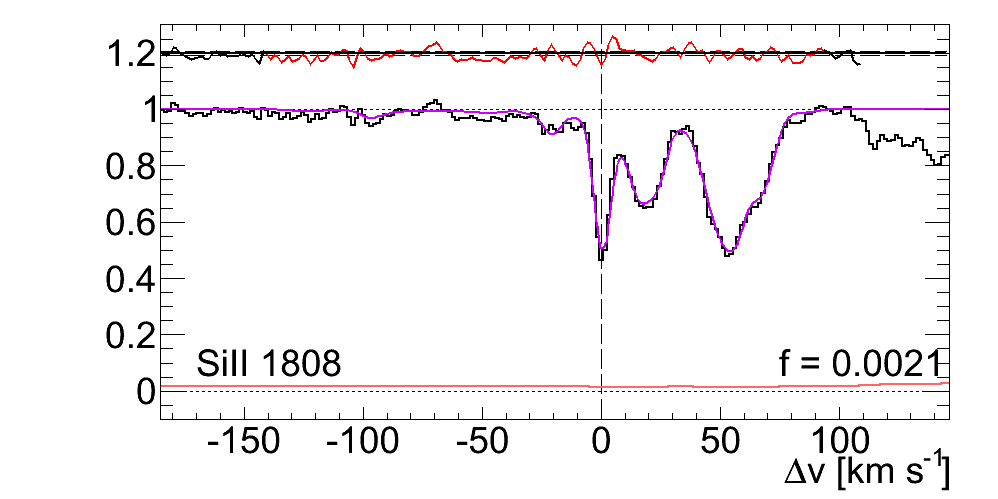}
\includegraphics[scale=0.122,natwidth=6cm,natheight=6cm]{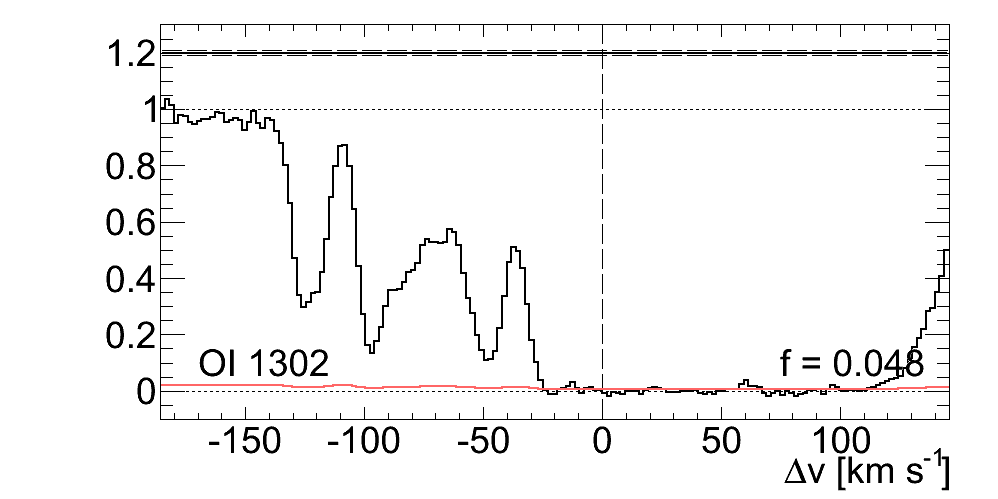}
\includegraphics[scale=0.122,natwidth=6cm,natheight=6cm]{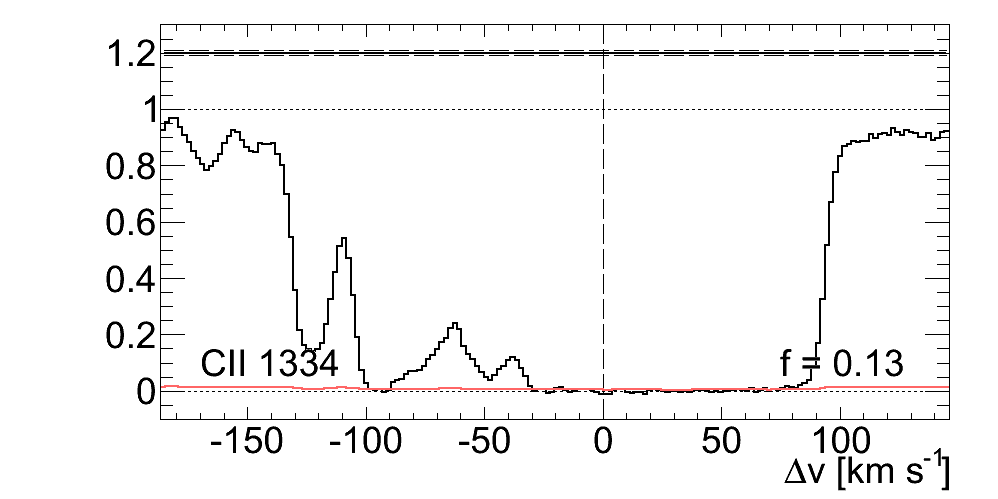}
\caption{Multiple-component Voigt profile fits to low ionisation element profiles. The fit to the data is represented by a line. Residuals of the fit are shown on top, in red when corresponding to the intervals used for the fit, in black otherwise. The observational error is shown at the bottom for reference. \cii\ and \oi\ are impossible to decompose because all features are heavily saturated.}
\label{fig:Metals}
\end{figure*}

In Fig.~\ref{fig:Metals} we present low ionisation metal absorption profiles, while Fig.~\ref{fig:Metals_h} shows highly ionised carbon and silicon lines.

\clearpage

\clearpage


\section{Transitions used for \dmm\ and their redshifts}
We summarize in Table~\ref{tab:h2lines} the \htwo\ lines used for the determination of \dmm. We give the rest frame wavelength and sensitivity coefficients we used, the redshifts we have measured with a two-component model and \dmm\ included as a free parameter of the fit, as well as the velocity shift with respect to the \htwo-bearing cloud redshift.

\begin{table*}
\label{tab:h2lines}
\caption{Laboratory wavelength of the set of \htwo\ transitions that are fitted along with the best redshift and errors from Vogit profile analysis. The uncontaminated (CLEAN) \htwo\ lines are highlighted in bold letters.}
\begin{center}
\begin{tabular}{lcccc}
\hline
\hline
Line ID & Lab wavelength$^a$ (\AA) & Redshift & Velocity (\kms) & $K$ coefficient$^{b}$ \\
\hline
    L14R0& 946.1693& 2.658594(041)& $-$0.58$\pm$0.34& $+$0.042  \\
    L10R0& 981.4387& 2.658604(056)& $+$0.19$\pm$0.46& $+$0.041  \\
    L8R0& 1001.8237& 2.658601(040)& $-$0.06$\pm$0.33& $+$0.035  \\
    L7R0& 1012.8129& 2.658591(039)& $-$0.86$\pm$0.32& $+$0.031  \\
    L4R0& 1049.3674& 2.658585(060)& $-$1.32$\pm$0.50& $+$0.017  \\
 {\bf L2R0}&{\bf 1077.1387}&{\bf 2.658602(040)}&{\bf $+$0.09$\pm$0.33}&{\bf $+$0.006 }\\
    L0R0& 1108.1273& 2.658600(015)& $-$0.12$\pm$0.12& $-$0.008  \\
 {\bf L17R1}&{\bf 924.6433}&{\bf 2.658605(083)}&{\bf $+$0.27$\pm$0.69}&{\bf $+$0.055 }\\
 {\bf L16P1}&{\bf 932.2662}&{\bf 2.658604(035)}&{\bf $+$0.24$\pm$0.29}&{\bf $+$0.053 }\\
 {\bf L15R1}&{\bf 939.1242}&{\bf 2.658604(044)}&{\bf $+$0.23$\pm$0.36}&{\bf $+$0.052 }\\
 {\bf L15P1}&{\bf 939.7067}&{\bf 2.658606(035)}&{\bf $+$0.37$\pm$0.29}&{\bf $+$0.051 }\\
    W3Q1& 947.4219& 2.658607(084)& $+$0.46$\pm$0.69& $+$0.021  \\
    L14P1& 947.5140& 2.658601(063)& $-$0.00$\pm$0.52& $+$0.050  \\
 {\bf L13R1}&{\bf 955.0658}&{\bf 2.658601(055)}&{\bf $-$0.04$\pm$0.46}&{\bf $+$0.048 }\\
 {\bf L13P1}&{\bf 955.7082}&{\bf 2.658600(044)}&{\bf $-$0.06$\pm$0.37}&{\bf $+$0.048 }\\
    L12R1& 963.6079& 2.658598(056)& $-$0.27$\pm$0.46& $+$0.046  \\
    L10P1& 982.8353& 2.658607(034)& $+$0.49$\pm$0.29& $+$0.040  \\
 {\bf L9P1}&{\bf 992.8096}&{\bf 2.658594(043)}&{\bf $-$0.63$\pm$0.36}&{\bf $+$0.037 }\\
    L8P1& 1003.2965& 2.658602(029)& $+$0.05$\pm$0.24& $+$0.033  \\
    L7R1& 1013.4369& 2.658605(030)& $+$0.34$\pm$0.25& $+$0.030  \\
 {\bf L5P1}&{\bf 1038.1570}&{\bf 2.658594(021)}&{\bf $-$0.62$\pm$0.17}&{\bf $+$0.021 }\\
    L4R1& 1049.9597& 2.658605(030)& $+$0.28$\pm$0.25& $+$0.016  \\
 {\bf L2R1}&{\bf 1077.6989}&{\bf 2.658599(021)}&{\bf $-$0.20$\pm$0.17}&{\bf $+$0.005 }\\
    L2P1& 1078.9254& 2.658601(018)& $-$0.05$\pm$0.15& $+$0.004  \\
 {\bf L1P1}&{\bf 1094.0519}&{\bf 2.658601(019)}&{\bf $-$0.00$\pm$0.16}&{\bf $-$0.003 }\\
    L0P1& 1110.0626& 2.658599(017)& $-$0.20$\pm$0.14& $-$0.010  \\
 {\bf L18P2}&{\bf 920.2432}&{\bf 2.658608(066)}&{\bf $+$0.55$\pm$0.55}&{\bf $+$0.053 }\\
 {\bf W4P2}&{\bf 932.6047}&{\bf 2.658608(080)}&{\bf $+$0.55$\pm$0.66}&{\bf $+$0.026 }\\
    L16R2& 933.2401& 2.658600(055)& $-$0.14$\pm$0.45& $+$0.051  \\
    L16P2& 934.1448& 2.658599(053)& $-$0.21$\pm$0.44& $+$0.051  \\
 {\bf L15P2}&{\bf 941.5992}&{\bf 2.658615(048)}&{\bf $+$1.11$\pm$0.40}&{\bf $+$0.050 }\\
    W3R2& 947.1117& 2.658606(052)& $+$0.39$\pm$0.43& $+$0.023  \\
    L13R2& 956.5799& 2.658581(056)& $-$1.62$\pm$0.47& $+$0.047  \\
 {\bf L13P2}&{\bf 957.6522}&{\bf 2.658603(029)}&{\bf $+$0.15$\pm$0.24}&{\bf $+$0.046 }\\
 {\bf L12P2}&{\bf 966.2754}&{\bf 2.658602(034)}&{\bf $+$0.04$\pm$0.28}&{\bf $+$0.043 }\\
 {\bf W2Q2}&{\bf 967.2811}&{\bf 2.658609(029)}&{\bf $+$0.63$\pm$0.24}&{\bf $+$0.013 }\\
 {\bf W2P2}&{\bf 968.2952}&{\bf 2.658594(034)}&{\bf $-$0.58$\pm$0.28}&{\bf $+$0.012 }\\
\hline
\end{tabular}
\end{center}
\begin{flushleft}
Column (1): Name of the \htwo\ fitted transitions. Column (2): The laboratory wavelengths. 
Columns (2) and (3): The best-fitting redshifts for \htwo\ lines and their errors. 
Column (5): Velocity offset between the redshift of a given \htwo\ transition and the weighted mean redshift 
of the all the \htwo\ lines. 
Column (6) Sensitivity coefficient of \htwo\ lines.\\
{$^a$} Wavelengths are from \citet{Malec2010}.\\
{$^b$} $K$ coefficient are from \citet{Ubachs2007}.\\
\end{flushleft}
\label{fitting_res1}
\end{table*}

\begin{table*}
\caption{Table \ref{fitting_res1} continued.}
\begin{center}
\begin{tabular}{lcccc}
\hline
\hline
Line ID & Lab wavelength$^a$ (\AA) & Redshift & Velocity (\kms) & $K$ coefficient$^{b}$   \\
\hline
 {\bf L11P2}&{\bf 975.3458}&{\bf 2.658603(029)}&{\bf $+$0.12$\pm$0.24}&{\bf $+$0.041 }\\
    L10R2& 983.5911& 2.658594(039)& $-$0.63$\pm$0.32& $+$0.039  \\
    L10P2& 984.8640& 2.658606(029)& $+$0.36$\pm$0.24& $+$0.038  \\
 {\bf W1R2}&{\bf 986.2441}&{\bf 2.658607(032)}&{\bf $+$0.50$\pm$0.27}&{\bf $+$0.006 }\\
    W1P2& 989.0884& 2.658601(031)& $-$0.04$\pm$0.26& $+$0.003  \\
 {\bf L9P2}&{\bf 994.8740}&{\bf 2.658601(032)}&{\bf $+$0.01$\pm$0.27}&{\bf $+$0.035 }\\
    L8R2& 1003.9854& 2.658596(021)& $-$0.46$\pm$0.17& $+$0.033  \\
 {\bf L8P2}&{\bf 1005.3931}&{\bf 2.658604(022)}&{\bf $+$0.20$\pm$0.18}&{\bf $+$0.031 }\\
 {\bf W0Q2}&{\bf 1010.9385}&{\bf 2.658600(025)}&{\bf $-$0.12$\pm$0.21}&{\bf $-$0.007 }\\
    W0P2& 1012.1695& 2.658600(022)& $-$0.13$\pm$0.19& $-$0.008  \\
    L7P2& 1016.4611& 2.658605(066)& $+$0.28$\pm$0.55& $+$0.028  \\
    L6P2& 1028.1059& 2.658599(023)& $-$0.18$\pm$0.19& $+$0.023  \\
 {\bf L5P2}&{\bf 1040.3672}&{\bf 2.658604(026)}&{\bf $+$0.23$\pm$0.22}&{\bf $+$0.019 }\\
 {\bf L4P2}&{\bf 1053.2842}&{\bf 2.658603(013)}&{\bf $+$0.14$\pm$0.11}&{\bf $+$0.013 }\\
  {\bf L3R2}&{\bf 1064.9948}&{\bf 2.658601(027)}&{\bf $-$0.00$\pm$0.22}&{\bf $+$0.010 }\\
 {\bf L2R2}&{\bf 1079.2254}&{\bf 2.658604(015)}&{\bf $+$0.21$\pm$0.13}&{\bf $+$0.004 }\\
 {\bf L2P2}&{\bf 1081.2659}&{\bf 2.658601(016)}&{\bf $-$0.00$\pm$0.13}&{\bf $+$0.002 }\\
    L0R2& 1110.1206& 2.658601(017)& $-$0.05$\pm$0.14& $-$0.010  \\
 {\bf L18R3}&{\bf 921.7302}&{\bf 2.658605(083)}&{\bf $+$0.29$\pm$0.68}&{\bf $+$0.052 }\\
 {\bf L17R3}&{\bf 928.4374}&{\bf 2.658591(071)}&{\bf $-$0.86$\pm$0.59}&{\bf $+$0.050 }\\
    W4P3& 934.7901& 2.658591(069)& $-$0.85$\pm$0.57& $+$0.023  \\
 {\bf L15R3}&{\bf 942.9642}&{\bf 2.658601(047)}&{\bf $-$0.01$\pm$0.39}&{\bf $+$0.048 }\\
 {\bf W3P3}&{\bf 951.6718}&{\bf 2.658603(039)}&{\bf $+$0.13$\pm$0.33}&{\bf $+$0.021 }\\
 {\bf L13R3}&{\bf 958.9466}&{\bf 2.658600(032)}&{\bf $-$0.07$\pm$0.27}&{\bf $+$0.044 }\\
    L13P3& 960.4506& 2.658607(048)& $+$0.51$\pm$0.40& $+$0.043  \\
 {\bf W2R3}&{\bf 966.7804}&{\bf 2.658608(032)}&{\bf $+$0.59$\pm$0.26}&{\bf $+$0.018 }\\
 {\bf L12R3}&{\bf 967.6770}&{\bf 2.658609(026)}&{\bf $+$0.64$\pm$0.22}&{\bf $+$0.037 }\\
    W2Q3& 969.0493& 2.658598(087)& $-$0.26$\pm$0.72& $+$0.011  \\
    L12P3& 969.0898& 2.658601(075)& $-$0.03$\pm$0.62& $+$0.040  \\
    W2P3& 970.5634& 2.658602(058)& $+$0.03$\pm$0.48& $+$0.010  \\
 {\bf L11P3}&{\bf 978.2180}&{\bf 2.658602(038)}&{\bf $+$0.05$\pm$0.32}&{\bf $+$0.038 }\\
    L10R3& 985.9628& 2.658607(047)& $+$0.43$\pm$0.39& $+$0.036  \\
 {\bf L9P3}&{\bf 997.8271}&{\bf 2.658595(033)}&{\bf $-$0.51$\pm$0.28}&{\bf $+$0.032 }\\
 {\bf W0R3}&{\bf 1010.1303}&{\bf 2.658602(028)}&{\bf $+$0.04$\pm$0.23}&{\bf $-$0.006 }\\
    W0Q3& 1012.6796& 2.658592(044)& $-$0.80$\pm$0.36& $-$0.009  \\
    W0P3& 1014.5043& 2.658602(028)& $+$0.04$\pm$0.23& $-$0.011  \\
    L7P3& 1019.5021& 2.658599(040)& $-$0.16$\pm$0.34& $+$0.025  \\
    L6P3& 1031.1927& 2.658595(035)& $-$0.54$\pm$0.29& $+$0.020  \\
 {\bf L5R3}&{\bf 1041.1588}&{\bf 2.658601(029)}&{\bf $-$0.04$\pm$0.24}&{\bf $+$0.018 }\\
 {\bf L4R3}&{\bf 1053.9761}&{\bf 2.658600(019)}&{\bf $-$0.09$\pm$0.16}&{\bf $+$0.013 }\\
 {\bf L4P3}&{\bf 1056.4714}&{\bf 2.658599(018)}&{\bf $-$0.17$\pm$0.15}&{\bf $+$0.011 }\\
 {\bf L3R3}&{\bf 1067.4786}&{\bf 2.658607(022)}&{\bf $+$0.49$\pm$0.18}&{\bf $+$0.007 }\\
    L3P3& 1070.1408& 2.658599(027)& $-$0.15$\pm$0.23& $+$0.005  \\
 {\bf L2P3}&{\bf 1084.5603}&{\bf 2.658606(020)}&{\bf $+$0.41$\pm$0.17}&{\bf $-$0.001 }\\
\hline
\end{tabular}
\end{center}
\label{fitting_res2}
\end{table*}


\end{document}